\documentclass[vecphys]{svmult}

\usepackage{makeidx}         % allows index generation
\usepackage{graphicx}        % standard LaTeX graphics tool
\usepackage{multicol}        % used for the two-column index
\usepackage[bottom]{footmisc}% places footnotes at page bottom
\usepackage{amssymb}
\usepackage{rotating}
\makeindex             % used for the subject index
\newcommand{\e}{{\rm e}}

\newcommand{\Ge}{\Gamma^{ex}}
\newcommand{\Ga}{\Gamma^{ad}}

\begin{document}

\title*{Self-organized surface nanopatterning by ion beam sputtering}


\author{Javier Mu\~noz-Garc\'{\i}a\inst{1}\and Luis V\'azquez\inst{2}\and Rodolfo
Cuerno\inst{1}\and Jos\'e A.\ S\'anchez-Garc\'{\i}a\inst{2}\and Mario Castro\inst{3}\and
Ra\'ul Gago\inst{4}}

\authorrunning{Mu\~noz-Garc\'{\i}a, V\'azquez, Cuerno, S\'anchez-Garc\'{\i}a, Castro, and Gago}
%\authorrunning{J Mu\~noz-Garc\'{\i}a \emph{et al.}}

\institute{Departamento de Matem\'aticas and Grupo Interdisciplinar de Sistemas
Complejos (GISC), Universidad Carlos III de Madrid, 28911 Legan\'es, Spain \and
Instituto de Ciencia de Materiales de Madrid (CSIC), 28049 Madrid, Spain \and
Escuela T\'ecnica Superior de Ingenier\'{\i}a and GISC, Universidad Pontificia
Comillas de Madrid, 28015 Madrid, Spain \and Centro de Micro-An\'alisis de
Materiales, Universidad Aut\'onoma de Madrid, 28049 Madrid, Spain.
\texttt{raul.gago@uam.es} }

\maketitle

To appear in Lecture Notes on Nanoscale Science and Technology, edited by Z. Wang (Springer, Heidelberg).

\section{Introduction}
\label{sec:intro}
% Always give a unique label
% and use \ref{<label>} for cross-references
% and \cite{<label>} for bibliographic references
% use \sectionmark{}
% to alter or adjust the section heading in the running head

Nanotechnology and Nanoscience are inducing a turning point in fields from
Condensed Matter Physics and Materials Science to Chemistry and Biology. This is
due to the new types of behaviors and properties displayed by nanostructures,
ranging across traditional disciplines \cite{Hodes_2007}. Interest in these
systems is also triggered by the possibility of characterizing them
morphologically by scanning probe microscopy techniques (SPM)
\cite{Binnig_1986} and by the advance in surface structure determination
methods. Within this more specific context, there is an increasing need for the
development of techniques and methods for the patterning of materials at the
nanoscale. Moreover, this goal must be achieved in an efficient, fast and
low-cost manner for eventual technological applications to be compatible with
mass-production. In addition, the processes involved should allow control over
the size, shape and composition of the nanostructures produced.

Within this trend, there is also a great interest in developing methods for producing confined
nanostructures, which would display quantum confinement effects, on surfaces. Of especial
interest would be those systems in which these surface nanostructures form an ordered
pattern.

Among the technological issues behind surface nanostructuring, we can highlight:

\begin{itemize}
\item New advances in magnetic storage technology \cite{Shen_2002}.

\item The production of surfaces and thin films with well controlled shape at
the nanoscale for applications in catalysis \cite{Zaera_2002}.

\item The design of nanoelectronic devices based on the use of quantum devices
\cite{Amirtharaj_2002}, for example for developing multispectral detector
arrays by exploring novel detection techniques, nanopatterning and control of
the production of self-assembled quantum structures.

\item The possibility of functionalization of these surface nanostructures that
allows the selective attachment of specific molecules \cite{Kasemo_2002}.

\item The design of nanosensors based on surface-enhanced Raman scattering
effects (SERS) or localized surface plasmon resonance \cite {Yonzon_2005}.

\item The use of these nanopatterned surfaces as templates to transfer these
patterns to highly-functional surfaces \cite{Azzaroni_2004} that can not be
patterned directly. Also, the use of these patterns as templates can be applied
to reduce fabrication steps or to increase productivity. \end{itemize}

There are different approaches for surface nanostructuring. Among the so-called
{\itshape top-down} methods we can mention lithographic techniques (nanoimprint
lithography \cite {Guo_2007}, nanosphere lithographic techniques \cite
{Yonzon_2005}, soft lithographic methods \cite{Xia_1997} and focused ion beam
(FIB) techniques \cite{Moore_1997}). Other approaches are based on the use of
SPM to induce nanostructures on a surface through different processes such as
tip induced oxidation \cite{Calleja_1999}, tip induced e-beam lithography
\cite{Soh_2001}, dip-pen nanolithography \cite{Mirkin_2001}, the application of
strong electric fields between tip and sample \cite{Li_1989}, and the mere use
of the probe as a stylus or pen to write at the nanoscale on the sample surface
\cite{Sohn_1995}. However, these methods present different limitations, such as
proximity effects, low resolution or the need of parallel processing because of
the (relatively) small processed area. In the case of SPM-based techniques,
mainly Scanning Tunneling Microscopy (STM) and Atomic Force Microscopy (AFM),
the use of an array of tips aims to compensate for this last limitation.

Potential alternatives to overcome the limitations of the {\itshape top-down}
approach are provided by {\itshape bottom-up} methods, which are mainly based
on self-organized processes. In this field, most efforts have been
traditionally focused on the production of self-organized nanostructures
occurring in semiconductor heterostructure growth \cite{Stangl_2004,
Teichert_2001} in which strain relief mechanisms take place. This simpler
method seems to be a highly cost-efficient route towards large-scale arrays of
nanostructures, although it presents some disadvantages such as enabling less
control on structure size and shape than other methods, and the need to work
under ultra-high vacuum conditions.

In recent years, special interest is being paid to the study of self-organized
nanopattern formation on surfaces by {\itshape ion beam sputtering} (IBS) techniques
\cite{Facsko_1999}. In general, two types of surface nanostructures can be induced by IBS:
(a) nano-ripples and (b) nanodots. In both cases, the pattern formed by these nanostructures
can have dimensions ranging from a few up to hundreds of nanometers. These patterns can be
produced on different materials, amorphous or crystalline, in just a few minutes and over
areas of several square millimeters. The diversity of materials processed and the similar
morphologies obtained indicate the universality of the process. In addition, it allows the
eventual control of the induced nanostructures by changing the sputtering parameters, such as
ion energy, dose, substrate temperature, ion incidence geometry, etc. Moreover, it can be used
to produce functional surfaces and isolated structures \cite{Valbusa_2002}. Thus, IBS
becomes a versatile, fast and cheap technique for surface nanopattering.

\begin{figure}
\center
\includegraphics[height=6cm]{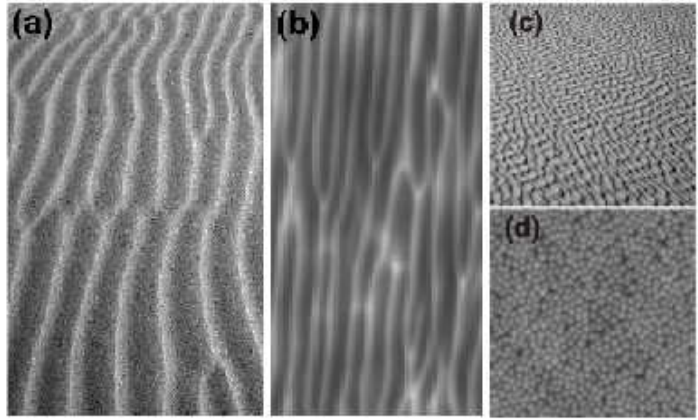}
\caption{(a) Ripples on a sand dune in Morocco. Photograph courtesy of J.
Rodr\'{\i}guez and E. Blesa. (b) $3.7 \times 6.7$ $ \mu \rm{m}^2$
top view AFM image of a Si surface immersed in argon plasma. (c) ``Dots'' on a
sand dune in New Mexico, USA. Copyright Bruce Molnia, Terra Photographics.
Image Courtesy Earth Science World Image Bank
http://www.earthscienceworld.org/images. (d) $1 \times
1$ $\mu \rm{m}^2$ top view AFM image of a GaSb surface irradiated by 0.7 keV Ar$^+$ ions
under normal incidence.} \label{Fig0_0}
\end{figure}

The ability of IBS to induce submicro-structures on surfaces was reported more
than forty years ago by Navez et al. \cite{Navez_1962}. In that work, they
reported the production of periodic nanoripples on glass surfaces. In Fig.\
\ref{Fig0_0}b we show ripples obtained on a silicon surface immersed in an
argon plasma. These ripples indeed remind us of those formed on sand dunes by
the wind or on the sand bed close to the water edge by the water flow (Fig.
\ref{Fig0_0}a). As we will show below, this similarity is more than merely
visual as the theories dealing with both types of (macro and nano) structures
share many conceptual aspects. In the pioneering work of Navez and coworkers the
authors did report important morphological observations, such as the change of the ripple orientation when the ion incidence polar
angle is larger than a threshold value. Later, similar findings were
reported for other materials, that have been reviewed elsewhere
\cite{Carter_2001, Makeev_2002, Murty_2002, Valbusa_2002}. Usually, the length
scales involved in these patterns lie in the sub-micrometer scale, rather than in the nanometer scale. Moreover, the traditional field for ripple applications
had been that of optical gratings \cite{Johnson_1979}. From the theoretical
point of view, the first advances for understanding IBS nanostructuring was
made by Sigmund \cite{Sigmund_1973} as he showed that local surface minima
should be eroded at a faster rate than local maxima (i.e. the sputtering rate
depends on the local surface curvature), leading to a surface instability,
which is the origin of the nanostructuring process. Based on this work, Bradley
and Harper proposed later the first continuum model describing ripple formation
\cite{Bradley_1988}. In recent years, other continuum models have been proposed
\cite{Cuerno_1995, Makeev_1997, Park_1999}. These models account for different
experimental behaviors, such as the presence of absence of saturation for the
ripple amplitude, ripple orientation, ripple dynamics as well as the existence
or not of kinetic roughening.

However, a clear turning-point occurred when in 1999 Fackso and coworkers
\cite{Facsko_1999} reported on the IBS production of GaSb nanodot patterns,
which also display short-range order. This work shifted the focus of the
research to the nanometer scale. The evident technological implications
\cite{Chen_2002, Lindner_2000} of the possibility to induce nanostructures on
semiconductor surfaces on a relatively large area (several square millimeters)
and in just a couple of minutes motivated further interest in these processes.
Thus, nanodot production has been reported in different materials such as InP
\cite{Frost_2000}, Si \cite {Gago_2001}, Ge \cite{Ziberi_2006}, etc. With
respect to the theoretical understanding of the nanodot IBS production, the
theories must share most of the concepts with those developed for IBS induced
ripples. An example is shown in Figures \ref{Fig0_0}c and d, in which ``dotted'' structures formed both on sand dunes and IBS nanodot structures are
displayed. Once more, this similarity appears that, rather than being just visual,
is, in fact, deeper as will be showed in Sec.\ \ref{coupled}.

The production of dot patterns has changed the focus of the theoretical
investigations. Now, for dot patterns the interest is more directed towards dot size
and pattern order because of their  technological implications. In the first case, it is
interesting to know how the dot size changes with the different IBS parameters,
specially ion energy, target temperature and sputtering time (i.e., fluence).
In the second case, it is necessary to know under which conditions the pattern
order is enhanced. Thus, issues such as pattern wavelength coarsening and order
enhancement have become relevant. Accordingly, new theoretical descriptions have been proposed
\cite{Castro_2005,Chen_2005,Facsko_2004,Kahng_2001, Makeev_2002, Munoz_2005, Vogel_2005}.

The aim of this chapter is to give an  overview of ripple and dot IBS
nanopatterning that focus specially on those issues that remain open or, at
least, ambiguous. This will be more evident for the case of nanodot patterns on amorphizable targets
due, probably, to their relative novelty. In fact, we will treat specially the
case of this type of patterns since previous reviews were devoted mainly either
to ripple patterns   \cite{Carter_2001, Makeev_2002} or to the case of metals
\cite{Valbusa_2002}. Stress will be paid to review the observed behaviors and
the predictions of the main existing continuum models. When possible, we will
contrast the experimental data with the theoretical predictions.

The chapter will be divided as follows. After  this introduction and for the sake of completeness, we will
present the basics of the Physics behind the ion sputtering process.
Then, we will give an overview of experimental findings on surface
nanopatterning by IBS on amorphous materials and single-crystal semiconductors.
We will review briefly the results for the special case of single-crystal metal surfaces as
they were already extensively reviewed by Valbusa and coworkers
\cite{Valbusa_2002}. In the next section a review of the theoretical studies of
these processes will be presented, with special emphasis on the various
continuum models proposed so far. In the last section of the chapter we will
propose possible applications of these nanopatterns and, finally, we will state
the issues that still remain open, from our point of view, both theoretically
and experimentally.

\section{Fundamentals of ion sputtering}

\subsection{Introduction to ion sputtering}

Physical sputtering is defined as the removal of atoms from a solid surface due to energetic
particle bombardment. This phenomenon was first reported by Grove \cite{Grove_1852} in
1852, although others had probably noted the effect while studying glow discharges. Grove
studied the discharge generated by the tip of a wire positioned close to a polished silver
surface and noted a ring-like deposit onto the silver surface when it was acting as an anode in
the circuit. However, it was not until the early 1900's when the effect was attributed to the
impingement of positive ions accelerated towards the cathode by the electrical field
\cite{Stark_1908}. The sputtering process resembles the macroscopic sandblasting process
but with ionized atoms instead of sand grains as bombarding species. In fact, the erosion
efficiency with the incident angle of the abrasive particles \cite{Finnie_1995} is analogous to
that observed in ion sputtering experiments.

The sputtering process is important both from a fundamental as well as a practical point of
view. On the one hand, the understanding of the process is relevant to describe the basic
interactions of ions with matter. On the other hand, as shown below, the process has found a
broad range of applications. The improvement in experimental methods, as well as in the
theoretical description of the process, has promoted the rapid maturation of the field during
the last decades.

\subsection{ Applications of ion sputtering }

The applications of ion sputtering to Surface Science and Technology are very diverse
\cite{Murty_2002}. First, the removal of atoms from the surface can be used as an effective
method for {\itshape surface cleaning} \cite{Taglauer_1990}. This may be aimed, for
example, at removing the undesirable contamination layer present on a surface (mostly
containing oxygen and carbon impurities), which may affect the intrinsic electrical or
optical properties. This cleaning process may be crucial for the optimal
performance of a semiconductor device \cite{Schubert_2005}.

The sputtered material from the target can also be transferred to a substrate for {\itshape thin
film deposition} (sputter deposition methods) \cite{Rossnagel_2003}. This application of
sputtering was first addressed by Pl\"ucker in 1958 \cite{Pluecker_1958} after he observed the
formation of a platinum film inside a discharge tube with a metallic mirror-like appearance.
Sputter deposition is a non-equilibrium process where material ejection occurs at
hyperthermal energies (in the few eV range) \cite{Stuart_1969}. This energy range is much
higher than in thermal evaporation methods, leading to coatings with better properties (higher
density, etc.). The sputtering phenomenon is the base of many industrial coating activities,
being scalable from small pieces up to processing of large areas (it is commonly used to coat
structural window panels). It is relevant for the development of magnetron sputtering techniques
\cite{Rossnagel_2003}, where high erosion/growth rates are obtained by confining the plasma
discharge close to the cathode through a magnetic field.

The ejected particles during sputtering can also be identified for
compositional analysis of the target material. This technique is known as
{\itshape secondary ion mass spectrometry} (SIMS) \cite{Zalm_1994} and is based
on the principle that a small fraction of the sputtered particles are ionized.
These particles are guided and detected by means of a mass spectrometer and
provides very high depth resolution (in the monolayer range) and elemental
sensitivity down to parts per billion (ppb).

The progressive etching of the target material through sputtering finds usage
for {\itshape sample thinning}. Such a process is often incorporated into
specimen preparation for transmission electron microscopy (TEM) where, after
mechanical milling, grazing incidence ions impinge onto the remaining part of
the sample until the desirable electron transparency is achieved (a few
hundreds of nm for the energies commonly used for imaging). The progressive
erosion of the surface is also an extended route for {\itshape depth profiling}
with surface analytical techniques by a sequence of sputtering and analysis
steps \cite{Oechsner_1984}. In this case, high depth-resolution can be achieved
by erosion depths of a few monolayers but, as may occur in SIMS, by-products of
ion bombardment such as atomic intermixing or preferential sputtering
deteriorate the resolution of the analysis.

The application of material etching through sputtering is more extended in the
fabrication of {\itshape integrated circuit devices}. Here, pattern or feature
formation taking part in implementing the design of the device is imprinted
onto the substrate in several steps of ion beam etching or ``drilling''. This
submicron architecture is defined by means of lithography, masks or FIB
methods.

Finally, as mentioned in the Introduction, sputtering can be used to alter the
{\itshape surface morphology} due to its dependence on the local surface
curvature \cite{Valbusa_2002}. In this case, depending on the conditions (angle
of incidence, energy, etc.), it can induce either surface {\itshape smoothening} or
{\itshape roughening}. These competing processes may be used to design the
surface geometry required for certain applications. For example, smooth
surfaces are needed in optical components such as mirrors or lenses
\cite{Frost_2004} or for tribological surfaces. For this purpose, sputtering
should preferentially erode prominent topographies by, for example, grazing
incidence bombardment, or promote ion-induced surface diffusion processes. On
the contrary, roughening of a surface may enhance, for example, the surface
activity (more effective surface is available), catalytic or other chemical
processes, as well as control the biological response of the surface. Also, the
competition between roughening and diffusion can lead to the evolution of
correlated features and induce the spontaneous formation of a regular pattern
\cite{Valbusa_2002}. This {\itshape self-organized pattern} formation is the
main subject of this chapter and will be described in
detail in subsequent sections.

\subsection{Quantification of the sputtering process}

The sputtering process depends on several parameters such as the nature of the target, type of
ion, ion energy, incidence angle, ion flux (rate of incoming ions towards the target per unit
area) and total fluence (or dose). For quantification of the ion-target interaction as a function
of the different parameters, early investigations focused on the average number of target
atoms ejected per incident ion. This magnitude is defined as the total sputtering yield
({\itshape S}) or, if measured in terms of a specific energy or angular interval, it is described
as differential yield. However, there are other observables such as the effect on surface
morphology or ion induced lattice damage that can be used for quantification of the sputtering
effect. The understanding of the latter effects has only been possible with the recent
development of proper characterization tools for surface imaging or structure determination.

For multicomponent targets, the yield for each
particular element should be taken into account. When the difference in {\itshape S} among species in the substrate is large, a {\itshape preferential sputtering} of the target elements
with greater {\itshape S} occurs. This implies that the composition of the sputtered particles
differs from that of the multi-component target and, as a consequence, induces an enrichment
of the low-{\itshape S} component at the outermost layers of the target.

\subsection{Experimental measurements of the sputtering yield}

\begin{figure}
\center
\includegraphics[height=6cm, width=11cm]{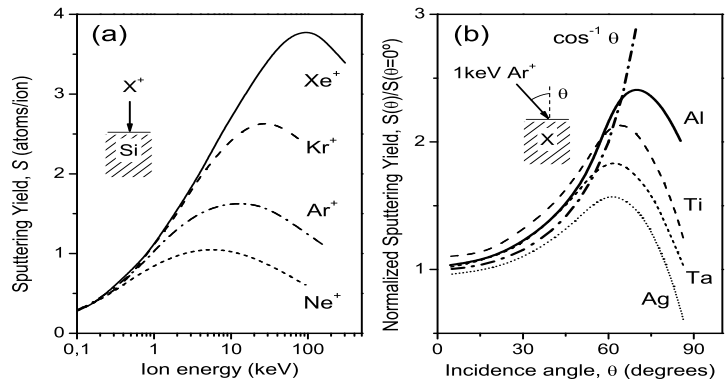}
\caption{(a) The observed energy dependence of the sputtering yield, {\itshape
S}, of Si for normal incidence bombardment with different ions (adapted from
Ref. \cite{Zalm_1986}). (b) Normalized sputtering yield as a function of the
polar angle of incidence for 1 keV Ar$^{+}$ on different materials (adapted
from Ref. \cite{Oechsner_1975}).} \label{Fig2_1}
\end{figure}

For a given ion and energy, there is no direct relation between the atomic
number of the target element and the resulting {\itshape S}
\cite{Andersen_1981}. However, there are general experimental observations on
the behavior of {\itshape S} as a function of other sputtering parameters,
mainly ion flux, energy and incidence angle. An important experimental
observation is that ion erosion occurs even in the limit of low ion flux and
that {\itshape S} is independent of the incoming flux. In addition, for a given
projectile, {\itshape S} varies rather smoothly with incident ion energy until
reaching a broad maximum (in the 10-100 keV range) and, then, gradually
dropping to zero for very high energies (in the MeV range) \cite{Zalm_1986}.
This behavior is found for almost all projectile-target combinations. The
variation of {\itshape S} with the ion incidence angle, $\theta$, which is
measured with respect to the surface normal, is considerable and increases from
normal to oblique incidence until reaching a maximum around 60-70$^{\circ}$, and
then sharply dropping for glancing incidence angles \cite{Oechsner_1975}. The
increase of {\itshape S} with $\theta$ is explained by the higher energy
deposited in the surface region, a simple geometrical projection giving a
$\cos^{-1}\,\theta$ relation. The reduction of {\itshape S} at glancing
incidence is due to the increase in the reflection of the incident ions
\cite{Witcomb_1974}. The variation of the sputtering yield with the angle
(relation between the maximum sputtering yield and the value at normal
incidence) depends strongly on the ion/target combination, being larger for
light ions. The general trends of {\itshape S} as a function of ion energy and
incidence angle are illustrated in Fig. \ref{Fig2_1}.

It has been reported that {\itshape S} can considerably increase with ion fluence
\cite{Andersen_1981}. The steady state is reached only after the target has been eroded to a
depth of the order of the projected ion range (maximum penetration depth of the ions in the
direction normal to the surface). Obviously, this observation could be explained by the ion
induced modification of the subsurface due to ion implantation. However, the analysis of the
ejected particles also suggests that the yield increase may come from a mechanism of gas trap
release \cite{Wittmaack_1984}.

Regarding other parameters, it is generally thought that the temperature, apart from small
changes in the sublimation energy, does not influence sputtering behavior significantly.
Regarding crystallinity, the value of {\itshape S} may differ significantly as a function of
energy among single crystal, polycrystalline or amorphous targets of the same element. In
single crystals, the yield may also depend on the surface orientation. A satisfactory
explanation for these behaviors relies on the concept of channeling (the collision probability is
reduced if the ion is guided through a crystal channel), the contribution of this effect
being less important in the low-energy range (below a few keV).

In the previous description, energetic monoatomic particles impinging on the
target surface have been considered. This is typically the case of sputtering
experiments with noble gas ions. However, the sputtering process can be further
complicated when the incoming ions react chemically with the target species or
by bombardment with polyatomic or molecular ions. In the first case, {\itshape
chemical sputtering} may also contribute to the increase of {\itshape S} by
formation of volatile compound products \cite{Roth_1983}. An example of this
process has been observed by the increase of {\itshape S} during bombardment of
Si with F or Cl ions as compared with the values obtained for noble gas ions
bombardment \cite{Tachi_1986}. Conversely, when a non volatile compound forms
by reaction of the incoming ion and the target element (for example, SiO
formation by O bombardment of Si), the value of {\itshape S} is generally
reduced. In the case of polyatomic or molecular ions, it should be considered
that the molecules fragment upon impact and {\itshape S} should account for the
relative contribution of the individual constituents. In this case, a fair
approximation considers that the individual ions have the same velocity as the
incoming molecule and transfer their energy independently to the surface.

Regarding the emitted particles, the spatial distribution of sputtered atoms is described
generally as a cosine distribution. However, as a function of energy, this distribution can
change into an under-cosine to over-cosine for low or high ion energy, respectively, or can be
somehow more directional if the incident particle is inclined at a high angle with respect to
the surface normal \cite{Okutani_1980}. Usually, the direction of the sputtered cone is
opposite to the incident direction and is thus called a forward-peaked distribution.

Because of the energetic nature of the sputtering process, emitted atoms tend
to have energies $E$ well exceeding the thermal energy that would be consistent
with the target temperature. The energy of the sputtered particles is typically
peaked at a few eV and has a long tail that falls as $E^{-2}$ and goes up to
the order of the incident ion energy \cite{Stuart_1969}. The peak or average
kinetic energy depend strongly on the mass of the target atom, as well as on
the energy and mass of the incident particle.

\subsection{Theory of sputtering}\label{Theory_sputtering}

The many-body considerations of the sputtering process pose a formidable challenge to
theoreticians. The first models to explain the sputtering process at atomic scale were
based in terms of evaporation from a hot spot, i.e., ``thermal spike'' \cite{Stark_1908}.
Although the angular distribution of sputtered atoms follows the distribution from a liquid
phase and the presence of thermal spikes has been described by Molecular Dynamics (MD)
simulations \cite{DiazdelaRubia_1987}, this mechanism gives very low {\itshape S} values
due to the short-lived nature of the spikes \cite{DiazdelaRubia_1987} and it can not account
for the flux independence in the sputtering yield.

The theoretical understanding of ion sputtering processes was considerably improved by the
introduction of collision cascades. This advance was achieved through a better description of
the ion-solid interaction, which was triggered by the building of particle accelerators in the
1950-60's. When an energetic ion penetrates the surface of a solid target, it travels through the
solid losing its energy through collisions with the nuclei and electrons of the target atoms.
The range (maximum penetration depth) of the ions can be determined to a good
approximation considering the nuclear ($S_n=[dE/dx]_n$) and electronic ($S_e=[dE/dx]_e$)
stopping independently \cite{Nastasi_1996}. Except for light ions (H, He), the nuclear
stopping dominates at low energies (below a few keV) since the ions get closer to the atom. If
the energy transfer to the target atoms is larger than the threshold given by the displacement
energy, lattice displacements take place, which precisely occurs mostly when nuclear stopping
dominates the energy losses.

The concept of nuclear stopping was thus used by Sigmund \cite{Sigmund_1969} to
develop a linear cascade model for sputtering. Here, the projectile shares its
kinetic energy with target atoms initially at rest in a series of random binary
collisions, a process in which fast recoils are created. The recoils knock on
other target atoms creating a collision cascade in which more atoms are in
motion, but progressively with slower speed. If the energy is deposited in the
near surface region, atoms located close to the surface may gain sufficient
energy and the appropriate momentum to leave the surface and be sputtered away.
The collision cascade occurs in a time interval of picoseconds after the
impact, when the recoil energies at the edges of the cascade have come below
the threshold of displacement energy (of the order of a few tens of eV). The
cascade is damped out by energy dissipation through several mechanisms, such as
creation of phonons.

By assuming a Gaussian
energy distribution centered at the ion range, an analytical expression for the
sputtering yield, {\itshape S}, can be deduced as follows

\begin{equation}
S= \alpha N S_{n}(E) / E_{B} \label{eq1}
\end{equation}

Here, $N$ is the target atom density, $E_{B}$ is the binding energy of a
surface atom and $\alpha$ contains material and geometric parameters. The
product $NS_{n}(E)$ represents the energy loss per unit length as the ion
travels in the target. Sigmund's theory accounts satisfactorily for many of the
observations in amorphous or polycrystalline elemental target materials, like
the angular and energy dependences of {\itshape S}, as well as the angular and
energy distributions of the sputtered particles. However, it is a continuum
theory and, therefore, it does not address inherent inhomogeneity of materials,
it can not be applied to crystalline structures since it neglects channeling
effects and, finally, it does not consider other relevant issues such as the
starting surface morphology. In addition, purely repulsive forces between the
ion and the target atoms are only valid for ion energies above a few keV.

The correlation of sputtering with the surface morphology is a crucial issue for the content of
this chapter. The effect of micro-rough surfaces was already posed by Sigmund in 1973
\cite{Sigmund_1973}. Differences in energy deposition, which are induced by topographical features, may
cause a significant reduction of the local sputtering yield at local maxima
and, inversely, an increase at local morphological minima. As a consequence, small
irregularities on a relatively smooth surface may result enhanced by ion bombardment. This
implies that microscopically flat surfaces are unstable under high-dose ion bombardment
unless atom migration acts as a dominating smoothing effect. Conversely, sharp cones appear
to be surprisingly stable under bombardment leading to surface nanostructuration by IBS. This prediction is the key-stone for the
theoretical understanding of roughening or pattern formation induced by IBS, as will be
shown in following sections.

Regarding the fundamental description of the sputtering process, the shortcomings of the
linear theory have been addressed by the development of computer simulations, providing a
direct view of the collision events and incorporating the role of intrinsic material properties.
In this case, discrete systems are considered where the ions and target atoms interact in a
deterministic way, like in MD simulations, or in a random nature, like in the case of binary
collision approximation (BCA). In these models, the interaction is driven by repulsion forces
that determine the ion trajectory in the solid. In the BCA method the interaction is treated
sequentially until the energy of the projectile and recoils are thermalized. This method was
implemented in the TRIM code \cite{Ziegler_1985,Ziegler_2006}, where a screened Coulomb potential is
considered for faster computation. The BCA is valid when the individual collisions are
separated in space (fulfilled only for energy above a few keV) and fails when there is
cascade overlap, as occurs for low ion energies. In this regime, attractive forces among the
particles become more relevant. Even at high energy, the BCA does not provide good
description of the collision cascade as the particles slow down.

In the low-energy range, a proper description should consider the simultaneous
movement of the projectile and recoils. Since typical processes during
sputtering occur in the picoseconds regime, they can be properly described by
MD simulations. In this way, experimental observations of crystalline targets
have been successfully reproduced, such as the yield as a function of energy or
the angular distribution of sputtered atoms \cite{Harrison_1988}. However, MD
calculations are time consuming in terms of computation and the interaction
potential may not be accurate. In addition, the rather small length scales as
well the short temporal window analyzed in MD simulations preclude a dynamical
description of the sputtering process by MD and make necessary the
implementation of Monte Carlo simulations.

\subsection{Experimental considerations for ion sputtering}

There are many considerations that should be taken into account for precise and
reproducible sputtering experiments. The large number of parameters involved
and the variety of experimental configurations may imply a large dispersion of
results that makes it difficult to present an unified picture. As an
illustration of this complexity, most experiments for the determination of
{\itshape S} before the 1970's are not reliable since they were not done under
sufficient vacuum conditions \cite{Murty_2002}. Most experiments of surface
nanostructuration by ion beam sputtering are done by means of low-energy broad
ion beam guns (Kaufman, cold cathode, hollow cathode, etc.), medium-energy ions
from ion implanters, focused ion beams (FIB) or directly by plasma immersion of
the target. In most cases, ion beams are generated by a glow discharge and then
extracted by means of proper beam optics. The nature of the discharge as well
as the beam optics will define the range of ion flux and the ion energy that
can be achieved. In all cases, several considerations should be taken into
account such as energy spread, ion flux, vacuum environment or the eventual
presence of multiple charge or mass ions. Moreover, a precise knowledge
of the irradiation dose is required (direct ion current measurement with a
Faraday Cup, collecting the sputtered material from the target or by measuring
the etched volume by masking the target). Other relevant issues may be the
unintentional temperature increase induced during the ion bombardment (target
cooling is desirable) or the control over the starting material conditions
(surface morphology, electrical properties, size, etc.) since small differences
among experiments may lead to inconsistent results.

Low-energy ion beams are difficult to handle in many cases due to space-charge
blow-up. This is normally solved by neutralizing the beam with thermal
electrons from a hot filament, by placing the ion source very close to the
target or decelerating the energetic ions in front of the target. The presence
of a neutralizer may also be necessary for the processing of insulating
materials. All these aspects may affect the beam divergence or beam focusing,
which has been shown to play a relevant role in surface nanostructuring
\cite{Cuenat_2002,Ziberi_2004}.

Finally, broad ion beams are normally used to incorporate large-area
processing. However, beam scanning in the case of small diameter beams, as in
the case of FIB or ion implanter, can also result in the formation of similar
patterns. In this case, the beam scanning may incorporate new effects in the
nanostructuring process \cite{Chini_2002,Cuenat_2002}, mainly driven by heating
effects on the target.

\section{Experimental observations of surface patterning by ion beam sputtering (IBS)}\label{sec:experiments}

As mentioned above, in the last years a relatively large amount of experimental
works have appeared on the production of nanostructures and nanopatterns by
IBS. Roughly, the experimental reports can be divided into two groups: (a)
those that merely report on the production of some IBS nanopattern, and (b)
those works in which some systematic pattern analysis is done, related with the dependence on some of the experimental variables,. In this section we
will try to summarize the most relevant results of the second type of works,
with special emphasis on these open issues described in the Introduction. We
will distinguish in this analysis two main categories depending on the nature
of target material. Thus, we will describe separately the experimental findings
for amorphous or amorphizable materials, such as single-crystalline
semiconductors, and for crystalline materials that are not amorphized by the
ion beam (i.e., metals). This division is grounded on the fact that continuum
models are mainly based on the random distribution of the target atoms, which is
the case for the first type of substrates. Although we will see that both types
of materials share many characteristics when they are subjected to IBS, the
physical processes that actually take place can naturally lead to important
differences in the pattern formation. An excellent recent review of IBS patterning
on metal surfaces can be found in the previous work by Valbusa and coworkers
\cite{Valbusa_2002}, the experimental section will be mainly focused on IBS
nanostructuring on amorphous and amorphizable materials.

\subsection{IBS patterning formation on amorphous or amorphizable surfaces}
\label{patterns_exp}

Under this label, we include all the target materials that are amorphous (i.e.,
glass). In addition, we consider crystalline targets whose atom layers close to
the surface become amorphous by the action of the impinging ions (i.e., Si,
GaSb, InP, etc). It is for these materials that Sigmund's theory
\cite{Sigmund_1969} of ion beam sputtering is strictly applicable.

In order to present the different results on the IBS nanostructuring of these
materials, we will divide them into two main groups: (a) nanoripple patterns and
(b) nanodot patterns. This is so because usually they are obtained under
different experimental conditions and because the work on ripple patterning
dates back to the 1960's whereas that performed on dot
patterns is only 8 years old.

\subsubsection{Ripple formation by off-normal ion incidence}

Ripple structures are induced by IBS under off-normal conditions. However, this
happens usually within a limited window of angles (for the Si case see
\cite{Carter_1996,Wittmaack_1990}). The ripple pattern has been reported in a huge amount
of materials. To mention just a few examples, we can quote observations on glass
\cite{Navez_1962}, SiO$_2$ \cite{Flamm_2001, Mayer_1994}, Si \cite{Carter_1996,
Datta_2004, Erlebacher_1999, Gago_2002,Habenicht_2002,Vajo_1996}, Ge \cite{Chason_1994,
Ziberi_2006}, GaAs \cite{Datta_2002, Karen_1991,Maclaren_1992}, InP
\cite{Maclaren_1992,Demanet_1995}, highly oriented pyrolytic graphite (HOPG)
\cite{Habenicht_2001}, and diamond \cite{Datta_2001}.

\hfill
\paragraph{Ripple morphology}

\begin{figure}
\center
\includegraphics[height=3cm]{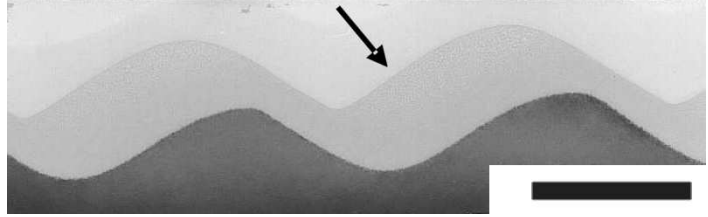}
\caption{(a) Cross-sectional TEM image of a Si surface sputtered with 120 keV
Ar$^+$ beam. The arrow indicates the ion beam direction. The horizontal bar corresponds
to 500 nm. Micrograph courtesy of T K Chini.}\label{Fig_TEM-Chini}
\end{figure}

The ripple morphology is usually characterized by AFM, STM and TEM techniques
because they provide enough resolution to reveal the ripple topography at the
nanometer scale. When the ripple dimensions are relatively large it can be
assessed also by Scanning Electron Microscopy (SEM). Whereas from SPM techniques we
can obtain morphological data such as ripple amplitude, surface roughness and
even pattern ordering (see below), with TEM we can access structural
properties such as the thickness of the ion-induced amorphized layer as well
any possible structure (for instance, bubbles) induced by the incorporation of
the bombarding ions into the target \cite{Chini_2003}. One example of the
information obtained by TEM is shown in Fig. \ref{Fig_TEM-Chini}. Here, we can
observe that the amorphized layer thickness is quite different if the ripple
slope faces the incoming beam or not, being larger for the face that is in front of
the ion beam. Besides, some cavities corresponding to bubble-like features, due
to the Ar$^+$ incorporation to the target are clearly visible on the front
slope.

\begin{figure}[!htmb]
\begin{center}
\begin{minipage}{0.37\linewidth}
\begin{center}
 \includegraphics[width=\linewidth]{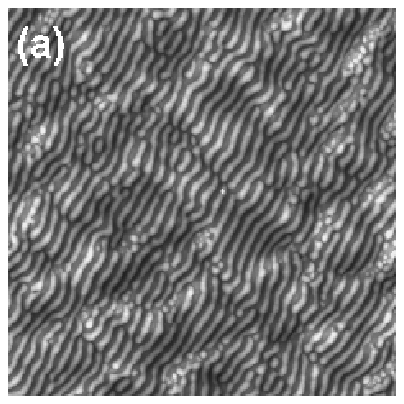}
\end{center}
\end{minipage}\hspace*{0.02\linewidth}
\begin{minipage}{0.55\linewidth}
\begin{center}
 \includegraphics[width=\linewidth]{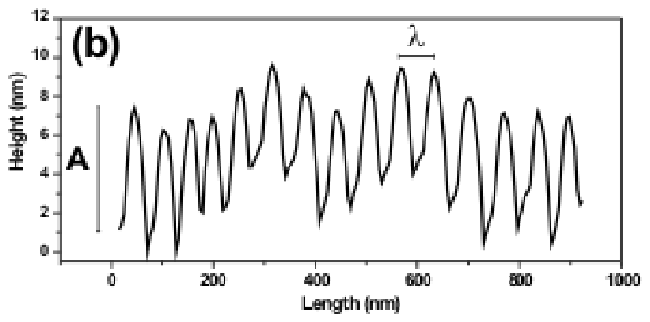}
\end{center}
\end{minipage}
\caption{(a) $2\times 2$ $\mu \rm{m}^{2}$ AFM image of a Si surface bombarded
by an Ar$^+$ 1 keV ion beam at $\theta=10^{\circ}$  for a fluence of $9 \times
10^{17}$ ions cm$^{-2}$. (b) Cross section taken perpendicularly to the ripples
in which $\lambda$ and $A$ are indicated.} \label{Fig3_1}
\end{center}
\end{figure}

In contrast, the data typically obtained by SPM-techniques are displayed in
Fig.\ \ref{Fig3_1}. In this figure we show a typical ripple pattern obtained on
a silicon surface by Ar$^+$ ion bombardment at $\theta=10^{\circ}$. In the
cross-section we define two magnitudes that are usually studied, namely, the
typical wavelength, $\lambda$, of the pattern and the amplitude of the ripple
structures, $A$. In particular, the analysis of the former is very interesting as
different models predict different behaviors of $\lambda$, for instance, with
the ion fluence. Thus, some models imply that $\lambda$ increases with
sputtering time, that is, that the pattern coarsens with time. Therefore, the
eventual coarsening of $\lambda$ is an object of study both experimentally
and theoretically.

%%%%%%%%%%%%%%Figure%%%%%%%%%%%%%%%%%%%%%%%
%%%%%%%%%%%%%%%%%%%%%%%%%%%%%%%%%%%%%%%%%%%%%%%%%%%%

\hfill
\paragraph{Ripple orientation}

\begin{figure}
\center
\includegraphics[height=4cm]{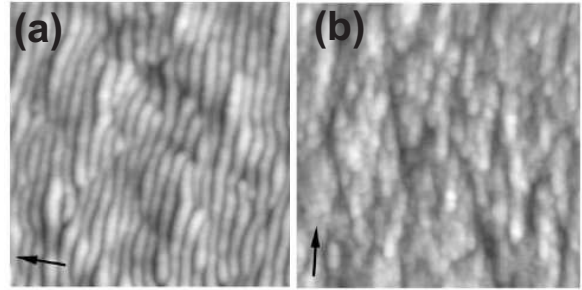}
\caption{(a) $1\times 1$ $\mu \rm{m}^{2}$ AFM image of a fused silica surface
bombarded by an Ar$^+$ beam with 0.8 keV with $\Phi = 400$ $\mu$A cm$^{-2}$:
(a) for 20 minutes at $\theta = 60^{\circ}$ and (b) for 60 minutes at $\theta =
80^{\circ}$. The dark arrow indicates the projection of the ion beam onto the
target surface. Figure reprinted with permission from \cite{Flamm_2001}.}
\label{Fig3_2}
\end{figure}

The dependence of the ripple orientation with
$\theta$ is one of the most relevant morphological properties of ripple formation by IBS. There is a critical
angle, $\theta_{c}$, such that for $\theta < \theta_{c}$ the ripples are
perpendicular to the beam direction while for $\theta > \theta_{c}$ the ripples
run parallel to it. This fact, which was already observed on glass in the seminal work by Navez et al.
\cite{Navez_1962}, is a classic behavior that was explained already by the theory proposed by Bradley and
Harper \cite{Bradley_1988}. This behavior has been observed for many target materials such as
glass, SiO$_2$ \cite{Flamm_2001} and HOPG \cite{Habenicht_1999}. One of
these examples is shown in Fig.\ \ref{Fig3_2} for the case of fused silica.

\hfill
\paragraph{Ripple pattern: dependence on ion energy and type}

\begin{itemize}
\item \emph{Ion type}--- Ripple patterns have been produced by bombarding the target surface by
different ions. The most frequently used species is Ar$^+$ due to its low cost, inertness
and relatively high mass. Among the noble gases, Kr$^+$ and Xe$^+$ have also
been employed \cite{Ziberi_2005}. Also, ripple patterns have been induced using
beams of Cs$^+$ \cite{Maclaren_1992}, Ga$^+$ \cite{Habenicht_2002}, O$_2^+$
\cite{Liu_2001, Smirnov_1999} and N$_2^+$ ions \cite{Smirnov_1999}. The two
last cases imply, as explained before, that reactive sputtering effects can
take place. Therefore, ripple formation by IBS is quite general a process virtually independent of the target materials and bombarding ions.
\\
\item \emph{Ion energy}--- The study of the dependence of the pattern wavelength with the ion energy, $E$, is
quite interesting because it can be used to further check the consistency of the experimental erosion system with the
assumptions of Sigmund's theory \cite{Sigmund_1969} on which the different continuum models proposed so far are based.

Regarding the energy of the ions,
the following distinction is usually made: (a) low-energy range, which is
normally applied to ion energies smaller than 2-3 keV, and (b) medium-energy
range, which is applied to quite a wide range of energies ranging from 10 keV
up to 100 keV or even higher values. Despite the wide range of energies,
most of the works addressing the influence of the energy on the ripple pattern
report, a qualitatively similar behavior is found, namely, that the typical wavelength
increases with energy
following a power dependence of the type $ \lambda \sim E^m$, where usually $0 < m
\leq 1$. Figure \ref{Fig3_3} shows two examples of ripple production on
silicon targets with low-energy (a) and medium-energy (b) Ar$^+$ ion beams.
Clearly, the wavelength is quite different but the surface morphology is nonetheless similar.

\begin{figure}
\center
\includegraphics[height=4cm]{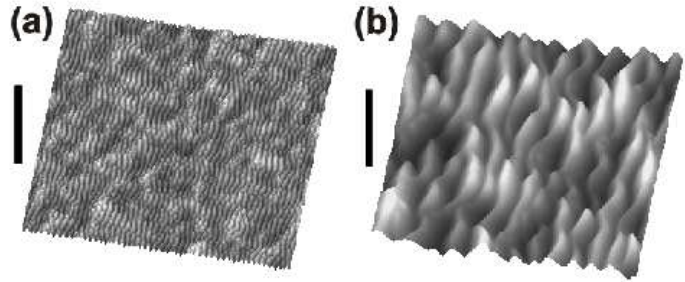}
\caption{ (a) $3\times 3$ $\mu \rm{m}^{2}$ AFM image of a Si surface bombarded by a
1 keV Ar$^+$ ion beam at $\theta=10^{\circ}$ for ion dose of $2 \times 10^{15}$ ions
cm$^{-2}$. (b) $4 \times 4$ $\mu \rm{m}^{2}$ AFM image of a Si surface bombarded by
a 40 keV Ar$^+$ ion beam at $\theta=50^{\circ}$ for ion dose of $2 \times 10^{18}$
ions cm$^{-2}$. The vertical bars indicate 150 nm and 900 nm, respectively.}
\label{Fig3_3}
\end{figure}

This behavior has been found for Si targets irradiated
at low-energy for different ion species: O$_2^+$ \cite{Alkemade_2001, Vajo_1996}, Ar$^+$
\cite{Ziberi_2005}, and, Kr$^+$ \cite{Ziberi_2005} and Xe$^+$ \cite{Ziberi_2006b}, and also for sputtering
 experiments at medium-energy employing O$_2^+$ ions \cite{Karmakar_2005} and Ar$^+$ ions under ion beam scanning
conditions \cite{Chini_2002}. Also, this dependence has been observed for
SiO$_2$ targets bombarded by low-energy Ar$^+$ ions \cite{Umbach_2001}, for
HOPG (bombarded by medium energy Ar$^+$
ions) \cite{Habenicht_2001} and diamond (bombarded by medium energy Ga$^+$
ions) \cite{Datta_2001}.

However, discrepancies arise when considering the values reported for the
exponent $m$. It should be noted that this study becomes especially difficult
for low-energy IBS experiments in which, usually, the sampled energy range is
quite narrow as recently pointed out in \cite{Ziberi_2005}. Furthermore,
sometimes it is difficult to assess whether the experimental data follow a linear
or, rather, a different power law behavior. Thus, for different targets irradiated by low-energies ions, values of $m$ in the
0.2-0.8 range have been reported \cite{Alkemade_2001, Umbach_2001, Vajo_1996, Ziberi_2005}.

For the medium ion energy experiments, the  energy range sampled is usually
wider due to the use of ion implanters. Therefore, the assessment of the
quantitative dependence between $\lambda$ and $E$ becomes more reliable.
However, discrepancies still exist among the different values reported since $m$ values in the 0.45-1 range
have been obtained \cite{Chini_2002, Habenicht_2001, Karmakar_2005}. Particularly interesting is the case of HOPG targets: when they were bombarded
by Ar$^+$ ions a linear relationship was found, but when the ion was Xe$^+$ a
power law behavior was observed with an exponent value $m$ $\approx$ 0.7
\cite{Habenicht_2001}.

Finally, quite a different behavior was
found for Si targets bombarded by low energy Ar$^+$ ions \cite{Brown_2005}. In
this work $\lambda$ was found to decrease with energy when the target
surface was held at 717 $^{\circ}$C, whereas no clear trend was observed for
ripples produced at 657 $^{\circ}$C. Also, an inverse relationship between
$\lambda$ and ion energy was reported by Chini et al. for ion beam
sputtering of Si surfaces without beam scanning \cite{Chini_2002}. Suppression or application of the ion beam scanning could influence the actual temperature at the target surface. This fact would be in agreement
with a similar inverse behavior observed
for the experiments performed by Brown and Erlebacher at higher temperature \cite{Brown_2005}.
\end{itemize}
\hfill
\paragraph{Ripple pattern evolution with time or ion fluence}

The study of the ripple morphological evolution with irradiation time is usually done through
the dynamics of two magnitudes, the ripple wavelength and the surface roughness. The former is related with the ripple lateral dimension and the latter with the vertical one (i.e., ripple amplitude). This study
is quite important because eventual control of the
pattern morphological properties would enable applications for technological
purposes.

\begin{itemize}

\item \emph{Ripple coarsening}--- The existence or not of
ripple coarsening can be, as mentioned above, a touchstone for the continuum models. In fact, ripple
coarsening is not predicted by the seminal BH theory \cite{Bradley_1988} or
some of its non linear extensions \cite{Cuerno_1995,Facsko_2004,Makeev_1997,Makeev_2002,Vogel_2005}. In contrast, it is
predicted by more recent theories \cite{Castro_2005, Munoz_2005}. Thus, the
analysis of ripple coarsening becomes a relevant issue. Moreover, since there are systems that display and others that do not show
ripple coarsening, it is necessary to assess experimentally the
differences between them in order to elucidate which physical phenomena are
behind the coarsening process. This coarsening process typically reflects in a
power law dependence such as $\lambda \sim t^n$ where $t$ is irradiation time
(fluence) and $n$ is a coarsening exponent.

An example of coarsening behavior (see Fig.\ \ref{Fig3_4}) shows two AFM images of
a Si(100) surface irradiated by 40 keV Ar$^+$ ions at 50$^{\circ}$ for
different times together with typical surface cross-sections of the surface
morphology for both cases. In all images, the ripples run along the
perpendicular direction with respect to the projected ion beam direction (which
runs along the horizontal axis of the top view AFM images). From the AFM images
coarsening of the typical ripple wavelength is already evident. This is better appreciated in panel (c), together with the clear
surface roughening that also implies a clear increase in the amplitude of
the ripple morphology.

%%%%%%%%%%%%%%Figure%%%%%%%%%%%%%%%%%%%%%%%
\begin{figure}[!htmb]
\begin{center}
\begin{minipage}{0.29\linewidth}
\begin{center}
 \includegraphics[width=\linewidth]{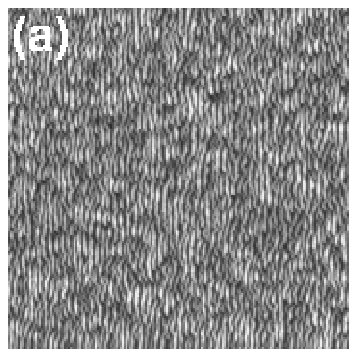}
\end{center}
\end{minipage}\hspace*{0.01\linewidth}
\begin{minipage}{0.29\linewidth}
\begin{center}
 \includegraphics[width=\linewidth]{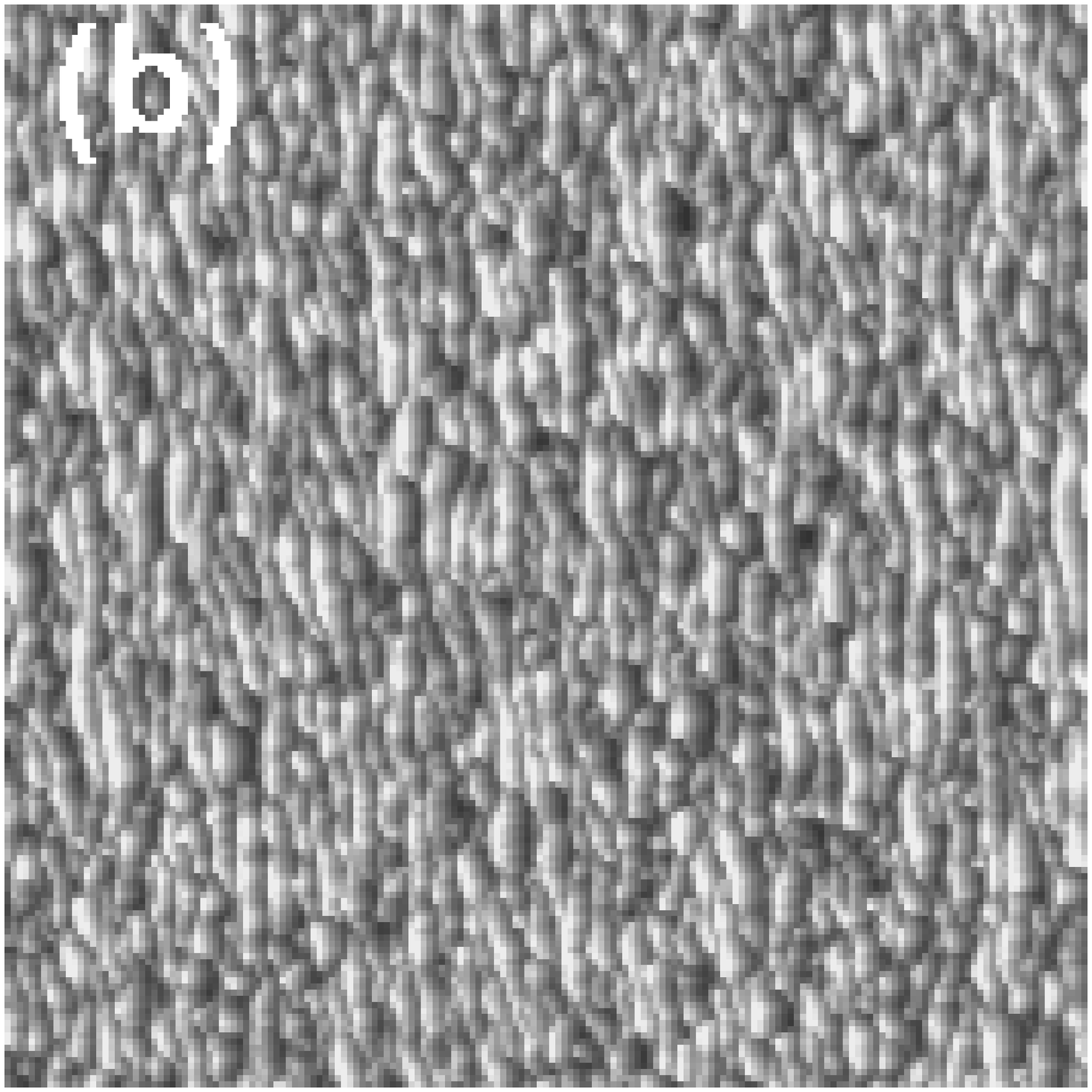}
\end{center}
\end{minipage}\hspace*{0.01\linewidth}
\begin{minipage}{0.4\linewidth}
\begin{center}
 \includegraphics[height=2.5cm, width=\linewidth]{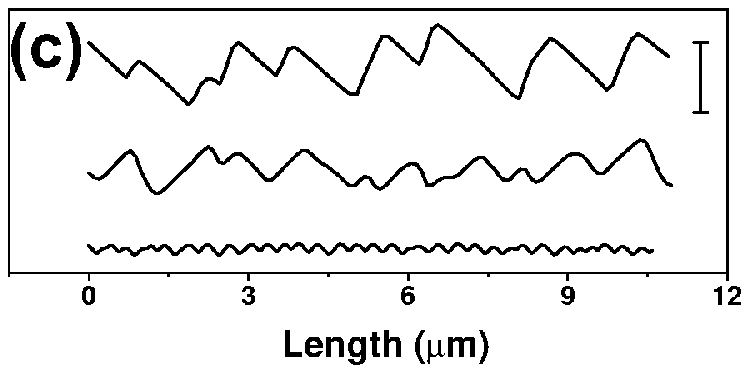}
\end{center}
\end{minipage}
\caption{ Top view AFM images of a Si(100) surface irradiated by 40 keV Ar$^+$
ions at $50^{\circ}$ for different times (ion fluences): (a) $20 \times 20$
$\mu \rm{m}^{2}$, irradiation time 2 h ($1.6 \times 10^{18}$ ions cm$^{-2}$) ; (b)
$50 \times 50$ $\mu \rm{m}^{2}$, irradiation time 16 h ($1.3 \times 10^{19}$ ions
cm$^{-2}$). (c) Typical surface profiles taken along the projection of the ion
beam of the samples irradiated for 2 h (bottom profile), 8 h (ion fluence of
$6.5 \times 10^{18}$ ions cm$^{-2}$; middle profile) and 16 h (top profile).}
\label{Fig3_4}
\end{center}
\end{figure}
%%%%%%%%%%%%%%%%%%%%%%%%%%%%%%%%%%%%%%%%%%%%%%%%%%%%

For the case of Si
targets, we find ripple coarsening for low-energy ion irradiation experiments
with 0.5 keV Ar$^+$ ions impinging at 60$^{\circ}$ \cite{Brown_2005b}.
Moreover, this coarsening process has been observed for a relatively wide
substrate temperature range, from 600 $^{\circ}$C up to 748 $^{\circ}$ C
\cite{Brown_2005}. Also, a coarsening process was observed for 1.5 keV Ar$^+$
ions impinging at 45$^{\circ}$ and room temperature \cite{Kulriya_2006} and 1
keV O$_2^+$ ions impinging at 52$^{\circ}$ \cite{Liu_2001} for both high and
low ion fluxes.

In contrast, coarsening was not observed for irradiation
experiments performed at 582$^{\circ}$C and at 67.5$^{\circ}$ with 0.75 keV Ar$^+$ ions
\cite{Erlebacher_1999}. This was also the case for experiments done by
bombarding the Si surface at $\theta=15^{\circ}$ with a 1.2 keV Ar$^+$ ion beam. The
same behavior was found when the ion species were Xe$^+$ or Kr$^+$
\cite{Ziberi_2005}.

For the medium-energy ion range, ripple coarsening has been
reported for Si surfaces bombarded at $\theta=45^{\circ}$ by 20 keV and 40 keV Xe$^+$
ions where target temperatures were maintained between 100-300 K
\cite{Carter_1996}. Similarly, it has been observed for Si targets irradiated
either by 60 keV Ar$^+$ ions at $\theta=60^{\circ}$ \cite{Datta_2004} or by 30 keV
Ga$^+$ ions at $\theta=30^{\circ}$ \cite{Habenicht_2002}.

In other materials, ripple coarsening has been observed for HOPG and diamond
surfaces. In the first case, the irradiation process was performed by 5 keV
Xe$^+$ ions at $\theta=60^{\circ}$ and $\theta=70^{\circ}$ \cite{Habenicht_1999} whereas in
the last case Ga$^+$ ions impinged at $\theta=60^{\circ}$ with energies of 50 keV and
10 keV \cite{Datta_2001}. Also, ripple coarsening has been reported for fused
silica bombarded by an 0.8 keV Ar$^+$ ion beam at $\theta=60^{\circ}$
\cite{Flamm_2001} and for glass targets irradiated by 0.8 keV Ar$^+$ ions,
which were generated in a defocused electron cyclotron resonance plasma, with
an angle of incidence of $35^{\circ}$ \cite{Toma_2005}. Finally, ripple
coarsening was also observed for InP targets irradiated at $\theta=41^{\circ}$ by
0.5 keV Ar$^+$ ions \cite{Demanet_1995} but not for GaAs surfaces bombarded at
$\theta=41^{\circ}$ by 10.5 keV O$_2^+$ ions \cite{Karen_1991}.

With regard to the value of the exponent $n$ different values have been
reported. Thus, on silicon targets a value of 0.5 for irradiation at 30
keV has been reported \cite{Habenicht_2002}. Also for high energies, 60 keV,
two regimes with values $n_1 = 0.64$ and $n_1 = 0.22$ were observed. In
contrast, for low-energy irradiation experiments an exponential dependence,
rather than a power-law, was found for a relatively wide range of substrate
temperatures \cite{Brown_2005}. When 0.8 keV Ar$^+$ ions were employed to
irradiate fused silica \cite{Flamm_2001} and glass surfaces \cite{Toma_2005}
coarsening exponents of 0.15 and 0.95 have been found, respectively.

Finally, only for 60 keV Ar$^+$ ion irradiation at $\theta=60^{\circ}$ of GaAs
surfaces, a disordered dot morphology has been reported previous
to, and later coexisting with, the ripple morphology
\cite{Datta_2002}. For long sputtering times (i.e., ion fluences of $3 \times
10^{18}$ ions cm$^{-2}$) only the ripple morphology remained, without any nanodot structure superimposed. \\

\item \emph{Surface roughening}--- The study of the surface roughness, $W$ ---defined as the mean square deviation of the local height with respect to its mean value--- of the patterns
can be very useful for both technological (e.g. developing metal surfaces for SERS applications) and fundamental
purposes. In principle, $W$ should be proportional to the ripple amplitude
(i.e., the peak to valley height difference), $A$, in case the patterns were
perfectly periodic. However, in real patterns there are height fluctuations
among ripples that imply that $A$ and $W$ are not completely equivalent. Although
most part of the studies deal with $W$, some of them analyze $A$ rather than $W$.

The most frequently observed behavior is that $W$ initially increases steeply
(usually increasing exponentially with ion fluence or sputtering time) to either
saturate or grow at a slower pace, usually following a power law dependence such as $W \sim
t^{\beta}$. In the latter case, it is interesting to measure the value of $\beta$
since it can be contrasted with predictions from theoretical models.

For the first case, i.e., exponential increase followed by saturation, we can
mention experiments on Si surfaces irradiated by low-energy Ar$^+$ ions
\cite{Erlebacher_1999}, \cite{Ziberi_2005}.

The second case, i.e., sharp increase followed by power law behavior, has
also been observed in many systems: Si targets bombarded by medium-energy
Ar$^+$ ions \cite{Datta_2004, Karmakar_2005} and also for HOPG surfaces
bombarded under similar conditions \cite{Habenicht_2001}. For Si targets irradiated by 60 keV Ar$^+$ ions
an initial value of $\beta_1 = 0.76$ was reported, although this relatively high value could be compatible with an
exponential increase, whereas for longer times $\beta_2 =
0.27$ was found \cite{Datta_2004}. In addition, for 16.7 keV O$_2^+$ ions $\beta = 0.38$ was observed after the initial sharp
increase of $W$ \cite{Karmakar_2005}. Also, for HOPG surfaces irradiated by 5 keV Xe$^+$
ions a $\beta$ value compatible with the Kardar-Parisi-Zhang \cite{Kardar_1986}
universality class was reported \cite{Habenicht_1999,Habenicht_2001}.

Finally, there are works where only power-law behaviors were reported. Most
of these systems studied present a $\beta$ value such that $0.45 \leq \beta \leq 1$
\cite{Carter_1996, Datta_2004, Flamm_2001, Toma_2005} (in the second case, for low ion flux conditions). These behaviors
could be due to partial analysis of the initial exponential increase that, when
analyzed in a limited temporal range, can be analyzed in terms of a power law
dependence with an exponent $\beta \gtrsim 0.5$ as remarked. \\

\item \emph{Shadowing effects}--- An important issue that should be taken into account for studying ripple
evolution is geometrical shadowing. These effects appear for long sputtering
times and when relatively large $\theta$ values are employed. The important role of
shadowing effects was highlighted in \cite{Carter_1996}; subsequently, Carter gave a simple estimation of the conditions under which these effects begin to
operate \cite{Carter_1999}. In particular, he proposed that shadowing operate
for ripples with an amplitude (here taken as proportional to $W$) to
wavelength ratio such that
\begin{equation}
W/\lambda \gtrsim 2\pi \tan(\pi-\theta). \label{eqshadow}
\end{equation}
In equation (\ref{eqshadow}) as $\theta$ increases (i.e., approaching grazing
incidence) the right hand side making shadowing effects more likely to appear. These effects have been considered recently for
Si irradiation by 60 keV Ar$^+$ ions at $\theta=60^{\circ}$ \cite{Datta_2004,
Datta_2005}. According to (\ref{eqshadow}), shadowing effects should appear when
$W/\lambda \gtrsim 0.09$, which occurs after 800 s of irradiation under the
sputtering conditions described in \cite{Datta_2004}. Interestingly, this value
is very close to the threshold time for which ripple coarsening begins to be
observed. Thus, shadowing effects can influence largely the ripple dynamics.
This fact can be important because usually shadowing is not incorporated into continuum models.
Moreover, the latter are usually derived under a small slope approximation and the
abrupt morphologies generated when shadowing processes appear cannot be
described under such approximation. A remarkable exception is the continuum equation proposed in \cite{Chen_2005} that seems to correctly describe steep surface features.
\end{itemize}

\hfill
\paragraph{Ripple pattern dependence on target temperature}

The study of the pattern evolution with target temperature can contribute to
further increase our knowledge of the main physical mechanisms governing the
IBS pattern formation. In particular, temperature can affect the surface
diffusivity, which can lead to changes in the morphology of the pattern.
However, studies are scarce due, probably, to their experimental complexity.
For low-energy ions there are two reports on silicon surfaces irradiated by
Ar$^+$. In the first one, in which 0.75 keV Ar$^+$ ions impinged at
$\theta=67.5^{\circ}$ onto the surface \cite{Chason_2001, Erlebacher_1999}, $\lambda$
coarsened with the target temperature in the 460-600$^{\circ}$ C range
following an Arrhenius law \cite{Makeev_2002}
\begin{equation}
\lambda \sim \frac{1}{T^{1/2}} \exp\left(\frac{-\Delta E}{2k_B T}\right). \label{eq2}
\end{equation}

\begin{figure}[!htmb]
\begin{center}
\begin{minipage}{0.5\linewidth}
\begin{center}
 \includegraphics[width=\linewidth]{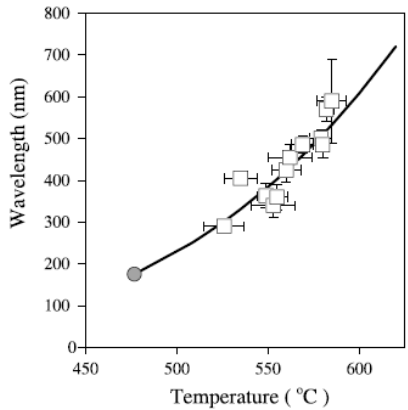}
\end{center}
\end{minipage}\hfill
\begin{minipage}{0.475\linewidth}
\begin{center}
\caption{Ripple wavelength $\lambda$ versus target temperature for
a Si(001) substrate irradiated at $\theta=67.5^{\circ}$ by 0.75 keV Ar$^+$ ions. Open
squares correspond to light scattering measurements, whereas the filled circle
was obtained by AFM. The solid line corresponds to a fit following the
Arrhenius law (\ref{eq2}). Figure reprinted from \cite{Erlebacher_1999} (http://link.aps.org/abstract/PRL/v82\newline/p2330) with permission}. \label{Fig3_5}
\end{center}
\end{minipage}
\end{center}
\end{figure}

These data are shown in Fig.\ \ref{Fig3_5}. The value of $\lambda$ obtained at the lowest
temperature was measured by AFM because it was out of range of the light
spectroscopy measurements. This analysis led to a value for the activation
energy, ($\Delta E =1.2 \pm 0.1$ eV) for surface mass transport on ion-bombarded Si(001).

%%%%%%%%%%%%%%Figure%%%%%%%%%%%%%%%%%%%%%%%
%%%%%%%%%%%%%%%%%%%%%%%%%%%%%%%%%%%%%%%%%%%%%%%%%%%%

The same behavior was found for Si(111) irradiated at 60$^{\circ}$ by 0.5 keV
Ar$^+$ ions in the 500-750$^{\circ}$ C range \cite{Brown_2005}. This study led
to $\Delta E = 1.7 \pm 0.1$ eV. Similarly, for SiO$_2$ targets irradiated by
0.5-2 keV Ar$^+$ ions an Arrhenius behavior was observed for $T > 200^{\circ}$
C \cite{Umbach_2001}.

Finally, there are two reports on the variation of the pattern with target
temperature for medium-energy IBS experiments. In the first one, HOPG surfaces
were irradiated by 5 keV Xe$^+$ ions in the 573-773$^{\circ}$ C range
\cite{Habenicht_2001}. For this system the ripple wavelength followed also the
Arrhenius law giving a value of $\Delta E = 0.14$ eV. The second experiment was
on bombardment of GaAs targets by 17.5 keV Cs$^+$ ions at $\theta=25^{\circ}$
\cite{Maclaren_1992}, in which the ripple wavelength was analyzed in the
$0-100^{\circ}$ C range. The data obtained have been later analyzed in
\cite{Malherbe_2003}. For $T>60^{\circ}$, the behavior was well described by
Eq.\ (\ref{eq2}) with an activation energy for surface self-diffusion of
$\Delta E=0.26$ eV. For lower temperatures, a slight decrease of $\lambda$ was
observed following a $T^{-1/2}$ law. However, within error bars, these
experimental data are also compatible with a temperature independent behavior,
in agreement with models for effective surface diffusion effects of erosive
origin \cite{Makeev_1997,Makeev_2002}. In any case, the present experiment
\cite{Maclaren_1992,Malherbe_2003} provides a clear example of the existence of
different temperature regimes for ripple formation under IBS.

In contrast, Carter and Vishnyakov did not find any change of $\lambda$ with
temperature in the sampled 100-300 K range when Si targets were irradiated by
10-40 keV \cite{Carter_1996}. These findings are consistent with the existing theories since for relatively
low temperatures ion-induced surface diffusion processes, which not depend on the target
temperature, dominate over thermally activated ones \cite{Makeev_1997}.

\hfill
\paragraph{Ripple pattern dependence on ion flux}

The ripple pattern morphology, in particular its wavelength, can depend also on
the ion flux, $\Phi$, i.e., the number of incoming ions per area and time
units.

Among the scarce studies of this behavior, most of them did not find any change
of the ripple wavelength with ion flux. This was the case for Si surfaces
bombarded by either 1.5 keV O$_2^+$ ions at $\theta=40^{\circ}$ \cite{Vajo_1996} or by
low-energy Ar$^+$ ions at different ion fluences \cite{Brown_2005} or by 2 keV
Xe$^+$ ions at $\theta=20^{\circ}$ \cite{Ziberi_2006b}. In addition, the same behavior
was found for fused silica targets irradiated at $\theta=60^{\circ}$ at different
angles by a low-energy Ar$^+$ ion beam \cite{Flamm_2001}. Also, no change of
$\lambda$ with the ion flux was obtained for diamond surfaces irradiated by 50
keV Ga$^+$ FIB at $\theta=57^{\circ}$ \cite{Datta_2001}. All these studies were done
at room temperature, except for the one on fused silica that was performed at
12$^{\circ}$ C.

In contrast, for Si surfaces bombarded by 0.75 keV Ar$^+$ ions at 67.5 degrees
to normal in the 500-600$^{\circ}$ C range \cite{Erlebacher_1999}, $\lambda$ was found to decrease with $\Phi$, as $\lambda \sim \Phi^{-1/2}$. Also, a
decrease of the ripple wavelength with ion flux was reported for Si surfaces
bombarded at room temperature by 1 keV O$_2^+$ ions at $\theta=52^{\circ}$, although in this case the
quantitative dependence was not addressed \cite{Liu_2001}.

\hfill
\paragraph{Ripple pattern order}

The size of ordered ripple domains is another essential property of these patterns
for potential technological applications. However, it is somehow difficult to
assess. In principle, there may be several methods to evaluate it. A first one
is based in the data obtained by AFM. From these data, it is straightforward to
obtain the Power Spectral Density (PSD) of the surface morphology,
$h(-\mathbf{r},t)$. The PSD is defined as ${\rm PSD}(k,t) = \langle
h(\mathbf{k},t) h(-\mathbf{k},t) \rangle$, where $h(\mathbf{k},t)$ is the
Fourier transform of $h(\mathbf{r},t) - \bar{h}(t)$ with $\bar{h}(t)$ being the
space average of the height
 and $k = |\mathbf{k}|$.
This PSD
curve usually presents a peak denoting the existence of a characteristic mode
whose associated length scale is identified with the ripple wavelength
$\lambda$. It has been proposed that the pattern lateral correlation length,
$\zeta$, which gives an estimation of the average size of the ordered
domains, can be obtained from the full width at half maximum of the PSD peak
\cite{Zhao_2001}. This method was employed by Ziberi et al. \cite{Ziberi_2005}
for estimating the range of order of ripple patterns produced by 1.2 keV
Ar$^+$, Kr$^+$ and Xe$^+$ ions at $\theta=15^{\circ}$  on silicon surfaces. Although in these experiments coarsening was not observed, (thus $\lambda$ was constant for all the ion fluences), in all
cases $\zeta$ was observed to increase with the ion fluence (especially, for the largest ion fluences $\zeta \geq 11\lambda$ was found).  In this work, the authors also studied the
change of $\zeta$ with the ion energy for the three different ions employed.
Whereas for Xe$^+$ and Kr$^+$ ions $\zeta$ and $\lambda$ were found to increase in the same way with ion energy, for the case of Ar$^+$ ions a
maximum for the $\zeta/\lambda$ ratio was observed for an ion energy of 1.2
keV. It should be noted, that for the largest energy employed in all cases, 2
keV, $\zeta \approx 10\lambda$, irrespectively of the ion species.

The previous method for assessing the pattern order degree has the disadvantage
of the local character of SPM techniques. However, there is another method, based on
grazing incidence diffraction (GID) or small angle scattering (GISAXS)
synchrotron techniques, that provides better sampling statistics. However, to our
knowledge, there is no published work using this technique for such purposes on amorphous materials where it has been used ripple crystallinity assessment \cite{Hazra_2004}.

\hfill
\paragraph{Ripple propagation}

The transversal collective motion of ripples has only been studied in two works by simultaneous real time
monitoring of the ion-induced ripple morphology by SEM. In the first one
\cite{Habenicht_2002} a Si surface was irradiated by 30 keV Ga$^+$ ions
and $\theta$ = 30$^{\circ}$, it was found that ripples initially propagated
along the ion beam projection direction with a velocity of $v$ = 0.33 nm
s$^{-1}$ to slow down later on as they coarsened. In the second case, in which
glass surfaces were irradiated by 30 keV Ga$^+$ ions \cite{Alkemade_2006},
ripples also propagated along the projection of the ion beam direction. In this
case,the ripple evolution was shown in real time. Moreover, ripples did not
initiate the propagation until most of them were completely formed, to finally
reach an uniform propagation velocity.

\subsubsection{Nanodot patterning in amorphous/amorphizable materials}

Dot IBS nanopatterns are produced when the anisotropy caused by the oblique
incidence of the ion beam is suppressed. Basically, there are two ways to
eliminate this anisotropy: (a) by IBS under normal incidence, which is the most
frequently employed technique \cite{Facsko_1999, Gago_2001}; (b) by IBS under
oblique incidence but with simultaneous rotation of the target
\cite{Frost_2000}. More recently, nanodot patterns have been produced on Ge
surfaces by a 2 keV Xe$^+$ ion beam impinging on the Ge surface at $\theta=20^{\circ}$
\cite{Ziberi_2006}, possibly related with the role of the critical angle
$\theta = \theta_c$ mentioned at the beginning of Sec.\ \ref{patterns_exp}.

These patterns are usually characterized by a highly uniform dot size
distribution and short-range in-plane ordering. These two properties make them very
interesting for potential technological applications. Although
the first report on the production of such IBS nanopatterns \cite{Facsko_1999}
is relatively recent, dating from 1999, many groups
are investigating the mechanisms leading to their formation. Thus, up
to now, these patterns have been produced in different materials: GaSb
\cite{Bobek_2003, Facsko_1999, Frost_2003, Xu_2004}, InP
\cite{Frost_2000}, InAs \cite{Frost_2004b}, InSb \cite{Facsko_2001}, Si
\cite{Gago_2001, Gago_2002, Ziberi_2005b}, and Ge
\cite{Ziberi_2006}.

As occurred for the ripple patterns, the formation of the dot patterns under
different experimental conditions and for a relatively wide range of materials
suggests that this process does not depend on the specific ion-target
interactions.

\begin{figure}[!htmb]
\begin{center}
\begin{minipage}{0.5\linewidth}
\begin{center}
 \includegraphics[width=\linewidth]{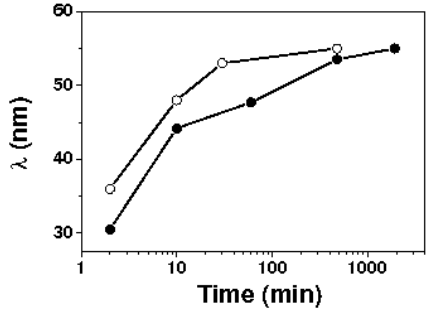}
\end{center}
\end{minipage}\hfill
\begin{minipage}{0.475\linewidth}
\begin{center}
\caption{ Dependence of $\lambda$ with sputtering time obtained by GID for
Si(001) ($\bullet$) and Si(111) (o) surfaces irradiated under normal incidence
by 1.2 keV Ar$^+$ ions. Note the faster coarsening dynamics for the IBS pattern
induced on Si(111) surface.} \label{Fig3_6}
\end{center}
\end{minipage}
\end{center}
\end{figure}

Another important issue regarding the target material is its crystallinity.
Thus, it has been proved that, for GaSb targets, IBS nanodot patterns can
be produced both on crystalline \cite{Facsko_1999} and amorphous surfaces \cite{Facsko_2002}. In
addition, these patterns have been produced by IBS on both Si(100) and Si(111)
surfaces \cite{Gago_2006}. In this work, it was found that, although the surface
crystallinity does not affect the pattern formation, it can affect to some
extent the pattern dynamics. In particular, it was observed that Si(111) surfaces have
a faster dynamics in terms of pattern coarsening (see Fig. \ref{Fig3_6}) and ordering than
Si(100) surfaces. Again, the fact that the surface crystallinity does not determine the formation of
the pattern is compatible with Sigmund's theory \cite{Sigmund_1969} for which the
surface is considered as amorphous, either originally or as induced by the ion beam action.

%%%%%%%%%%%%%%Figure%%%%%%%%%%%%%%%%%%%%%%%
%%%%%%%%%%%%%%%%%%%%%%%%%%%%%%%%%%%%%%%%%%%%%%%%%%%%

In the following, we will review the main experimental findings on the
properties of these patterns depending on the different experimental
parameters. One of the most studied issues is the variation with physical
parameters of the basic pattern length scale, $\lambda$, which now corresponds to the
average dot-to-dot distance, usually proportional to the dot size. It is worth
noting that the interest in these studies on the dot shape
and size lies in the required control of these parameters, particularly the dot
size, for developing technological applications.

\begin{figure}[!hmbt]
\begin{center}
\begin{minipage}{0.275\linewidth}
\begin{center}
 \includegraphics[width=\linewidth]{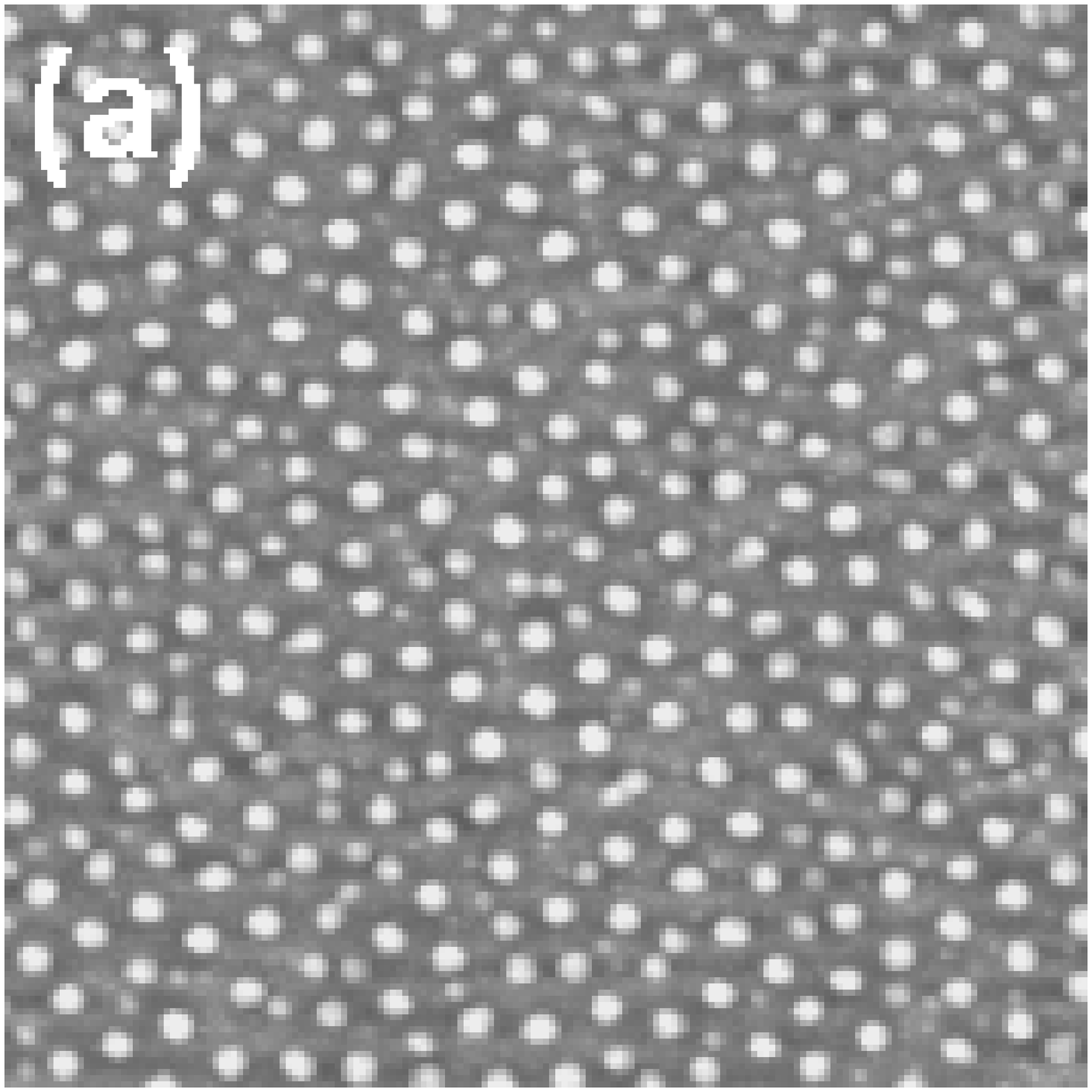}
\end{center}
\end{minipage}\hspace*{0.01\linewidth}
\begin{minipage}{0.275\linewidth}
\begin{center}
 \includegraphics[width=\linewidth]{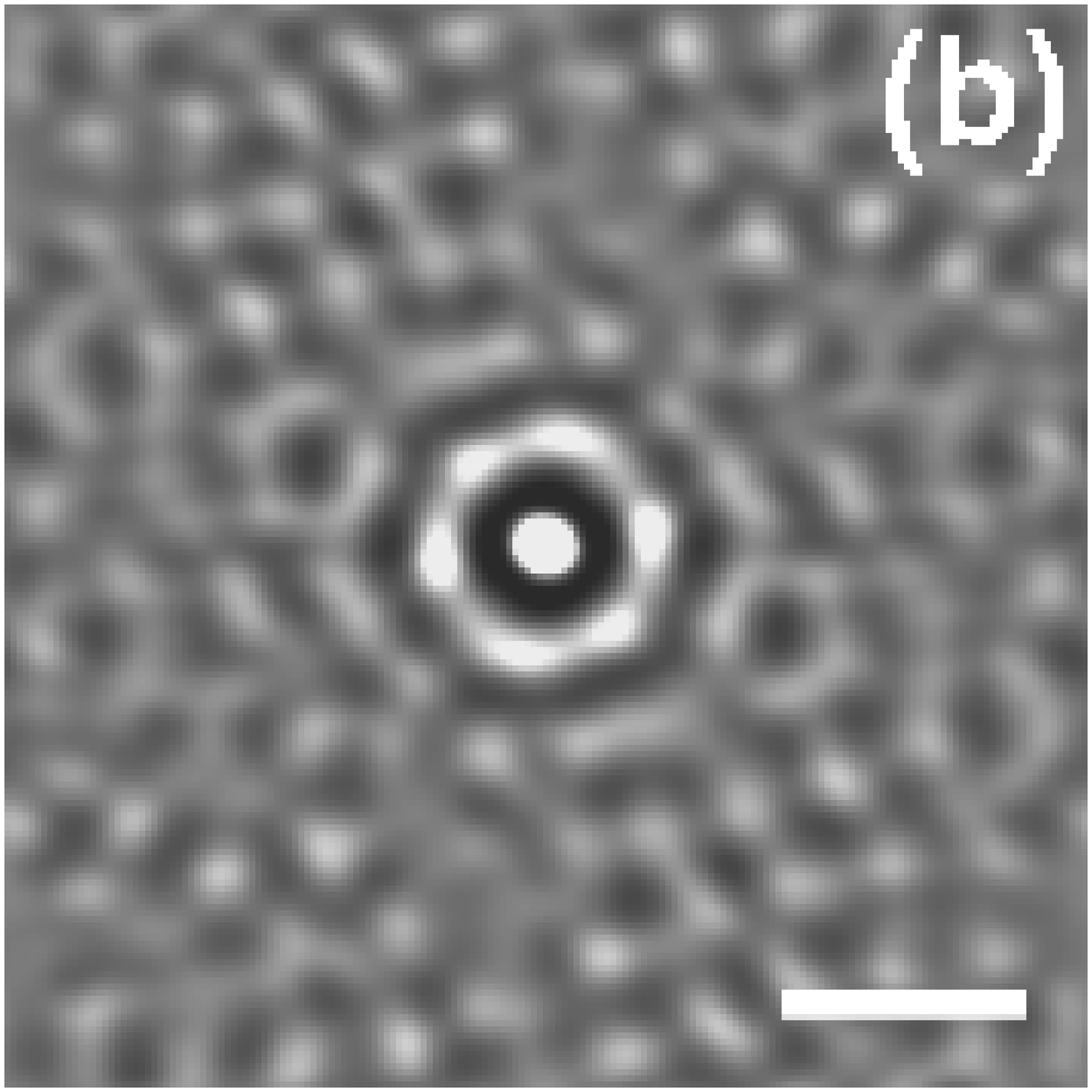}
\end{center}
\end{minipage}\hspace*{0.01\linewidth}
\begin{minipage}{0.41\linewidth}
\begin{center}
 \includegraphics[height=3.25cm, width=\linewidth]{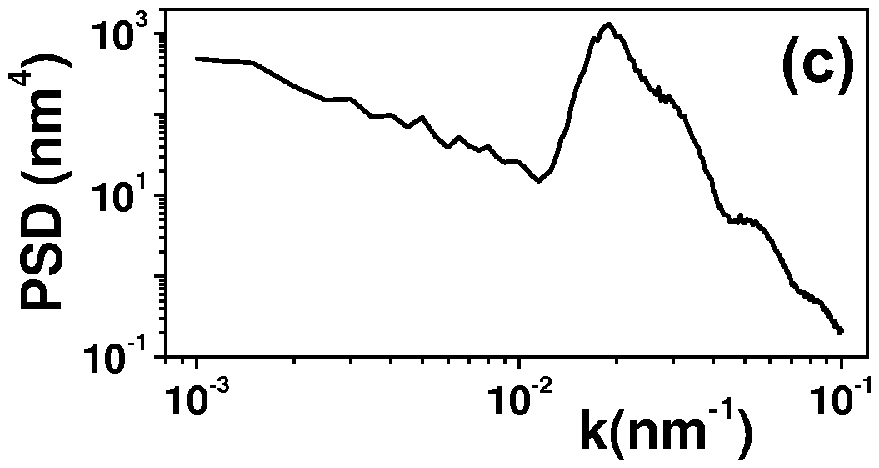}
\end{center}
\end{minipage}
\caption{(a) $1\times 1$ $\mu \rm{m}^{2}$ AFM image of a Si(001) substrate
irradiated under normal incidence by 1.2 keV Ar$^+$ ions. (b) Two-dimensional
auto-correlation function in which the short-range hexagonal order is observed.
The horizontal bar corresponds to 68 nm. (c) Radially-averaged PSD for image of
(a). The main peak corresponds to the $k = 2\pi / \lambda$ value corresponding to
the dominant pattern wavelength.} \label{Fig3_8}
\end{center}
\end{figure}

It should be noted that, whereas the dot size is usually affected by the AFM
tip (AFM being the routine technique for characterizing the surface
morphology), the dot-to-dot distance is not usually affected by tip convolution
effects \cite{Frost_2001}. The pattern wavelength is usually determined from
the radially-averaged PSD of AFM images as is shown in Fig. \ref{Fig3_8}. In
panel (a) of this figure we display a typical AFM image of a nanodot pattern
induced onto a Si(100) surface. Panel (c) shows the radially-averaged PSD
function corresponding to this image, in which we can observe a dominant peak
corresponding to the basic wavelength $\lambda$. At higher $k$ values we can
detect other minor peaks or shoulders indicating the high lateral ordering and
size homogeneity of dots \cite{Ziberi_2006c}. Conversely, panel (c) also illustrates the power-law behavior of the $PSD$ at small $k$, which signals height disorder between dots at long distances. This behavior is very frequently found for this type of experiments.

%%%%%%%%%%%%%%Figure%%%%%%%%%%%%%%%%%%%%%%%
%%%%%%%%%%%%%%%%%%%%%%%%%%%%%%%%%%%%%%%%%%%%%%%%%%%%

\begin{figure}[!htmb]
\begin{center}
\begin{minipage}{0.525\linewidth}
\begin{center}
 \includegraphics[width=\linewidth]{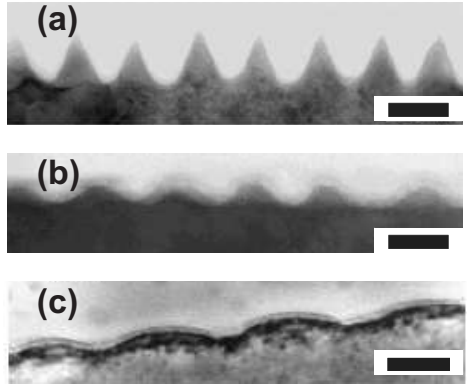}
\end{center}
\end{minipage}\hfill
\begin{minipage}{0.45\linewidth}
\begin{center}
\caption{Cross-sectional TEM images of (a) GaSb dots nanopatterns obtained
under 0.5 keV Ar$^+$ irradiation with a fluence of $10^{19}$ ions cm$^{-2}$ at
normal incidence with target rotation; (b) GaSb dots nanopatterns obtained
under 1.2 keV Ar$^+$ irradiation with a fluence of $10^{19}$ ions cm$^{-2}$ at
$\theta = 75^{\circ} $ with target rotation; (c) Si dot nanopatterns obtained
under 1.2 keV Ar$^+$ irradiation with a fluence of $9 \times 10^{17}$ ions
cm$^{-2}$ at normal incidence. The horizontal bars correspond to 50 (a), 25 (b)
and 30 nm (c), respectively. Figures (a) y (b) taken from \cite{Frost_2004b}
with permission.} \label{Fig3_TEM_dots}
\end{center}
\end{minipage}
\end{center}
\end{figure}

Similarly to the case of ripples, TEM analysis can provide us useful information regarding the morphology and structure of the dot patterns.
Thus, different dot morphologies obtained by TEM are presented in Fig.
\ref{Fig3_TEM_dots}. In (a) we observe the conical morphology of crystalline
GaSb dots produced under normal
irradiation and target rotation \cite{Frost_2004c}. In contrast, in (b) we
observe GaSb dots with a sinusoidal shape obtained under irradiation at $\theta
= 75^{\circ}$ and target rotation. In these cases, the amorphous layer thickness
was $\simeq 4$ nm \cite{Frost_2004c}. Finally, in (c) we observe Si nanodots
produced under normal irradiation without target rotation displaying, rather, a
lenticular shape with an amorphous layer thickness of $\simeq 2$ nm \cite{Gago_2001}.

%%%%%%%%%%%%%%Figure%%%%%%%%%%%%%%%%%%%%%%%
%%%%%%%%%%%%%%%%%%%%%%%%%%%%%%%%%%%%%%%%%%%%

\hfill
\paragraph{Nanodot pattern dependence on ion energy and type}

\begin{itemize}
\item \emph{Ion type}--- To date, all the experiments in which nanodot patterns have been reported are done with
low-energy Ar$^+$ ion beams, except for one work in which Ne$^+$, Kr$^+$ and Xe$^+$ ion
beams were also employed \cite{Ziberi_2006c}. In this work oblique ion beam
incidence and simultaneous target rotation were employed. Although some
characteristics of the pattern differed, the authors concluded that there was
almost no difference in the morphological evolution of the mean size of dots
when using different ions. In contrast to the experimental findings of the same
group on ripple formation, the use of Ne$^+$ ions did lead to the production of
nanodot patterns with an experimental behavior similar to that observed when
Ar$^+$ ions were employed. Recently, experiments have been performed using 1
keV O$_2^+$ ions \cite{Tan_2006}. However, in this work a sort of nanodot
chains was
produced that seem more similar to nanoripples with a superimposed nanodot morphology.\\

\item \emph{Ion energy}--- Regarding the energy range used in the nanodot experiments, there is a marked
difference with respect to the studies realized on nanoripple IBS production
(see above). Namely, for nanodot experiments only low-energy
ions have been used, with $E \leq 2$ keV. As remarked earlier in Sec. \ref{patterns_exp}, working in
this energy range can lead to relatively large errors in the determination of
the power-law dependence $\lambda \sim E^m$ , especially in the value of $m$
\cite{Frost_2003}. However, this was not the case for the study by Facsko et
al. for IBS nanodot patterns on GaSb since they sampled 15 energy values within
this range \cite{Facsko_2001}. They obtained a value of $m = 0.5 \pm
0.02$. In the same study the authors estimated, a
similar $m$ value for InSb surfaces from three experimental data points.

In Fig. \ref{Fig3_8new} we present our results on GaSb surfaces irradiated at
different ion energies. Clearly, the dot size coarsens with ion energy. In
fact, our experimental data are consistent with an exponent of $m = 0.5$ as in
the case of Facsko and coworkers \cite{Facsko_2001}. Also we must note the evident
surface roughening with ion energy given that the vertical scale is the same in all
three AFM images.

Another experimental study was done by the
group of Frost and coworkers for Si surfaces with Ne$^+$, Ar$^+$, Kr$^+$ and Xe$^+$ oblique
ion beams onto rotating Si targets \cite{Ziberi_2006c}. In all cases, they found
an increase of $\lambda$ with ion energy. Unfortunately, they did not estimate
the value of $m$. However, from a visual inspection of Fig.\ 6 of
\cite{Ziberi_2006c} it is clear than $m_{\rm Ne} \approx m_{\rm Ar} > m_{\rm
Kr} > m_{\rm Xe}$.

\begin{figure}[!hmbtp]
\center
\includegraphics[height=8cm]{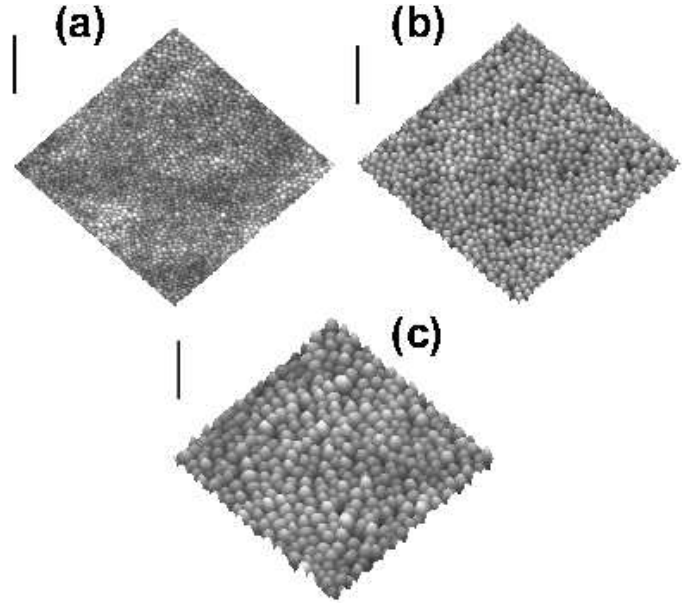}
\caption{(a) $1\times 1$ $\mu \rm{m}^{2}$ AFM images of a GaSb substrate
irradiated with Ar$^+$ ions under normal incidence with an approximate fluence of
$10^{17}$ ions cm$^{-2}$ at: (a) 0.3 keV; (b) 0.7 keV; (c) 1.2 keV. The vertical
bars correspond to 200 nm. Note the surface roughening and dot-size coarsening
with increasing ion energy.} \label{Fig3_8new}
\end{figure}
\end{itemize}

\hfill
\paragraph{Nanodot pattern evolution with sputtering time or ion fluence}

The study of the influence of sputtering time (i.e., ion fluence) on the
pattern morphology has attracted the interest of the researchers already since
the seminal work by the group of Facsko and coworkers \cite{Facsko_1999}. As occurred for the ripple morphologies, two main dynamics are studied: that of the
pattern wavelength and the evolution of surface roughness.

\begin{itemize}

\item \emph{Dot nanopattern coarsening}--- Basically, the same behavior is observed for many of the various experimental
systems: initially, $\lambda$ increases to saturate afterwards. However, the
dynamics of this process is quite different depending on the target material
and ion current density. Thus, for GaSb surfaces \cite{Bobek_2003, Xu_2004} and
InP \cite{Frost_2000, Frost_2003} the saturation regime is attained for ion
dose close to $10^{18}$ ions cm$^{-2}$. It should be noted that this saturation
was attained for an ion dose of $1.7 \times 10^{17}$ ions cm$^{-2}$ when InP targets
were irradiated under normal incidence by Ar$^+$ ions without rotation, but with
a flux six times smaller \cite{Tan_2006b} than that employed in
\cite{Frost_2000}. For Si surfaces saturation takes  place for a considerably
larger ion dose, at $4 \times 10^{19}$ ions cm$^{-2}$ \cite{Gago_2006}. In another study
where the Si surface was intentionally seeded with molybdenum, $\lambda$ can be
estimated to saturate at an ion dose of $7 \times 10^{17}$ \cite{Ozaydin_2005}. The
different dynamics of GaSb and Si surfaces under normal incidence Ar$^+$
irradiation and rotating InP targets under oblique Ar$^+$ bombardment is shown
in Fig.\ \ref{Fig3_9}, where results obtained on these systems are displayed.

%%%%%%%%%%%%%%Figure%%%%%%%%%%%%%%%%%%%%%%%
\begin{figure}[!htmb]
\begin{center}
\begin{minipage}{0.55\linewidth}
\begin{center}
 \includegraphics[width=\linewidth]{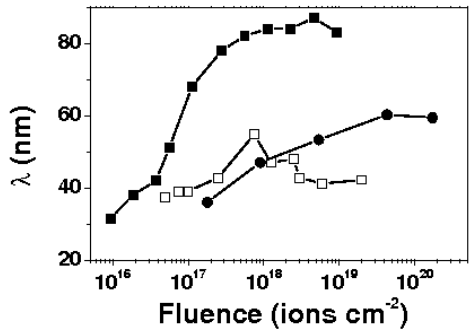}
\end{center}
\end{minipage}\hfill
\begin{minipage}{0.45\linewidth}
\begin{center}
\caption{Dependence of the characteristic pattern wavelength with ion fluence for GaSb, $\Phi= 0.8$ mA cm$^{-2}$ ($\square$), InP at $\Phi = 0.15$ mA cm$^{-2}$
($\blacksquare$) and Si at $\Phi = 0.24$ mA cm$^{-2}$ ($\bullet$) surfaces. The data for GaSb and InP have been adapted from
\cite{Bobek_2003} and \cite{Frost_2000}, respectively.} \label{Fig3_9}
\end{center}
\end{minipage}
\end{center}
\end{figure}
%%%%%%%%%%%%%%%%%%%%%%%%%%%%%%%%%%%%%%%%%%%%

Measurement of the exponent $n$ in the power law dependence $\lambda \sim t^n$
before saturation has been done for four  systems: (a) for Si
surfaces irradiated under normal incidence by 1.2 keV Ar$^+$ ions with $\Phi = 0.24$ mA cm$^{-2}$ Gago et al.\ found $n \simeq 0.2$ \cite{Gago_2001}.
(b) For GaSb targets sputtered by 0.5 keV Ar$^+$ ions under normal incidence
with $\Phi = 0.8$ mA cm$^{-2}$ Xu et al.\ reported $n = 0.14 \pm 0.03$
\cite{Xu_2004}. (c) For rotating InP targets irradiated by oblique 0.5 keV
Ar$^+$ ions with $\Phi = 0.15$ mA cm$^{-2}$, Frost et al.\ reported a
value of $n = 0.26 \pm 0.04$ \cite{Frost_2000, Frost_2003}. (d) For InP targets
irradiated under normal incidence by 1 keV Ar$^+$ ions with $\Phi=0.0233$ mA cm$^{-2}$ , Tan et al.\ reported a value of $n = 0.23 \pm 0.01$
\cite{Tan_2006b}.

In addition, for Si
surfaces irradiated by 0.5 keV Ar$^+$ ions Ludwig et al.\ did observe a
coarsening process with sputtering time but they did not estimate the value of
the coarsening exponent \cite{Ludwig_2002}. However, they did not observe the
saturation regime up to ion doses of $4.8 \times 10^{17}$, which agrees with
other experimental reports on Si surfaces. Besides, the authors did find that the
coarsening exponent value increased with the ion energy in the 100-200 eV
range. However, the opposite behavior, i.e., absence of coarsening, was
observed by the group of Frost for rotating Si targets irradiated by oblique
beams of Ne$^+$, Ar$^+$, Kr$^+$ and Xe$^+$ ions \cite{Ziberi_2005b, Ziberi_2006c}.\\

\item \emph{Surface roughening}--- The surface roughness gives us a measure of the nanodot height as well as of
the height fluctuations among dots. For GaSb surfaces \cite{Bobek_2003}, $W$
initially increases rapidly with sputtering time to  reach later a maximum
value. In this first region Xu and Teichert obtained $\beta = 0.87 \pm 0.12$
\cite{Xu_2004}. Again, this high value could be an indication that the surface
roughness increases exponentially rather than follow a power law. For longer
times the surface roughness decreases to attain a saturation value.

For the case of InP targets irradiated under normal incidence conditions a
first regime for which $\beta_1 = 0.74 \pm 0.03$ was reported \cite{Tan_2006b}.
This regime led to another one in which $\beta_2 = 0.09 \pm 0.03$, also roughly compatible with roughness saturation. In contrast, for rotating InP targets
\cite{Frost_2000}, the surface roughness
increases for the whole temporal range that was sampled. A first value $\beta_1 = 0.8 \pm 0.1$ was measured, once more
possibly compatible with an exponential time dependence, while for longer times
$\beta_2 = 0.27 \pm 0.06$ was obtained.

For Si targets, both fixed and rotating, a similar behavior of the roughness
has been reported, namely a sharp initial increase followed by saturation
\cite{Gago_2001, Ozaydin_2005, Ziberi_2006c}. In the latter study, the behavior observed for Ar$^+$, Kr$^+$ and Xe$^+$ ions was found to be
consistent with an initial regime during which the roughness increased
exponentially.

The main findings of the above works are displayed in Fig.\ \ref{Fig3_15} where
we plot the evolution of $W$ with sputtering time for the GaSb,
InP and Si systems. In this plot we have represented the $x$-axis in
logarithmic scale in order to display the three systems in a single graph,
which have different dynamics. It can be appreciated that, although the
temporal evolution is different for each system, there is always a sharp
initial increase of the roughness before reaching either a stationary value or
a regime with a slower growth. In the inset we display the same plot only for the initial
stages of the sputtering process (i.e., $t < 170$ s); now the $y$-axis is the
one in logarithmic scale so that for all three systems the initial
roughness seems to increase exponentially with time as the straight lines suggest in
the plot .

%%%%%%%%%%%%%%Figure%%%%%%%%%%%%%%%%%%%%%%%
\begin{figure}[!htmb]
\begin{center}
\begin{minipage}{0.54\linewidth}
\begin{center}
\includegraphics[height=5.5cm,width=\linewidth]{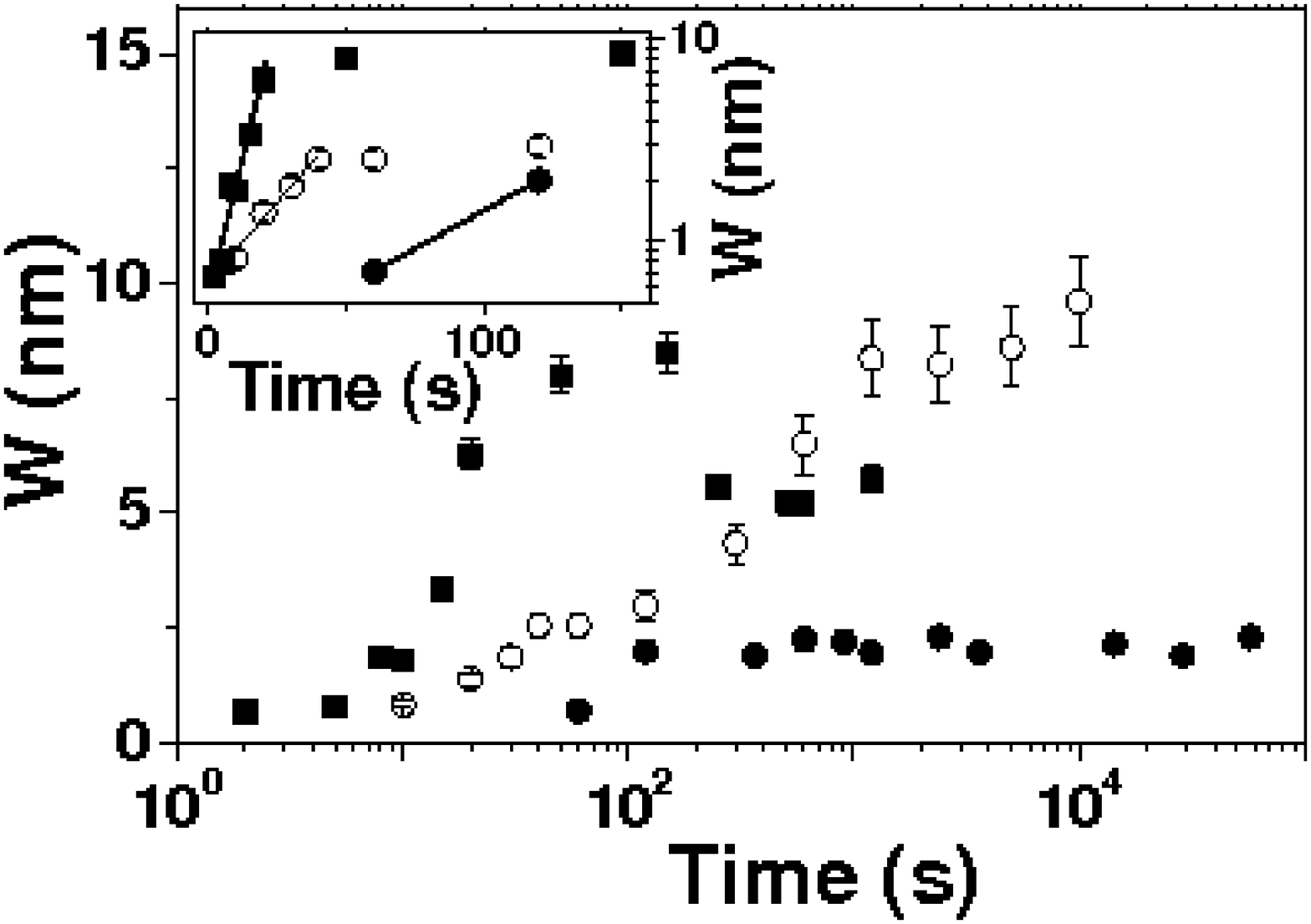}
\end{center}
\end{minipage}\hfill
\begin{minipage}{0.46\linewidth}
\begin{center}
\caption{ Surface roughness vs sputtering  time for IBS dot patterning
of GaSb irradiated under normal incidence ($\blacksquare$), InP bombarded under
oblique incidence and simultaneous target rotation (o) and Si irradiated under
normal incidence ($\bullet$). Note that the temporal axis is in
logarithmic scale. Inset: same graph but restricted to short
sputtering times, i.e., $t < 170$ s. Note that now the $y$-axis is in
logarithmic scale. The solid straight lines are guides to eye to indicate the
exponential dependence of the initial roughness increase for the GaSb, InP and
Si systems, respectively.} \label{Fig3_15}
\end{center}
\end{minipage}
\end{center}
\end{figure}
%%%%%%%%%%%%%%%%%%%%%%%%%%%%%%%%%%%%%%%%%%%%

Finally, we have also studied \cite{Gago_2001}, how the surface roughness changes with the
substrate temperature for Si targets irradiated under normal incidence. We observed that $W$,
which was constant for temperatures up to 400 K, decreased to reach a saturation value at 550
K where the pattern vanished \cite{Gago_2006b}.
\end{itemize}

\hfill
\paragraph{Nanodot pattern dependence on target temperature}

To the best of our knowledge, there are four studies on the dependence of the pattern morphology on the
target temperature, each one on a different material. Thus, we will describe
separately the main findings of these studies.

\begin{enumerate}
\item GaSb surfaces irradiated under normal incidence by 0.5 keV Ar$^+$ ions
\cite{Facsko_2001}: for this system the pattern wavelength did not change with
target temperature between  $-60^{\circ}$ C and $60^{\circ}$ C.
This behavior was interpreted as a confirmation that the main relevant
smoothing process is nonthermal under these experimental
conditions; the main relaxation process is, rather, due to ion-induced
effects \cite{Makeev_1997}.

\item Rotating InP surfaces irradiated by 0.5 keV Ar$^+$ ions impinging at
30$^{\circ}$ \cite{Frost_2003}: in this case, quite a striking complex behavior
was found since pattern symmetry changed in the temperature range between 268 K
and 335 K from short-range hexagonal to square patterns symmetry. In addition, the characteristic wavelength increased with temperature
\cite{Frost_2004b}. This behavior is shown in Fig.\ \ref{Fig3_10}.

%%%%%%%%%%%%%%Figure%%%%%%%%%%%%%%%%%%%%%%%
\begin{figure}[!htmb]
\begin{center}
\begin{minipage}{0.322\linewidth}
\begin{center}
\includegraphics[width=\linewidth]{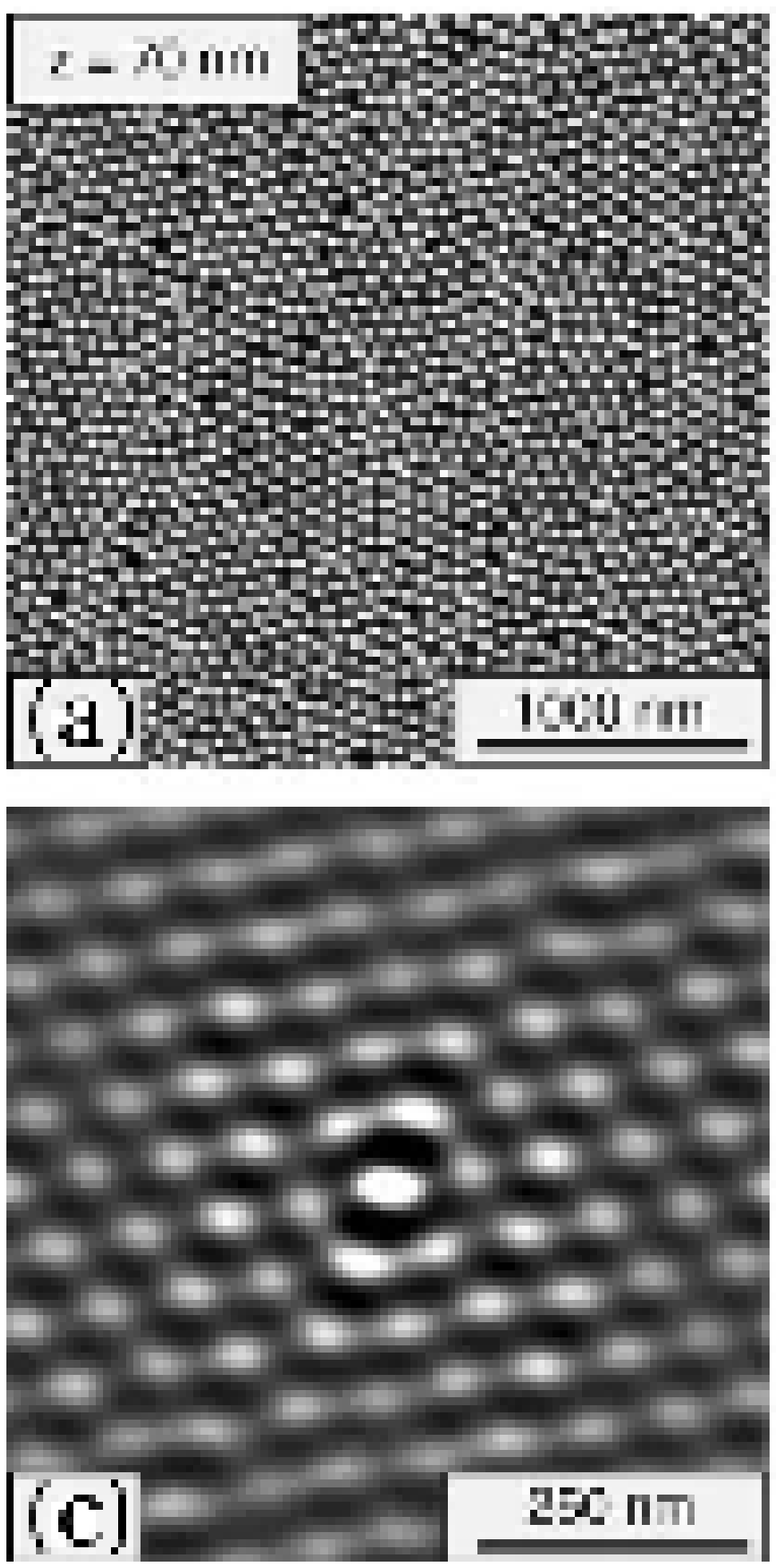}
\end{center}
\end{minipage}\hspace*{0.01\linewidth}
\begin{minipage}{0.4\linewidth}
\begin{center}
\includegraphics[width=\linewidth]{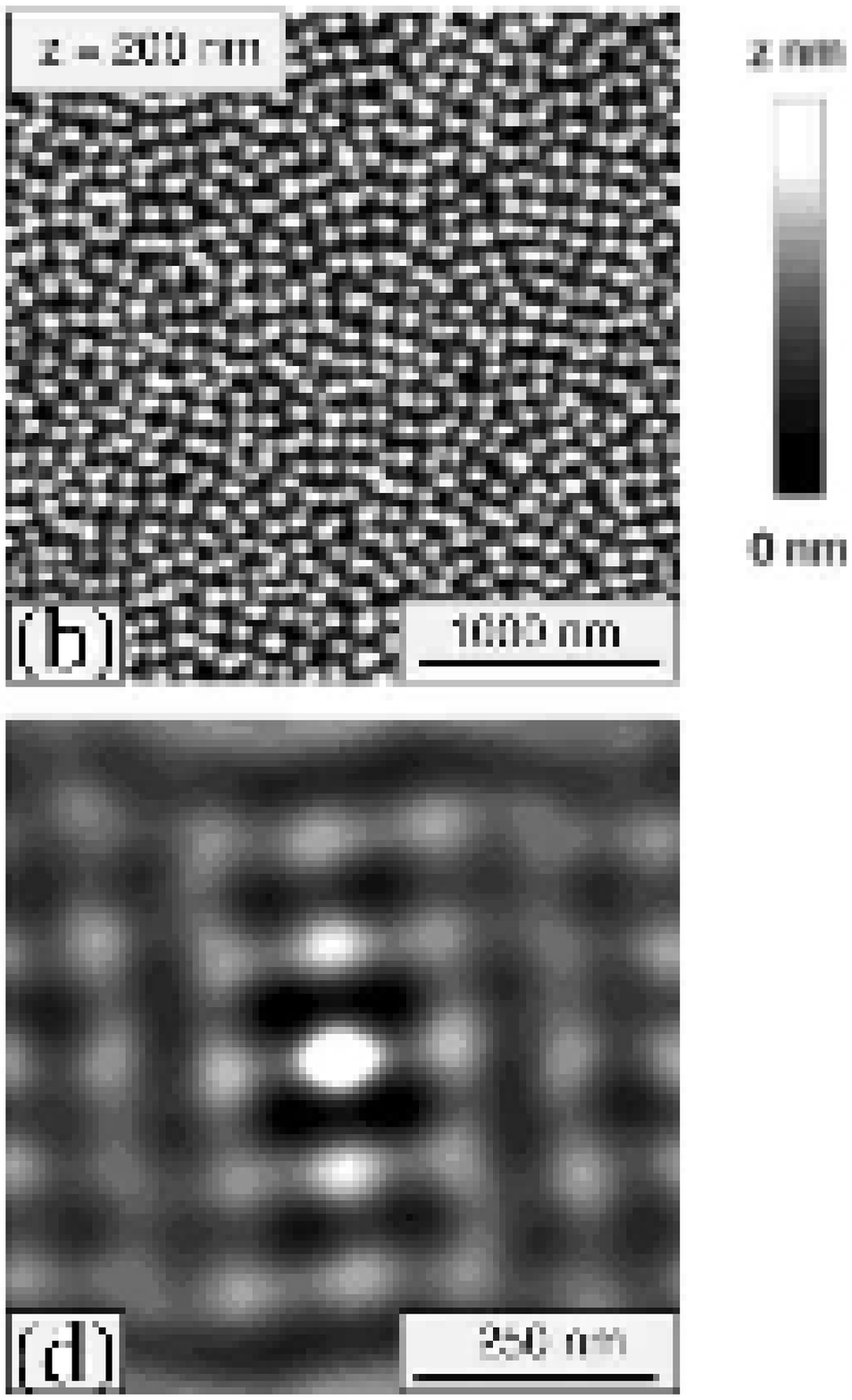}
\end{center}
\end{minipage}
\caption{AFM images of a rotating InP target irradiated at $\theta=30^{\circ}$ by 0.5
keV Ar$^+$ at $\Phi = 300$ $\mu$A cm$^{-2}$ at different target
temperatures: (a) 268 K, (b) 335 K. The corresponding two-dimensional
autocorrelation functions are displayed in (c) and (d), respectively. Figure
reprinted from \cite{Frost_2004b} with permission.} \label{Fig3_10}
\end{center}
\end{figure}
%%%%%%%%%%%%%%%%%%%%%%%%%%%%%%%%%%%%%%%%%%%%

\item For Si targets irradiated by 1.2 keV Ar$^+$ ions \cite{Gago_2006b} was
observed, by both AFM and GISAXS, the nanopattern wavelength to be a constant up to 425 K, and then to decrease in the 425-525 K range. For higher
temperatures the pattern vanished and the surface became featureless. This
behavior is not explained by any of the existing theories on IBS
nanostructuring. In Fig. \ref{Fig3_11} we display AFM images of IBS induced
dot patterns at 300 K (a) and 425 K (b) showing qualitatively how both
$\lambda$ and the dot size become smaller with increasing substrate
temperature. This shrinking process becomes evident in panels (c) and (d) in
which we show the PSD and GID curves measured on both patterns, respectively.
In both cases, the main peak shifts to higher $k$ and $q$ values, i.e., the
characteristic length scale diminishes as substrate temperature increases
\cite{Gago_2006b}.

\begin{figure}[!htmb]
\center
\includegraphics[height=6cm]{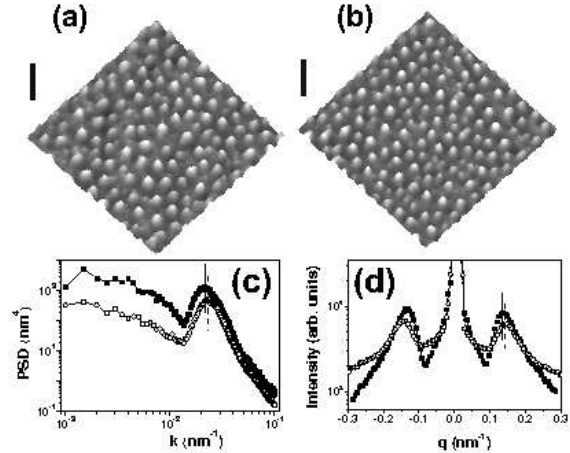}
\caption{$500 \times 500$ nm$^2 $AFM images of a Si(100) substrate irradiated
at normal incidence by 1.2 keV Ar$^+$ at different target temperatures: (a) 300
K, (b) 425 K. The vertical bars indicate 25 nm. PSD (c) and GID (d) curves for
the patterns produced at 300 K ($\blacksquare$) and 425 K (o). The vertical
solid and dashed lines indicate the $k = 2\pi/\lambda$ (PSD) and
$q=2\pi$/$\lambda$ (GID) values for the 300 K and 425 K systems, respectively.
} \label{Fig3_11}
\end{figure}

\item Fixed InP surfaces irradiated under normal 
incidence \cite{Tan_2006b} by 1 keV Ar$^+$ ion 
beam, which was scanned over the target surface, 
at $\Phi = 0.0233$ mA cm$^{-2}$ with a 
fluence of $1.05 \times 10^{18}$ ions cm$^{-2}$ at 
three temperatures, namely -110$^{\circ}$ C, 
23$^{\circ}$ C and 36$^{\circ}$ C. Under these 
conditions the pattern was only obtained at 
$\theta=23^{\circ}$. In parallel, surface roughness 
increased markedly with temperature from 0.5 nm up to 76.8 nm.

\end{enumerate}

In addition, there is one study \cite{Fan_2005} where it is found that for low
ion fluxes the dot size changes with increasing temperature according with
the existence of Ehrlich-Schwoebel energy barriers \cite{Ehrlich_1966,
Schwoebel_1966} whereas for high ion fluxes the dot size decreases with
temperature. However, Si dots seem to arrange differently from other IBS
nanodot structures induced on Si surfaces, specifically a clear pattern can not be visualized.

\hfill
\paragraph{Nanodot pattern dependence on ion flux }

For GaSb \cite{Facsko_2001, Frost_2003} and InP \cite{Frost_2003} surfaces it
was found that the pattern wavelength was independent of the ion current
density or ion flux for the different ranges sampled, namely $10^{15}-4 \times
10^{15}$ cm$^{-2}$ s$^{-1}$ and $6.2 \times 10^{14}- 5\times10^{15}$ cm$^{-2}$
s$^{-1}$, respectively. For fixed InP targets irradiated at normal 
incidence by 1 keV Ar$^+$ ions at fixed fluence 
and 23$^{\circ}$, Tan and Wee \cite{Tan_2006b} 
found that at low ion fluxes there was not any 
dot pattern but it appeared at j$_{\rm ion}$ = 
0.0174 mA cm$^{-2}$. The pattern in-plane 
hexagonal order increased when ion flux was 
increased up to j$_{\rm ion}$ = 0.0233 mA cm$^{-2}$. In another study, where dot structures did not form a
clear pattern, it was proposed that for ion fluxes below 220 $\mu$A cm$^{-2}$
Ehrlich-Schwoebel energy barriers \cite{Ehrlich_1966, Schwoebel_1966}
dominate while for higher flux values the dot size decreased as $ \sim
1/\Phi^{1/2}$ \cite{Fan_2005}.

\hfill
\paragraph{In-plane order of nanodot pattern}

The nanodot patterns usually present short-range hexagonal in-plane order. The analysis
usually employed for assessing their degree of order is made through the
two-dimensional auto-correlation function $C(\mathbf{r},t)$ of the AFM images 
\begin{equation}\label{funcion_autocorrelacion}
	C(\mathbf{r},t)=\left\langle \frac{1}{L^d}\int \left[h(\mathbf{x}+\mathbf{r},t)h(\mathbf{x},t)-\bar{h}^2(t) \right] d\mathbf{x}  \right \rangle,
\end{equation}
which is a measure of how well a structure matches a space-shifted version of itself \cite{Zhao_2001}.
An example is shown in Fig. \ref{Fig3_8} where an AFM image of the dot pattern is displayed in panel (a)
together with its corresponding 2-D auto-correlation (panel b). Here, six
bright spots are clearly observed indicating the short-range hexagonal ordering of the dot pattern.

When the symmetry of the pattern is large, it can be also assessed through the
two-dimensional Fourier Transform (FFT) of the AFM images \cite{Ziberi_2006},
although this is not the most common case, as IBS patterns usually give ringed
FFTs \cite{Ziberi_2006}.

Besides the symmetry of the pattern, a further issue is to quantify its degree of order in the pattern. As
mentioned for the ripple patterns, two main approaches exist: (a) through size
of the mean peak of the PSD of AFM images, and (b) using synchrotron
techniques such as GISAXS and GID.

Based on PSD data, Bobek and coworkers observed that the range of order of the
nanopattern for GaSb surfaces increased appreciably after 100-200 s of Ar$^+$
ion irradiation \cite{Bobek_2003}. Also with this type of analysis the group of
Frost found that the range of in-plane order decreased with increasing temperature, in
the 268-335 K range, for rotating InP surfaces sputtered at $\theta=30^{\circ}$  by
0.5 keV Ar$^+$ ions \cite{Frost_2004b}.

Two studies have been performed for Si surfaces. For rotating Si substrates
bombarded at $\theta=75^{\circ}$ by either 0.5 keV Ar$^+$ ions, by 1 keV Kr$^+$
ions or by 1 keV Xe$^+$ ions \cite{Ziberi_2006c}; Ziberi and coworkers obtained
that the lateral correlation length, $\zeta$, increased with sputtering time up
to a value close to 170 nm, in the window sampled, when Ar$^+$ ions were
employed. In contrast, for both Kr$^+$ and Xe$^+$, $\zeta$ saturated at $\simeq
145 $ nm, and $\simeq 120 $ nm, respectively. Saturation took place earlier for
Xe$^+$. Moreover, the change of $\zeta$ with ion energy depended on the ion
species. Thus, for Kr$^+$ and Xe$^+$ the order increased with ion energy
until the normalized correlation length, $\zeta/\lambda$, saturated for 1
keV. In contrast, for Ar$^+$ ions $\zeta$ had a maximum value at 0.5 keV.

\begin{figure}[!htmb]
\begin{center}
\begin{minipage}{0.525\linewidth}
\begin{center}
 \includegraphics[width=\linewidth]{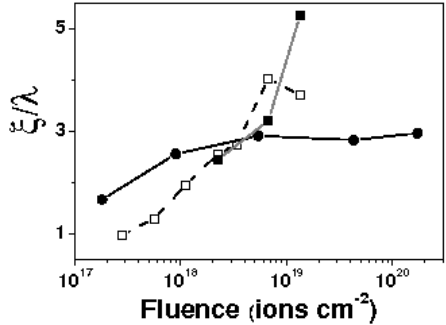}
\end{center}
\end{minipage}\hfill
\begin{minipage}{0.45\linewidth}
\begin{center}
\caption{Normalized ordered domains size, $\zeta/\lambda$, vs ion
dose for: fixed Si(001) target sputtered by 1.2 keV Ar$^+$ ions under normal
incidence with $\Phi = 240$ $\mu$A cm$^{-2}$ ($\bullet$); rotating Si target
sputtered by 0.5 keV Ar$^+$ ions ($\blacksquare$) and 1 keV Kr$^+$
 with $\Phi = 300$
$\mu$A cm$^{-2}$ (o). Data for the rotating substrates have been adapted from
\cite{Ziberi_2006c}. } \label{Fig3_13} 
\end{center}
\end{minipage}
\end{center}
\end{figure}

The second system consists of a fixed Si target bombarded at normal incidence
by 1.2 keV Ar$^+$ ions. In this work, we used both AFM and synchrotron
techniques to study the pattern dynamics in terms of coarsening and ordering as
functions of sputtering time. In Fig. \ref{Fig3_13} we display the data
obtained on Si by our group \cite{Gago_2006} and  by the group of Frost
\cite{Ziberi_2006c} in terms of the normalized ordered domain size (i.e, the
ratio of the lateral correlation length to the pattern wavelength) versus
ion dose. For the rotating Si target the order seems to
increase with ion dose irrespective of the ion species,
although saturation is observed for 1 keV Kr$^+$ ions but not for 0.5 keV Ar$^+$. In principle, it appears that the order of the pattern is larger for the rotating
substrate configuration than for the fixed configuration when Ar$^+$ ions are
employed. However, it should be noted that for 1 keV Ar$^+$ ions the order for
the rotating configuration dropped to less than half that obtained for 0.5 keV
Ar$^+$ ions.

%%%%%%%%%%%%%%Figure%%%%%%%%%%%%%%%%%%%%%%%
%%%%%%%%%%%%%%%%%%%%%%%%%%%%%%%%%%%%%%%%%%%%

\begin{figure}[!hmbt]
\begin{center}
\begin{minipage}{0.525\linewidth}
\begin{center}
 \includegraphics[width=\linewidth]{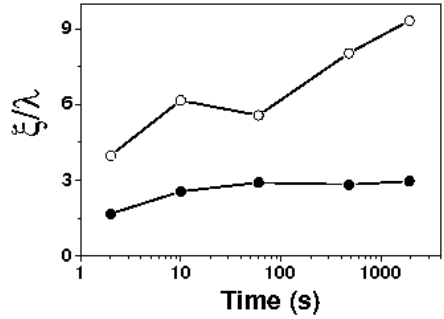}
\end{center}
\end{minipage}\hfill
\begin{minipage}{0.45\linewidth}
\begin{center}
\caption{ Normalized ordered domain size, $\zeta/\lambda$ obtained by
AFM ($\bullet$) and GID (o) versus sputtering time for a Si(001) target sputtered by
1.2 keV Ar$^+$ ions under normal incidence with $\Phi = 240$ $\mu$A cm$^{-2}$
} \label{Fig3_14}
\end{center}
\end{minipage}
\end{center}
\end{figure}

Regarding the quantification of the range of order in the pattern, SPM-based
techniques pose a problem due to their locality. These surface characterization
techniques sample a relative small area of the surface and lack enough
statistics to a certain extent. Thus, in a recent work \cite{Gago_2006}, we
have used AFM and synchrotron based techniques (namely, GISAXS and GID) to
quantify the ordering of the same IBS patterns. The corresponding results for
the normalized ordered domain size are presented in Fig. \ref{Fig3_14}.
Clearly, AFM data saturate at a value very close to 3 while GID data roughly
increase monotonously in the sampled temporal window up to a value close to 10.
This difference is due to the improved sampling statistics of GID with respect
to AFM \cite{Gago_2006}. Thus, although AFM is a well suited technique for
routine characterization of IBS nanopatterns, GID and GISAXS can provide us
with more reliable quantitative data.

%%%%%%%%%%%%%%Figure%%%%%%%%%%%%%%%%%%%%%%%
%%%%%%%%%%%%%%%%%%%%%%%%%%%%%%%%%%%%%%%%%%%%

\subsubsection{Nanohole or nanopit patterning}

In principle, hole nanopatterns can also be induced by IBS. The fact that nanodots or nanoholes are produced on the target surface depends only on the anisotropies of the collision cascades \cite{Kahng_2001}. After
the ion sputtering theory by Sigmund \cite{Sigmund_1969} this shape depends
only, for a given target material and ion species, on the ion energy. However, despite such a
theoretical prediction, there is not clear evidence, up
to now, of hole or pit pattern production by IBS. Two main pit-structures have been produced by
IBS: the so-called cellular structures \cite{Chen_1997, Chey_1995,
Frost_2004c, Hofer_2004} and hole structures on semiconductor
heterostructures \cite{Lau_2003}. The former look like network of hole-like structures, with a relatively wide distribution of hole
sizes, resembling those obtained in plasma etching of silicon \cite{Zhao_1999}.
In contrast, those induced on semiconductor heterostructures display a more
homogeneous hole size distribution with typical hole diameter of $ 170 \pm 30$
nm and hole spacing of $ 190 \pm 40$ nm.

\begin{figure}[!hmbt]
\begin{center}
\begin{minipage}{0.75\linewidth}
\begin{center}
 \includegraphics[width=\linewidth]{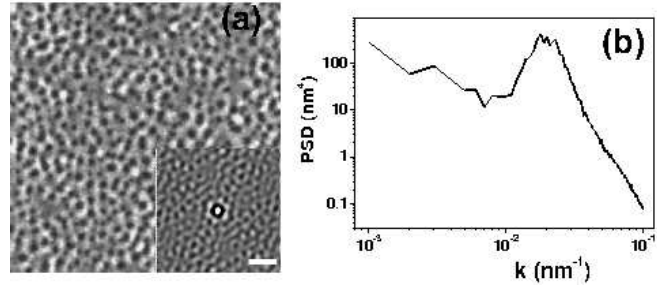}
\end{center}
\end{minipage}
\caption{ (a) $1 \times 1$ $\mu$m$^2 $AFM image of a nanohole pattern induced
on Si surfaces irradiated by 1 keV Ar$^+$ ions under normal incidence for
a fluence of $4 \times 10^{17}$ ions cm$^{-2}$. Inset: two-dimensional
auto-correlation function corresponding to a surface area of $740 \times 740$
nm$^2$ in which the short-range hexagonal order is observed. The horizontal bar
corresponds to 165 nm. (b) Radially averaged PSD function corresponding to the
image of panel (a). The peak observed corresponds to an average inter-hole
distance of $\approx 50$ nm.} \label{Fig3_16}
\end{center}
\end{figure}

Recently, we have achieved the production of nano-hole patterns by IBS on
silicon surfaces. These patterns display a characteristic wavelength similar to
that obtained on the standard IBS experiments on dot pattern production. As an
example we present in Fig.\ \ref{Fig3_16}a a typical AFM image of a nanohole
IBS pattern. The hole structures are 2-3 nm deep and have a lateral size in the
30-40 nm range. It should be noted that both the hole depth and lateral size can be underestimated
because of tip convolution effects. The inset of Fig.\ \ref{Fig3_16}a displays
the auto-correlation of an AFM image taken over an area of  $740 \times 740$
nm$^2$. Although there is some distortion, the short-range hexagonal order is
clear. Finally, in  panel (b) the radially averaged PSD function of image
(a) is presented. Similarly to the results obtained above for nanodot patterns,
the PSD presents a clear peak that is associated with the characteristic pattern
wavelength that in this case is close to 50 nm.

%%%%%%%%%%%%%%Figure%%%%%%%%%%%%%%%%%%%%%%%
%%%%%%%%%%%%%%%%%%%%%%%%%%%%%%%%%%%%%%%%%%%%

\subsubsection{General considerations}

In general, the behaviors found for ripple and dot  nanopattern formation
are analogous. Nevertheless, the ripple patterns present specific properties,
due to the anisotropy of the process, such as  ripple orientation, ripple
propagation and shadowing. Analyzing the behaviors reported for the changes of
both pattern morphologies with other parameters, the following map of trends can
be drawn (we collect the main experimental observations in tables
\ref{table_exp1} and \ref{table_exp2}):

\begin{itemize}

\item IBS patterns are produced with different ion species, even with those implying reactive sputtering.

\item IBS patterns are produced on a large variety of substrates: amorphous, semiconductors and metals.

\item The pattern symmetry reflects that of the experimental set-up (i.e., anisotropic or isotropic).

\item In general, the pattern in-plane order increases with fluence (i.e,
sputtering time).

\item Frequently, the pattern wavelength coarsens with time to finally
saturate.

\item The surface roughness initially increases sharply to later on either
saturate or increase at a lower rate.

\item At low temperatures, the pattern wavelength does not depend on target
temperature and ion flux, but it increases with ion energy.

\item Ripple patterns:

\begin{itemize}

\item Ripples run perpendicular or parallel to the ion beam projection direction depending on the ion beam incidence angle.

\item Ripples propagate with a non-uniform velocity.

\item Shadowing effects appear in the ripple temporal evolution for long sputtering times and large $\theta$ values.

\item Ripples are produced by ions with energies ranging from just a few hundreds of eV up to $10^5$ eV.

\item  At high temperatures, $\lambda$ increases with target temperature following the Arrhenius behavior and decreases with ion flux and ion energy.

\end{itemize}

\item Dot patterns:

\begin{itemize}

\item Patterns are produced mainly with in-plane hexagonal order but also with square symmetry (only for oblique incidence and rotating targets).

\item To date, dot patterns are only produced by low-energy ions.

\item Different dependences with target temperature have been found implying either no change, increase or decrease of $\lambda$ with increasing target temperature. In addition, for InP the pattern symmetry changes from hexagonal to square with increasing temperature.

\end{itemize}

\end{itemize}

In general, the behaviors for ripple patterns seem to be better established
than for dot patterns. This may be related to the fact that dot pattern production
by IBS is relatively recent. In particular, the dependence of the morphological
properties of these patterns with temperature, ion energy and fluence has to be
systematically addressed in order to obtain a more general picture of the
nanopatterning process. On the other hand, both types of IBS patterns share
many behaviors, which is consistent with the corresponding theoretical models.

Also, there is a further experimental issue to be addressed: the
possible influence of technical parameters on IBS pattern production.
This important problem has been only studied by Ziberi and coworkers
\cite{Ziberi_2004}. They obtained that the settings of the Kaufman ion-gun, the
one that they employed, can affect the IBS nanopattern. In particular, they
found that the divergence of the ion beam as well as the angular distribution
of the ions within the ion beam influenced the pattern formation. If this happens already for a given ion-gun, we should likely expect some differences also when using other ion-gun types. Thus, it is convenient to perform systematic experiments on different targets
and under different experimental conditions with the same equipment. This
approach would allow a more direct comparison and contrast of the different
experimental findings, which also would contribute to improve contrast between these results
and the theoretical predictions.

\begin{center}
\begin{sidewaystable}[!htbp] \label{table_exp1}
\begin{tabular}{|c|c|c|c|c|}\hline\hline

\multicolumn{5}{|c|}{Oblique incidence}\\\hline

&Temperature&Flux&Time&Energy\\\hline

$\lambda$ 

& $\begin{array}{c} \textrm{const. (low }
T) \\ \textrm{~\cite{Carter_1996,Mayer_1994}} \\ \\
 T^{-1/2}e^{-\Delta T/2K_BT}\textrm{ (high
}T) \\ \textrm{~\cite{Habenicht_2001,Makeev_2002,Mayer_1994,Vajo_1996}}
\end{array} $ 

& $\begin{array}{c} \textrm{ const. (low
$\Phi$)} \\~\cite{Brown_2005,Datta_2001,Flamm_2001,Vajo_1996,Ziberi_2006b}
\\  \\%\sim \Phi^{-1/2}
\textrm{decreases (high
$\Phi$)} \\ ~\cite{Chason_2001,Erlebacher_1999} \end{array} $

&$\begin{array}{c}
 \textrm{const. (short $t$)}\\  \\  t^n\textrm{ (intermediate $t$)
}\\ \left\{\begin{array}{l} n=0\textrm{(i.e. no coarsening)} \\ ~\cite{Erlebacher_1999,Ziberi_2005}\\ \\
0.15\leq n\leq
1\\ \textrm{~\cite{Brown_2005,Carter_1996,Datta_2004,Flamm_2001,Habenicht_2002,Karmakar_2005,Toma_2005}}
\end{array}\right.\\ \\ \textrm{const. (large $t$)}\\
\end{array}$

&$ E^m \left\{\begin{array}{l}0.2\leq m \lesssim 1 \\ \textrm{~\cite{Alkemade_2001,Chini_2002,Flamm_2001,Habenicht_2002,Karmakar_2005,Umbach_2001,Vajo_1996,Ziberi_2005}}\\ 
\\m<0\textrm{~\cite{Brown_2005,Chini_2002}}\end{array}
\right.$
\\\hline
$W$
&decreases~\cite{Habenicht_2001}
& $\begin{array}{c} \textrm{ const. } \\~\cite{Flamm_2001,Ziberi_2006b} \\ \\
\textrm{decreases} \\ ~\cite{Liu_2001} \end{array}$

&$\begin{array}{c}
 \e^{\omega t}\textrm{ (short $t$})\\ \\
 t^\beta\textrm{(intermediate $t$)}\\
\\ \textrm{const. (large $t$)}\\
\end{array}$
&$\begin{array}{l}\textrm{const. (with beam scanning)}
\\\textrm{increases (without beam scanning)}
\end{array}$~\cite{Chini_2002}\\\hline
$v$& NR & NR & $\begin{array}{c} t^{-0.75}\textrm{ (short
}t)\\\textrm{const. (large }t) \end{array}$~\cite{Habenicht_2002}& NR\\
\hline\hline
\end{tabular}
\caption{Summary of experimental pattern behaviors for oblique incidence IBS onto amorphizable targets. NR stands for {\em Not reported}}
\end{sidewaystable}
\end{center}

\begin{center}
\begin{sidewaystable}[!htbp]\label{table_exp2}
\begin{tabular}{|c|c|c|c|c|}\hline\hline
\multicolumn{5}{|c|}{Normal incidence}\\\hline
&Temperature&Flux&Time&Energy\\\hline
$\lambda$ &$ \begin{array}{c} \textrm{const.~\cite{Facsko_2001}} \\ \\
\textrm{decreases~\cite{Gago_2006b}} \end{array}$ & const.~\cite{Facsko_2001}
&
$t^n$ (i.e. coarsening)~\cite{Bobek_2003,Gago_2001,Gago_2002,Ozaydin_2005,Tan_2006,Xu_2004}&$
 E^{0.5}$~\cite{Facsko_2001}\\ \hline $W$ & decreases~\cite{Gago_2006b} &
NR & $\begin{array}{c}  e^{\omega t}\textrm{ (short }t) \\ 
t^\beta\textrm{ (intermediate } t)\\ \textrm{const. (large }t) \end{array}$& NR \\ \hline
Order & Hexagonal & Hexagonal~\cite{Facsko_2001}& Hexagonal (enhancement with
time)~\cite{Bobek_2003,Gago_2006}&Hexagonal~\cite{Facsko_2001} \\ \hline
\hline
\multicolumn{5}{|c|}{Oblique incidence with rotating substrate}\\\hline
&Temperature&Flux&Time&Energy\\\hline
$\lambda$ &increases with $T$~\cite{Frost_2000} &NR
&$\begin{array}{c}t^{0.26}\textrm{ (intermediate $t$)~\cite{Frost_2000}}\\\textrm{const. (i.e. no coarsening)
}~\cite{Ziberi_2005b,Ziberi_2006c}
\end{array}$ & Increases~\cite{Frost_2000,Ziberi_2006c} \\ \hline
$W$ &increases with $T$~\cite{Frost_2000}
&increases~\cite{Frost_2000,Ziberi_2006} &$t^\beta\textrm{ with
}\left\{\begin{array}{c}
\beta=0.8\textrm{, short }t \\\beta=0.27\textrm{, large }t
\end{array}\right.$~\cite{Frost_2000}
& Increases~\cite{Frost_2000,Ziberi_2006c}\\ \hline
Order & $\begin{array}{c} \textrm{Hexagonal (low }T) \\ \textrm{Square 
(high }T)
\end{array}$~\cite{Frost_2004}  & NR & $\begin{array}{c}\textrm{Disordered
at short }t \\ \textrm{Hexagonal (enhanced order)  (large
$t$)~\cite{Ziberi_2005b,Ziberi_2006c}}
\end{array}$&Increases~\cite{Ziberi_2006c} \\ \hline
\end{tabular}
\caption{Summary of experimental pattern behaviors for normal incidence and rotating target IBS of amorphizable targets. NR stands for {\em Not reported}}
\end{sidewaystable}
\end{center}

\subsection{Pattern formation in single-crystal metals by ion beam sputtering}

The general observation of ripple and nanodot formation by off-normal and
isotropic ion beam irradiation is only valid in the case of amorphous materials
or for those that amorphize upon ion bombardment, such as
single-crystal semiconductors. The case of metals should be treated through a
different approach since the above properties do not occur. For
example, ripples can be produced at normal ion incidence, or isotropic patterns
may be formed under off-normal irradiation \cite{Rusponi_1997}. A large
number of experimental results reported in metals has been compiled by Valbusa
et al. \cite{Valbusa_2002}.

The peculiar behavior of metal surfaces relies in the nature of the metallic
bond. Due to its non-directional character, metals do not amorphize upon ion
bombardment, at least for low fluence and up to energies of a few keV
\cite{Valbusa_2002}.  Therefore, the surface retains its properties, the ion
bombardment being only responsible for producing vacancies or vacancy
aggregates at the surface \cite{Costantini_2001}, which increase the already
large surface diffusion in these systems. Another relevant constraint for
surface diffusion in metals comes from the presence of Ehrlich-Schwoebel (ES)
energy barriers for adatoms to descend steps \cite{Ehrlich_1966,
Schwoebel_1966}. These contributions add more complexity to surface
nanostructuring by IBS in the case of metals.

\begin{figure}
\center
\includegraphics[height=6.25cm]{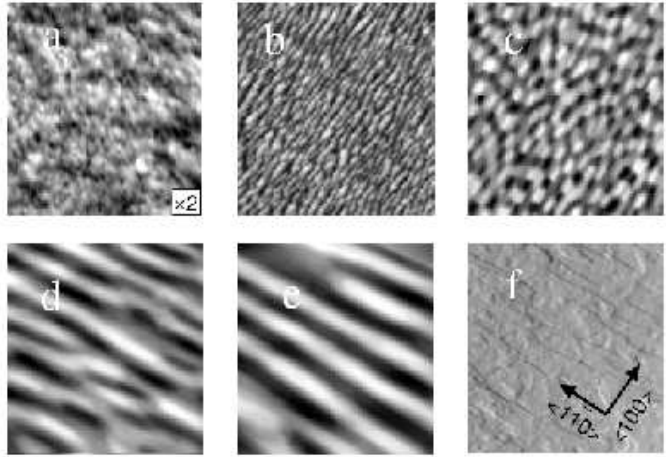}
\caption{STM images ($350\times 350$ nm$^{2}$) of Ag(110) surfaces bombarded at
normal incidence with 1 keV Ar$^{+}$ for 15 min. at (a) 160 K, (b) 230 K, (c)
270 K, (d) 290 K, (e) 320 K and (f) 350 K (taken from Ref.\
\cite{Valbusa_2002}).} \label{Fig3-X} \end{figure}

The {\itshape diffusive regime} \cite{Valbusa_2002} for pattern formation in metals appears when
the ES energy barrier or a preferential diffusion path determines the final
pattern characteristics. In this regime, the erosion process is limited to
incorporating a larger number of diffusive particles (such as atoms or
vacancies) that align along the more thermodynamically favorable directions.
This means that, in contrast with amorphous or semiconductor materials, the
intrinsic properties of the surface reflect onto the resulting surface pattern.
For example, the pattern will reflect the intrinsic isotropic characteristics
of the material as in the case reported for Pt(111) by Michely et al.
\cite{Michely_1991}. In the presence of diffusion anisotropy, as shown in Fig.\
\ref{Fig3-X} for Ag(100) \cite{Rusponi_1997}, ripples may be observed even
under normal ion incidence, or the ripple orientation may vary with
temperature. In this context, temperature governs thermal activation of
diffusion pathways, leading under certain conditions to isotropic or
anisotropic configurations. Additional relevant conclusions extracted from
Fig.\ \ref{Fig3-X} are that the pattern wavelength increases with temperature
as a result of enhanced surface diffusion and that the pattern coarsens with
sputtering time. It has also been observed that the pattern wavelength under
this diffusive regime depends only slightly on the ion energy.

\begin{figure}
\center
\includegraphics[height=3.5cm]{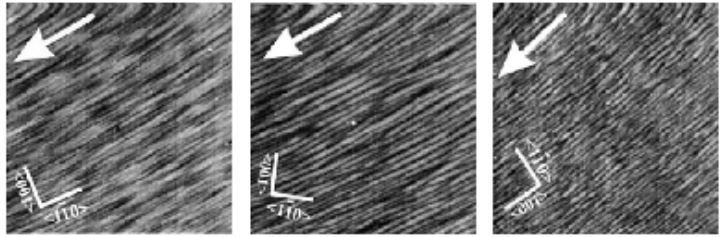}
\caption{STM images ($400\times 400$ nm$^{2}$) of Cu(110) surfaces bombarded at
$70^{\circ}$ with respect to the surface normal and 1 keV Ar$^{+}$ for 15 min.
at different angles between the beam projection and the surface orientation
(adapted from Ref.\ \protect\cite{Valbusa_2002}).} \label{Fig3-Y} \end{figure}

The {\itshape erosive regime} in metals \cite{Valbusa_2002} is attained only for near glancing
incidence angles and at low temperatures in order to inhibit thermal surface
diffusion. In this regime the pattern formation can not be correlated with any
symmetry of the surface, such as the orientation of the crystal. This fact is
shown for Cu(110) surfaces in Fig.\ \ref{Fig3-Y} where the ion beam projection
is modified with respect to the surface orientation without relevant change on
the pattern characteristics \cite{Rusponi_1998}. This implies that IBS patterns
can be aligned in directions that are not thermodynamically favorable, this
fact being one of the major potentials of surface nanostructuring by IBS in
contrast with other techniques such as MBE. In
this regime, the ripples are parallel to the beam direction, their wavelength
depends linearly with the ion energy and both, wavelength and roughness,
increases with fluence \cite{Valbusa_2002}.

Finally, ion bombardment of metals does not always results in the formation of
a regular pattern on the surface. In this case, the surface may additionally display
kinetic roughening or not depending on the conditions. An example of these
results was obtained by Krim et al. \cite{Krim_1993} on Fe surfaces bombarded with 5 keV
Ar$^{+}$ at $\theta = 25^{\circ}$.

The different parameter
windows for pattern formation, smoothening or roughening have been studied by Chason
et al. \cite{Chason_2006} for Cu(001) surfaces as a function of ion dose and
target temperature. In this scheme, the surface remains flat for low ion fluxes
and high temperatures whereas it roughens for high fluxes and low temperatures.
For intermediate temperatures (200-350 K) the diffusive regime dominates and,
finally, high ion fluxes and high temperatures imply an erosive regime.

\subsection{Pattern formation in thin metal films by ion beam sputtering}

In the previous section, we have seen that ripples can be also produced by IBS
on single-crystal metals for which ES energy barriers \cite{Ehrlich_1966,
Schwoebel_1966} play such a crucial role, influencing the stability of the
ripple morphology even at room temperature \cite{Valbusa_2002}. However, a
different scenario has to be considered for IBS nanostructuring of thin
metallic films. These systems are usually polycrystalline and mainly formed by
grains that are randomly oriented, and with sizes that are usually much smaller
than the film thickness \cite{Jeffries_1996, Wei_1998}. For such systems, the
existence of ES barriers becomes improbable due to the lack of well-defined
atomic steps at the surface \cite{Jeffries_1996}. These facts support the
effective existence of an isotropic surface diffusivity rather than an
anisotropic one. In this sense, thin metal films subject to IBS nanostructuring
processes would be akin to the case of amorphous or amorphizable surfaces. Two
groups have reported their findings for such systems.

Karmakar and Ghose have produced ripple structures on Co, Cu, Ag, Pt and Au
thin film (thickness in the 30-200 nm range) surfaces irradiated at
80$^{\circ}$ by 16.7 keV Ar$^+$ ions \cite{Karmakar_2004}. They found that the
ripple wavelength as a function of the target element was qualitatively
consistent with the behavior predicted by Makeev et al.\ \cite{Makeev_2002}.
Under this approach $\lambda$ depends mainly on the lateral and longitudinal
straggling widths of the ion cascade, as well as on the angle of incidence.
Karmakar and Ghose obtained these values from the computer code SRIM
\cite{Ziegler_2006}, and found qualitative agreement between the theoretical
and experimental behaviors. However, there are sizeable quantitative
differences. This is not new, quantitative disagreement between calculated and
measured $\lambda$ values having been previously pointed out by different
groups \cite{Kim_2003, Facsko_2001, Gago_2006b}. Another interesting finding
refers to the stability of the induced ripple patterns at room temperature
\cite{Karmakar_2004} in comparison with those produced on single crystals
\cite{Valbusa_2002}. This fact is an indication that the thermally activated
diffusion energy barriers in thin polycrystalline films are comparatively
higher than in single crystal metal surfaces, which agrees with the assumption on the absence of ES barriers. In fact, measurements of activation
energies for adatom surface diffusion on various polycrystalline materials are
consistent with this result \cite{Rossnagel_1982}. The studies of Stepanova and
coworkers consisted in the irradiation by a 1.2 keV Ar$^+$ ion beam of a 50
nm-thick Cu film deposited on glass or silicon substrates \cite{Stepanova_2005,
Stepanova_2006}. When the irradiation angle was oblique, at $\theta=82^{\circ}$, a
clear ripple pattern was induced. These experiments allowed to discard any
possible influence of the underlying substrate (either glass or silicon) in
terms of composition and crystalline structure on the production of ripple
patterns on the top metal thin film. Another set of experiments was
performed under the same conditions but at normal incidence. In this case, a
self-assembled network of $\approx$ 20 nm-sized ring-like Cu features was
observed on a SiO$_2$ substrate after 1.2 keV Ar$^+$ beam etching of the
Cu/substrate interface \cite{Stepanova_2005}.

The previous studies suggest that IBS of interfaces might be a reasonable
process to fabricate metal nanopatterns on nonmetallic substrates. Also, the
specific properties of these thin film polycrystalline metal systems make them
suitable of study through to the same approach that is employed to understand
IBS nanopatterning of amorphous/amorphizable surfaces.

\section{Theoretical approaches}
\label{sec:theo}
% Always give a unique label
% and use \ref{<label>} for cross-references
% and \cite{<label>} for bibliographic references
% use \sectionmark{}
% to alter or adjust the section heading in the running head

As seen in the previous sections, typically the IBS induced nanopatterns fully
evolve in macroscopic time and length scales (minutes and several microns,
respectively). It is at these scales where, e.g., interaction among ripples can
be seen to lead to order improvement with fluence or, rather, to eventual
disorder in heights. Although detailed knowledge on the phenomenon of
sputtering is rapidly and consistently developing through (microscopic)
Molecular Dynamic type of studies (see e.g.\ \cite{Bringa_2002,Nord_2003} and
references therein, and Sec.\ 2), the scales that are currently reachable to
these methods remain in the 1 ms and 50 nm ranges. Monte Carlo (MC) and
continuum methods (CM) can probe larger scales so that these are the approaches
that we will consider in what follows.

\subsection{Sigmund's theory of sputtering}

\begin{figure}[!htmb]
  \vspace*{2mm}
  \begin{center}
    \begin{minipage}{0.45\linewidth}
   \includegraphics[width=\linewidth]{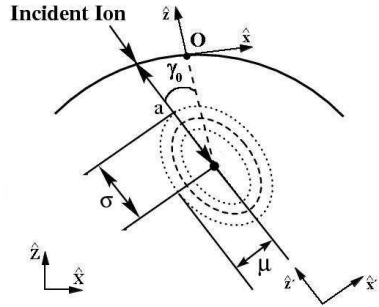}
    \end{minipage}\hfill
    \begin{minipage}{0.5\linewidth}
      \caption{Two-dimensional profile of the distribution of energy deposition, Eq.\ (\ref{distrib1}).
      $\sigma$ and  $\mu$ are the transversal and longitudinal widths of the Gaussian, respectively.
      The reference frames considered in the text are also shown. Adapted from
      \protect\cite{Makeev_2002}.}\label{ejes}
    \end{minipage}
  \end{center}
\end{figure}

As mentioned in section \ref{Theory_sputtering}, in a classic work Sigmund
analyzed the kinetic transport theory of the sputtering process
\cite{Sigmund_1969}. Assuming an infinite medium, he found that, in the elastic
collision regime at the energies of a few keV where electronic stopping is not
dominating, the deposited energy can be approximated by a Gaussian distribution
near its maximum. Specifically, the density of energy spread out in the bulk by
an ion with kinetic energy $E$ is given by
\begin{equation}
    \varepsilon_s(\mathbf{{ r'}} )= E N_s \, e^{-\frac{x'^2+y'^2}{2\mu^2}}
e^{-\frac{(z'+a)^2}{2\sigma^2}},    \label{distrib1}
\end{equation}
where the origin of the $\mathbf{{ r'}}=(x',y',z')$ coordinate system is placed
at the impact point of the ion within the surface; $\mathbf{\hat{z'}}$ is
aligned along the ion beam direction, and $\mathbf{\hat{x'}}$ and
$\mathbf{\hat{y'}}$ belong to the perpendicular plane to $\mathbf{\hat{z'}}$,
see Fig.\ (\ref{ejes}).
%%%%%%%%%%%%%%%%%%%%%%%%%%%%%%%%%%%%%%%%
%%%%%%%%%%%%%%%%%%%%%%%%%%%%%%%%%%%%%%%%
In Eq.\ (\ref{distrib1}), $N_s=\left[(2\pi)^{3/2}\sigma\mu^2\right]^{-1}$ is a
normalization constant and, due to the initial momentum of the ion, the maximum
of energy deposition occurs at a distance $a$ along the ion trajectory inside
the bulk. The longitudinal and transversal straggling widths of the
distribution are $\sigma$ and $\mu$ respectively. If mass differences between
substrate and ion are not too large, the quality of the approximation is
reasonable and expression (\ref{distrib1}) provides a good approximation for
polycrystalline and amorphous targets \cite{Sigmund_1969}. Despite the fact
that $a$ depends on the microscopic details of the interactions between the ion
and the bulk, at intermediate energies ($10-100$ keV), it is usually considered
to be proportional to $E$. Within this range of energies, $\sigma$, $\nu$ and
$a$ are of the order of a few nanometers. This point will be relevant when
information about the evolution of the topography is obtained.

As we noted above, correlation of the sputtering yield with the surface
geography is a crucial issue. In Ref.\ \cite{Sigmund_1973}, Sigmund showed that
the topography of the surface can indeed influence the magnitude of the rate of
erosion and, provided an analytical description that describes the increase of
yield for geometries different from the flat morphology. It is assumed that the
speed of erosion at a point $O$ on the surface is proportional to the amount of
energy deposited there by the ions, with a proportionality constant $\Lambda$
that is characteristic of the substrate and depends on the atomic density of
the target $n_v$, the atomic energy of connection in the surface $U_0$, and a
proportional constant $C_0$ which is related to the square of the effective
radius of the potential of effective interaction according to
\begin{equation}
\Lambda= \frac {3} {4 \pi^2 n_v U_0 C_0}.
\end {equation}

Under these hypotheses, it is possible to obtain the mathematical expression
for the rate of volume eroded in $O$. This is simply given by
\begin{equation}\label{vO1}
V_O = \Lambda \int_\mathcal{R} \Phi({r'})  \varepsilon_s({ r'}) d \mathcal{R}.
\end{equation}
The integral extends to the region $\mathcal{R}$ where the impact of the ions
contributes to energy deposition at $O$. The term $\Phi({r'})$ represents the
local flux. We have specified its dependence with ${r'}$ to show the
corrections due to the local geometry to the homogenous flux $\Phi_0$. From
(\ref{vO1}), Sigmund derived the rate of erosion for diverse artificial
geometries, such as a flat surface followed of an inclined plane or a vertex,
in the case of normal incidence.

\begin{figure}[!htmb]
   \begin{center}
    \begin{minipage}{0.45\linewidth}
        \includegraphics[width=\linewidth]{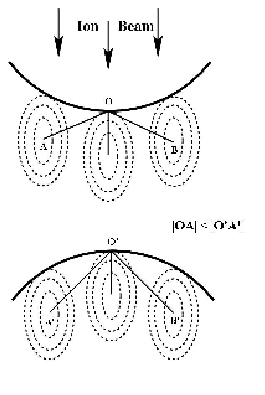}
    \end{minipage}\hfill %\hspace{0.025\linewidth}
    \begin{minipage}{0.50\linewidth}
      \caption{Sketch of deposited energy for two different profiles.
      The energy deposited at $O$ is larger than at $O'$.
      This induces more erosion in surface valleys than in crests,
      which produces a morphologic instability in the system.
       Adapted from \protect \cite{Makeev_2002}.} \label{inestabilidad}
    \end{minipage}
  \end{center}
\end{figure}

The previous description, as already observed by Sigmund \cite{Sigmund_1973},
and briefly noted in Sec.\ \ref{Theory_sputtering} above, implies the
occurrence of a morphological instability. Let us suppose that we irradiate a
certain surface with an homogenous flow, as shown in Fig.\ \ref{inestabilidad}.
We can verify that distances $ \overline{OA} $ and $ \overline{OB} $ are
smaller than $ \overline{O'A'}$ and $ \overline{O'B'}$ due to the geometry of
the interface. This implies that, for this energy distribution, the
penetrations of ions at $A$ and $B$ induce large energy deposition at $O$ than
the impacts on $A'$ and $B'$ at $O'$. As the rate of erosion is proportional to
the deposited energy, erosion is faster at $O$ than at $O'$. Thus, valleys are
excavated more quickly that crests amplifying initial differences in heights.
Sigmund suggested that an alternative process that flattens the surface must
exist and he proposed atomic migration as a mechanism to correct this
instability.

The theory of Sigmund is the basis for most of the later continuum approaches
to the dynamics of surfaces undergoing by IBS. In principle, as a \emph{theory
of sputtering}, and as mentioned in Sec.\ \ref{patterns_exp}, it is known to
have limitations. Within the range of energies we are considering, two of the
most conspicuous ones refer to: (i) the behavior of the sputtering yield for
large incidence angles; (ii) crystalline targets. Related with the former,
Sigmund's theory predicts an steady increase of the yield with incidence angle,
not being able to account for the well-known decrease of $S$ at glancing
incidence due to more efficient projectile reflection and channeling effects at
the surface. Nevertheless, recent continuum studies \cite{Chen_2005} have shown
that improved distributions accounting for this effect (see in \cite{Chen_2005}
and \cite{Nastasi_1996}) do not modify qualitatively the morphological
predictions to be derived below from the energy distribution (\ref{distrib1}).
Similarly, channeling effects limit in principle the applicability of Sigmund's
theory to crystalline solids [point (ii)], by inducing sizeable departures from
the energy distribution as described by (\ref{distrib1}). Thus, different
distributions must be considered, as done e.g.\ in \cite{Feix_2005} for the
particular case of Cu targets. However, as seen in this work, again these
modifications do not alter the qualitative morphological implications of the
continuum theories to be described below, which justifies (at least for
temperatures at which Ehrlich-Schwoebel barriers do not dominate surface
diffusion \cite{Valbusa_2002}) the strong similarities in nanopattern formation
by IBS on metals, as compared with amorphizable targets.

As a \emph{morphological theory}, again Sigmund's \cite{Sigmund_1973} has some
limitations: it does not predict the alteration of the morphology during the
process for scales much larger than penetration depth for ions; as surface
diffusion is not considered, the wavelength of the pattern needs to be of the
order of the length scales appearing in the energy distribution Eq.\
(\ref{distrib1}); the effects of surface shadowing or redeposition are not
considered; it does not predict the time evolution of the morphology and how it
affects the rate of erosion. Thus, an additional physical mechanisms and a more
detailed description of the surface height are needed in order to derive a
morphological theory with an increased predictive power.

\subsection{Monte Carlo type models}

(Kinetic) MC methods have a long and successful tradition in the context of IBS
\cite{Ziegler_1985,Ziegler_2006}. For our morphological purposes, they have
found increased application during the last decade. Their main advantage is
that, in principle, they allow to make a more direct connection to microscopic
relaxation mechanisms than e.g.\ continuum approaches. In addition, they do not
require analytical approximations, lending themselves, rather, to numerical
simulations that can reach reasonably large scales (around 10 s and around 100
monolayers of erosion). However, as compared with CM, dependencies of
observables with physical parameters are less direct, accessible scales are
shorter and universal (i.e.\ substrate/ion independent) properties are harder
to assess unambiguously. Actually, the two types of approaches ---MC and CM---
to some extent complement each other so that, although our focus will be mostly
on CM, we consider shortly some of the main results of the most recent MC
studies.

In order to access large scales as mentioned, MC methods consider physical
relaxation rules in which microscopic details ---such as specific interactions
among ions and target atoms--- are coarse-grained out and replaced by effective
dynamical rules. These have to represent, on the one hand, the events leading
to erosion by sputtering and, on the other hand, competing relaxation
mechanisms such as surface diffusion. For the sputtering processes, starting
with the study of \cite{Cuerno_1995b} many MC models consider an erosion rule
that is either a discretization \cite{Cuerno_1995b} or, rather, a direct
implementation
\cite{Hartmann_2002,Tok_2004,Brown_2005b,Yewande_2005,Yewande_2006,
Chason_2006b} of Sigmund's continuum law for energy deposition due to linear
collision cascades (with corrections for large local incidence angles
\cite{Stepanova_2004,Stepanova_2006}). Actually, this energy deposition rule
has been extended even to MC studies of different physical processes, such as
e.g.\ laser induced jet chemical etching \cite{Mora_2005}. Nevertheless, other
MC models do consider more microscopic (albeit still approximate) rules for
erosion, such as the binary collision approximation similar to that employed in
{\tt SRIM} simulations \cite{Ziegler_1985,Ziegler_2006}, as done in the works
by Koponen and collaborators
\cite{Koponen_1996,Koponen_1997,Koponen_1997b,Koponen_1997c,Koponen_1997d}. Or
else, still different approaches to the sputtering events consider dynamical
creation of surface adatoms/advacancies
\cite{Chason_1997,Murty_1998,Strobel_2001}.

As for the additional surface relaxation mechanisms, most of the works
mentioned restrict their nature to surface diffusion (there being additional
physical possibilities, such as redeposition, viscous flow, etc.) for adatoms
and/or advacancies. The rates for diffusion vary among the different studies,
being either dependent on the energies of the initial and final states
---within a local equilibrium approximation--- or, rather, following
temperature-activated Arrhenius laws; see a recent overview in
\cite{Chason_2006b}.

Although there are some differences in the morphological dynamics predicted
by the various MC models, there is agreement in a number of basic properties,
such as: (i) the occurrence of a morphological instability at small/intermediate
times, that is well described by CM; (ii) at oblique ion incidence, the
non-linear evolution of the ripples as reflected in their non-uniform
transverse motion; (iii) the saturation (stabilization) of the surface
roughness at sufficiently long times. Differences among various models do occur
in their predictions: for instance, in some models the dominant pattern
wavelength remains fixed in the course of time
\cite{Cuerno_1995b,Koponen_1996,Koponen_1997,Koponen_1997b,Koponen_1997c,Koponen_1997d,Chason_2006b},
while in other models a consistent wavelength coarsening phenomenon occurs
\cite{Hartmann_2002,Stepanova_2005,Yewande_2005,Yewande_2006,Chason_2006,Stepanova_2006}
with properties (like the coarsening exponent value) that depend on physical
parameters like the substrate temperature. Then, there are properties
---such as the the kinetic roughening of the interface height fluctuations at
long times--- about which some of these studies do not extract conclusions (if
they do not focus on such large scale issues), while other works reach
different quantitative results (regarding e.g.\ the values of the scaling
exponents, although most predict their occurrence).

\subsection{Continuum descriptions}

The wide separation in scales between the microscopic events that take place in
the target during the irradiation process, and the large-scale response of its
surface morphology recall a similar between microscopic and collective motion in the case of e.g.\ fluid dynamics. Thus, it is natural to expect \cite{Cuerno_2007} that some continuum description in the spirit of,
say, the Navier-Stokes equations might be appropriate in the case of
nanopattern formation by IBS. The advantage of continuum descriptions, if
available, is that they provide compact descriptions of complex physical
phenomena. Moreover, these frameworks are frequently more efficient
computationally for the study of large scale properties, whose generic
properties can be faithfully described by them \cite{Krug_2004}. Nevertheless, for the
type of non-equilibrium phenomena which are our present focus, the relevant
continuum equations are most times non-linear, their study requiring either
analytical approximations or numerical simulations.

\subsubsection{Dynamics of the surface height}

In order to put forward a continuum description of these non-equilibrium
systems, Sigmund's theory \cite{Sigmund_1969,Sigmund_1973} constitutes an ideal
starting point. This is so because Eq.\ (\ref{vO1}) already provides an explicit
expression for the most relevant kinetic quantity describing the sputtering
problem, namely, the local velocity of erosion. In turn, one may add different
contributions to the right hand side of this equation if additional physical
mechanisms contribute to the local variation of the target height, and this has
indeed been done in order to improve the physical description. Still, for
theoretical work, (\ref{vO1}) is technically a very complex expression. The
insight of Bradley and Harper (BH) \cite{Bradley_1988} is that, under
reasonable physical approximations, (\ref{vO1}) becomes a closed time $(t)$
evolution equation for a single physical field, the target height $h(x,y,t)$
above point $(x,y)$ on a reference plane. It is within such (simple)
single-field approach that most of the recent continuum descriptions have
circumscribed themselves. However, we will see in this section that this
program encounters consistency problems ---that question its physical
applicability--- unless the dynamics of additional independent fields are taken
into account. The evolution of these fields ---e.g.\ the density of species
that diffuse on the surface--- also modifies the local height velocity. This
does not mean that a description by a closed equation for the height field is
inappropriate but, rather, that to correct such description, one has to
necessarily take into account for its derivation physical mechanisms that complement
Sigmund's formula. In this section we recall the main results obtained along
BH's pioneering ``single-field'' height approach, while the next section will
(mostly) consider recent developments that incorporate the dynamics of
additional physical fields to the continuum description.

\hfill
\paragraph{Bradley and Harper's theory}

Bradley and Harper (BH) \cite{Bradley_1988} derived a partial differential
equation to describe the evolution of the morphology assuming that the
variation in the height of the surface is smooth when seen at a scale that is
comparable with the average penetration length of the ions. By expanding the
target height as a function of the surface geometry up to linear order in the
surface curvatures, they obtained a closed evolution equation for the surface
height. In the following we recall the main steps of the derivation.

Define a local coordinate system with the $\mathbf{\hat{z}}$ axis oriented
along the surface normal at $O$ and the $\mathbf{\hat{x}}$ and
$\mathbf{\hat{y}}$ axes lying on the perpendicular plane to $\mathbf{\hat{z}}$
(i.e.\ in the tangent plane through $O$) where $\mathbf{\hat{x}}$ is aligned
along the projection of the ion beam onto this plane as shown in Fig.\
\ref{ejes}.

Assuming that the function which represents the height of the surface,
$z(x,y)$, is single valued and varies smoothly, we can approximate the height
in the neighborhoods of $O$ as
\begin{equation} \label{alturalocal}
z(x,y)=-\frac{1}{2}\left(\frac{x^2}{R_x}+\frac{y^2}{R_y}\right),
\end{equation}
where $R_x$ and $R_y$ are the principal radii of curvature. The simple formula
(\ref{alturalocal}) is enough to our purposes for two different reasons. First,
for smooth surfaces the average penetration length $a$ is smaller than the
radii of curvature and only ions penetrating the target a distance smaller than
$a$ from the point $O$ contribute appreciably to erosion there (since energy
deposition decays with distance as a Gaussian). Second, it is assumed that the
surface principal directions ---along which curvature is maximized or
minimized--- are aligned along the $ \mathbf {\hat x} $ and $ \mathbf {\hat y}$
axes. This holds at least for the cases in which structures are aligned either
parallel or perpendicular to the projection of the incident beam (which is the
usual case, as observed in previous sections).

We can obtain the local speed of erosion at $O$ using Eq.\ (\ref{distrib1}). To
this end, one needs to change coordinates in Eq.\ (\ref{distrib1}) to the
$(x,y)$ frame in Fig.\ \ref{ejes} and take into account that the local flux of
ions at any point of the surface, $\mathbf{r}=(x,y,z)$, is related to the
homogenous flux, $\Phi_0$, by $\Phi(\mathbf{r})=\Phi_0 \cos \gamma(
\mathbf{r})$, where $\gamma( \mathbf{r})$ is the angle between the beam
direction and the local normal direction. By Taylor expanding to linear order
in the curvature radii and using Eq.\ (\ref{vO1}), BH obtained an expression for the erosion rate at $O$ that reads
\begin{equation}\label{eqvo}
V_O= {S}_0 \left[\Gamma_0(\gamma_0)+\frac{\Gamma_x(\gamma_0)}{R_x}+
\frac{\Gamma_y(\gamma_0)}{R_y}\right],
\end{equation}
where the functions $\Gamma_0$, $\Gamma_x$ and $\Gamma_y$ only depend on
$\Phi_0, E, \Lambda$, the local angle of incidence $\gamma_0$ and the
distances, $a$, $\sigma$ and $\mu$ of  Eq.\ (\ref{distrib1}).

Once the dependence of the erosion rate with the local morphology has been
obtained, we can obtain a continuum equation for the local surface height,
$Z=h(X,Y)$, in the laboratory system of reference $( \mathbf {\hat {X}},
\mathbf {\hat {Y}}, \mathbf {\hat {Z}}) $ sketched in Fig.\ \ref{ejes}. Thus,
we define the component $ \mathbf {\hat {Z}} $ to be perpendicular to the
initial surface, $\mathbf {\hat {X}}$ as parallel to the projection of the ion
beam onto the initial flat surface, and as $\mathbf {\hat {Y}}$ perpendicular
to $\mathbf {\hat {X}}$ and $\mathbf {\hat {Z}}$. Within our linear
approximation (see \cite{Makeev_2002}) we can assume $\partial_t h \simeq
-V_O$ and $\theta=\gamma_0+\partial_X h$, where $\theta$ is the angle between the
ion beam and the normal to the uneroded surface, and using (\ref{eqvo}) one
obtains \cite{Bradley_1988}
\begin{equation}\label{varh2}
\frac{\partial h}{\partial t} = -v_0+ \gamma_x \frac{\partial h}{\partial
X}+\nu_x \frac{\partial^2 h}{\partial X^2}+\nu_y \frac{\partial^2 h}{\partial
Y^2},
\end{equation}
with
\begin{equation}\label{relacion_varh2}
v_0={S}_0\Gamma_0(\theta),\quad \gamma_x=S_0\partial_{\theta}
\Gamma_0(\theta),\quad \nu_x=S_0 \Gamma_x(\theta), \quad \nu_y=S_0
\Gamma_y(\theta),
\end{equation}
where any nonlinear terms have been neglected and we have used
${R_{x}^{-1}}=-\partial_X^2 h$ and ${R_{y}^{-1}}=-\partial_Y^2 h$. Here, $S_0$
and $\Gamma_0$ are always positive and, therefore, the surface height decreases
at constant speed ${S} _0 \Gamma_0$.

In order to get information about the evolution of the surface, we will assume
the following perturbation with spatial frequencies $k_X$ and $k_Y$, and
amplitude $A$ at time $t$:
\begin{equation}\label{perturbation}
h(X,Y,t)=A\, {\rm e}^{{\rm i}(k_XX+k_YY)+ \omega t}.
\end{equation}
If we substitute this expression into (\ref{varh2}), we obtain the real part of
the dispersion relation, $\omega$. This reads
\begin{equation}\label{RewBH}
\mathcal{R}e\,\omega(\mathbf{k})= -\nu_x k_X^2-\nu_y k_Y^2,
\end{equation}
and the imaginary part is
\begin{equation}\label{ImwBH}
\mathcal{I}m\,\omega(\mathbf{k})=\gamma_x k_X.
\end{equation}
The imaginary part of $\omega(\mathbf{k})$ indicates the magnitude and
direction of the speed of transverse in-plane motion of the disturbances. Since
$\Gamma_0$ is an increasing function of $\theta$ \cite{Bradley_1988}, the
perturbation moves in the $-\mathbf {\hat {X}} $ direction (that is, upstream
with respect to the projection of the beam). On the other hand,
$\mathcal{R}e,\omega(\mathbf{k})$ describes the rate at which the amplitude of
a perturbation with wave-vector $\mathbf{k}$ grows or decays with time. Given
that at small angles of incidence \cite{Bradley_1988} $\nu_x(\theta)<0$ and
$\nu_y(\theta)<0$, any spatial perturbations on the initial surface grow
exponentially in time. Since $\nabla^2 h$ is positive at the bottoms of the
valleys, at these points the rate of erosion is larger than at the top of the
crests where $\nabla^2 h<0$. Therefore, the negative signs of $\nu_x$ and
$\nu_y$ are the mathematical expression of Sigmund's morphological instability.
At these small incidence angles, moreover, $\nu_x(\theta)<\nu_y(\theta)$ and
perturbations grow faster along the $\mathbf {\hat {X}} $ axis than along the
$\mathbf {\hat {Y}}$. Hence, at these angles, the ripples crests are
perpendicular to the projection of the ion beam onto the surface as observed
experimentally. For the critical angle $\theta_c$, one precisely has
$\nu_x(\theta_c)=\nu_y(\theta_c)$ while for larger angles
$\nu_y(\theta)<\nu_x(\theta)$ and the ripple crests align with the ion beam
projection.

These results predict some of most basic experimental features of IBS ripples
but, without an additional mechanism which stabilizes the system, disturbances
of arbitrarily small length-scales would increase exponentially without bound.
On the other hand, from Eq.\ (\ref{varh2}) one would expect the ripple
wavelengths to be of the order of the distances involved in the description,
that is to say, of the order of the penetration length $a$. This does not
happen, rather, the experimental ripple wavelengths are frequently almost two
orders of magnitude larger than $a$. In order to solve these problems, BH
incorporate the effects of the surface diffusion of thermal origin by
introducing an analogous term to that derived by Mullins \cite{Mullins_1957}
for  isotropic surfaces. With this aim, a term $-B\nabla^2 \nabla^2 h$ is
included into (\ref{varh2}) to obtain the following linear equation which
describes the evolution of $h$
\begin{equation}\label{varh2a}
\frac{\partial h}{\partial t} = -v_0+ \gamma_x \frac{\partial h}{\partial
X}+\nu_x \frac{\partial^2 h}{\partial X^2}+\nu_y \frac{\partial^2 h}{\partial
Y^2}-B\nabla^2 \nabla^2 h.
\end{equation}

Proceeding as in Eqs.\ (\ref{perturbation})-(\ref{ImwBH}), the real part of the dispersion relation modifies into
\begin{equation}\label{RewBH2}
\mathcal{R}e\,\omega(\mathbf{k})= -\nu_x k_X^2-\nu_y k_Y^2-B \mathbf{k}^4,
\end{equation}
that is negative for wave vectors of sufficiently large $k$ (i.e. short length scales). However, there is a band of unstable long-wavelength modes for which (\ref{RewBH2}) is positive. The orientation and wavelength of the observed ripple structure corresponds to that mode which maximizes (\ref{RewBH2}); this is $2\pi(2B/\nu_x)^{1/2}$ if the
ripples are oriented along the $\mathbf {\hat {X}} $ axis as a result of
$\nu_x<\nu_y$, or $2\pi(2B/\nu_y)^{1/2}$ if the ripples are oriented along the
$\mathbf {\hat {Y}} $ axis as a consequence of $\nu_y<\nu_x$.  With the aim of obtaining an order of magnitude estimation of this quantity, Bradley and Harper deduced that for
normal incidence, a sputtering yield of 2 atoms per ion, $a\simeq \sigma \simeq
\mu \simeq 10$ nm, and a typical value $B\simeq 2\times 10^{-22}$ cm$^4$ s$^{-1}$,
the value of the linear wavelength is $\lambda_l\simeq 5$ $\mu$m, in reasonable agreement with the typical distance between ripples observed experimentally.

If surface diffusion is thermally activated, as effectively occurs for high
temperatures and small flows, it is easy to obtain how the wavelength of the
pattern varies with the temperature, $T$. Since $\nu_{x,y}$ do not depend on
$T$ and assuming that $B$ verifies $B \sim (1/T)\exp(-\Delta E /k_B T )$, where
$\Delta E$ is the energetic barrier to activate surface diffusion and $k_B$
is the Boltzman's constant, the linear wavelength $\lambda_l$ must verify ${\lambda_l} \sim
(1/T^{1/2})\exp(-\Delta E /2k_B T )$. On the other hand, it is also possible to
obtain a relationship between ripple wavelength and the kinetic energy and flux of
ions, $E$ and $\Phi_0$, respectively. Assuming that $a$, $\sigma$ and $\nu$ are
proportional to $E$ and independent of $\Phi_0$, one gets ${\lambda_l}\sim
E^{-1/2}$ and ${\lambda_l}\sim \Phi_0^{-1/2}$.

Working still within a linear approximation to the surface height in a similar spirit to the introduction of surface diffusion in (\ref{varh2a}), some alternative physical mechanisms have been proposed,
specially in order to account for the lack of pattern (ripple) formation at
either low temperatures \cite{Chason_1994} or small angles of incidence
\cite{Carter_1996}. Thus, a term of the form $- F|k| h(k,t)$ added to the
(Fourier transformed) right hand side of Eq.\ (\ref{varh2a}) can describe the
effect of viscous flow in the bulk onto the surface dynamics, with $F$ being a
coefficient that depends on the surface free energy and bulk viscosity
\cite{Mullins_1957,Mullins_1959}. For low enough temperatures at which thermal
surface diffusion is hampered, bulk viscous flow could dominate the surface
dynamics even to the extent of preventing ripple formation for oblique angle
incidence \cite{Chason_1994} or dot formation for rotating substrates
\cite{Cirlin_1991}. Alternatively, the fact that a fraction of the sputtered
atoms that move close to the surface are recaptured has led to arguing
\cite{Carter_1996} for the effective description of such a process by inclusion of a term with the form $+\nu_{recc} \partial^2_x h$, where
$\nu_{recc}$ would be a positive, angle dependent coefficient. We will see
later that also direct knock-on sputtering events would have effectively an
analogous continuum description, thus corresponding to stabilizing mechanisms
that can counterbalance Sigmund's instability.

\hfill
\paragraph{Higher order corrections}

As we have seen, BH's linear theory already predicts many important features of
the nanopatterning process, but leaves out a number of additional properties.
For instance, it cannot predict saturation of the ripple amplitude or their
non-uniform lateral motion. In order to account for these features, one
needs to generalize the BH theory, the most natural procedure being the
extension of their perturbative expansion in surface derivatives in (\ref{distrib1})-(\ref{vO1}) up to higher order. In this process one needs to approximate the surface height
locally through a Taylor expansion, such as BH did, only that the order of
approximation is higher. Along the way, not only non-linear contributions do arise, but also corrections appear to previous linear terms. These thus provide
contributions of erosive origin to linear relaxation mechanisms such as
transverse pattern motion, and surface diffusion. This program has been
carried out in \cite{Cuerno_1995,Makeev_1997,Makeev_2002}. The result is the
following equation (the coefficients $\lambda$ with various indices appearing in this and the following height equations are not to be confused with the pattern wavelength)
\begin{eqnarray}\label{Eq.complete}
\frac{\partial h}{\partial t} & = & -v_0 + \gamma_x \frac{\partial h}{\partial
x} + \Omega_1 \frac{\partial^3 h}{\partial x^3} + \Omega_2 \frac{\partial^3
h}{\partial xy^2} + \xi_x\frac{\partial h}{\partial x} \frac{\partial^2
h}{\partial x^2} + \xi_y\frac{\partial h}{\partial x} \frac{\partial^2
h}{\partial y^2} \nonumber \\
 & + & \nu_x \frac{\partial^2 h}{\partial x^2} + \nu_y \frac{\partial^2 h}{\partial y^2}
+ \frac{\lambda_x}{2} \left(\frac{\partial h}{\partial x}\right)^2 +
\frac{\lambda_y}{2} \left(\frac{\partial h}{\partial y}\right)^2 \label{kdv_ani}\\
 & - & D_{xy} \frac{\partial^4 h}{\partial x^2y^2} - D_{xx}\frac{\partial^4 h}{\partial x^4}
 - D_{yy} \frac{\partial^4 h}{\partial y^4}, \nonumber
\end{eqnarray}
where all coefficients depend on the experimental parameters from Sigmund's
theory \cite{Makeev_2002}, like in BH's equation (\ref{varh2}), which now
becomes a linear low order approximation of Eq.\ (\ref{kdv_ani}). In the first
line of (\ref{kdv_ani}), and except for the constant average velocity $v_0$,
all terms provide linear and non-linear contributions to the transverse ripple
motion. The second line reflects the dependence of the sputtering yield with
the local curvatures (as in BH's equation) and slopes, while the third line
contains (linear) surface diffusion terms that include both thermal (as in BH)
and erosive contributions. This type of contributions break in general the
$x\to -x$ symmetry while respecting symmetry under $y\to -y$, dependence of all
parameters on the incidence angle being such that for normal incidence
rotational in-plane symmetry is restored, Eq.\ (\ref{kdv_ani}) becoming the
celebrated Kuramoto-Sivashinsky (KS) equation:
\begin{equation}
\frac{\partial h}{\partial t} = -v_0 + \nu \nabla^2 h - D \nabla^4 h +
\frac{\lambda_0}{2}(\nabla h)^2 , \label{ks}
\end{equation}
where we have used that, for $\theta=0$, $D_{xx}=D_{yy}=D_{xy}/2$,
$\nu_x=\nu_y\equiv\nu$, $\lambda_x=\lambda_y\equiv\lambda_0$,
$\gamma_x=\xi_x=\xi_y=\Omega_1=\Omega_2=0$ \cite{Makeev_2002}.

The main morphological implications of Eqs.\ (\ref{kdv_ani}) and (\ref{ks})
have been discussed elsewhere \cite{Makeev_2002}. Here, we briefly mention the
main conclusions. First of all, from the point of view of these equations, the
behavior predicted by BH appears as a short-time transient, and all BH
predictions ---on the dependence of e.g.\ the ripple structure with incidence
angle--- still carry over here. Moreover, for temperatures at which surface
diffusion is activated, the dependences of the ripple wavelength with energy,
temperature and flux are as in BH, at least for short to intermediate times. If
$T$ is low enough that the only contribution to surface diffusion is from
erosive origin, these dependencies are modified and compare better with
experiments at low temperatures, see \cite{Makeev_2002} and Sec.\
\ref{sec:experiments} above. For intermediate to long times, the nonlinear
terms in Eqs.\ (\ref{kdv_ani}), (\ref{ks}) are such that the exponential growth
of the ripple amplitude is stabilized, yielding to a much slower power-law
increase of e.g.\ the surface roughness with time. At normal incidence, the
surface displays kinetic roughening for long time and length scales, and this
is also the case at oblique incidence, as long as the nonlinearities
$\lambda_{x,y}(\theta)$ have the same signs. Otherwise, Eq.\ (\ref{kdv_ani})
features {\em cancellation modes} as first identified in the anisotropic KS
equation, that is the following particular case of Eq.\ (\ref{kdv_ani}) in
which the propagative terms with coefficients $\xi_{x,y}$ and $\Omega_{1,2}$
(being arguably irrelevant to the asymptotic limit) are simply neglected
\cite{Cuerno_1995,Rost_1995}:
\begin{eqnarray}
\frac{\partial h}{\partial t} & = & -v_0 + \gamma_x \frac{\partial h}{\partial
x} + \nu_x \frac{\partial^2 h}{\partial x^2} + \nu_y \frac{\partial^2
h}{\partial y^2} + \frac{\lambda_x}{2} \left(\frac{\partial h}{\partial
x}\right)^2 +
\frac{\lambda_y}{2} \left(\frac{\partial h}{\partial y}\right)^2 \label{ks_ani}\\
 & - & D_{xy} \frac{\partial^4 h}{\partial x^2y^2} - D_{xx}\frac{\partial^4 h}{\partial x^4}
 - D_{yy} \frac{\partial^4 h}{\partial y^4}. \nonumber
\end{eqnarray}
Cancellation modes are height Fourier modes with wave vector in the unstable
band, for which the non-linear terms cancel each other, leaving the system
non-linearly unstable, and inducing ripples which are oriented in an oblique
direction that is parallel neither to the $x$ nor to the $y$ axis
\cite{Rost_1995}.

Eq.\ (\ref{ks}) has, on the other hand, the capability (beyond linear equations
such as that of BH) of predicting both ``dot'' and ``hole'' production,
depending on the sign of the non-linear parameter $\lambda$, which in turn
depends on the characteristics of Sigmund's distribution \cite{Kahng_2001}.
However, conspicuous non-linear features still remain beyond description by the
KS equation (\ref{ks}) and its generalizations (\ref{kdv_ani}), (\ref{ks_ani}).
First, the ``dot'' or ripple structures described by these non-linear equations
are characterized by a dominant wavelength that remains fixed in time and does
not coarsen for any parameter values, so that experiments in which coarsening
occurs cannot be accounted for. A stronger limitation is that the patterns
described by (\ref{kdv_ani}), (\ref{ks}), (\ref{ks_ani}) disorder in heights to
the extent that the size of ordered domains of ripples or dots essentially
restricts to a single structure. There is thus no lateral ordering of the dots
or ripples. Actually, the KS equation is known as a paradigm of spatio-temporal
chaos in the wider field of Non-Linear Science \cite{Mori_1997}.

Following the program sketched at the beginning of this paragraph, a natural
step is to carry on still further the perturbative study of Sigmund's local
velocity of erosion pioneered by BH. At the present stage, one could close Eq.\
(\ref{ks}) (we now consider normal incidence for simplicity) by including non-linear terms
that are quadratic in the height field and fourth order in space derivatives,
reaching an equation of the form (in reference frame comoving with the eroded surface)
\begin{equation}
\frac{\partial h}{\partial t} =  \nu \nabla^2 h - D \nabla^4 h +
\lambda_1(\nabla h)^2 + \lambda_2 \nabla^2(\nabla h)^2 . \label{ks_mixta}
\end{equation}
Indeed, such a higher order generalization has been performed \cite{Kim_2004}.
However, the expressions of $\lambda_1$ and $\lambda_2$ in terms of Sigmund's
parameters are such that the coefficients of the two nonlinearities in Eq.\
(\ref{ks_mixta}) have the same signs for all physical parameter values.
Unfortunately, as shown shortly after \cite{Castro_2005b,Kim_2005}, this
introduces cancellation modes that seriously question the mathematical validity
of (\ref{ks_mixta}) for our physical system: writing the equation in Fourier
space, we get
\begin{equation}
\frac{\partial h(\mathbf{k})}{\partial t} = (-\nu k^2 - D k^4) h(\mathbf{k}) +
(\lambda_1 - \lambda_2 k^2) FT\left[(\nabla h)^2\right] , \label{ks_mixta_Four}
\end{equation}
where $FT$ denotes Fourier Transform. For the unstable mode $\mathbf{k}_0$
(that indeed occurs physically, see the experiments of IBS of Pd(001) in
\cite{Kim_2004,Castro_2005b}) such that $k_0^2=\lambda_1/\lambda_2$, the
nonlinear terms cancel each other, so that the amplitude of this mode explodes
exponentially. We are seemingly left with a matter-of-principle limitation,
namely, the simpler theoretical approach that is limited to studying Sigmund's
local velocity of erosion (at sufficiently high linear and non-linear orders)
meets mathematical limitations before being able to cover for the various
experimental features of the patterns we wish to study.

In the following paragraph we will take a wider viewpoint in which the dynamics
is more complete, in the sense that the surface height will be coupled to an
additional physical field describing the flux of adsorbed material that
diffuses on the near-surface layer. This procedure will be seen to provide an
improved description solving some of the above physical and mathematical
shortcomings. Before doing so, we note that additional studies exist that focus
on the evolution of the height field only. Thus, the standard result in the
field of Pattern Formation (see references in \cite{Paniconi_1997}) that a
linear damping in the KS equation induces the appearance of ordered patterns,
has directly led to the proposition in \cite{Facsko_2004} of (a modified)
damped KS equation to describe the experiments described in Sec.\
\ref{sec:experiments}. This modified equation actually becomes the standard damped
KS equation after an appropriate non-local time transformation \cite{Vogel_2005}. Although the natural anisotropic generalization has been
duly proposed \cite{Vogel_2006}, these equations unfortunately do not allow
much improvement in the continuum description of nanopatterning by IBS for
several reasons: (i) there is no connection to phenomenological parameters (the
equations are not {\em derived} from any model but are, rather, argued for on a
phenomenological basis); (ii) damped generalizations of the KS equation are
known (see \cite{Castro_2007} and references therein) {\em not} to allow for
wavelength coarsening for any parameter values, which leaves out of these
descriptions many of the experimentally observed patterns; (iii) similarly, at
long times the fluctuations of the PSD function predicted by any of the
damped generalizations of the KS equation are cut-off and do not show the type
of power-law like behavior that is seen, even in experiments in which there is
no coarsening \cite{Ziberi_2005,Ziberi_2006c}. In contrast, a more successful
extension of the BH-type approach has been its generalization in
\cite{Chen_2005} to surfaces with steep slopes. Recall that starting with BH's,
all continuum approaches mentioned (in contrast e.g.\ with MC models) work
within a small-slope approximation that allows neglection of higher order terms in Eqs.\ such as (\ref{Eq.complete})-(\ref{ks_mixta}). Remarkably, a suitable generalization for
arbitrary slope values has been seen in \cite{Chen_2005} to lead to a
non-linear equation whose traveling wave (shock) solutions compare well with
experiments on the motion of the walls of pits excavated by a FIB.
These results possibly provide important clues for a more complete model of
nanopatterning by IBS.

\subsubsection{Coupling to diffusive surface species}\label{coupled}

Due to physical and mathematical considerations, the previous section leaves us
with the need to enlarge the continuum description of nanopatterning by IBS.
The expectation is that, by incorporating the dynamics of additional physically
relevant fields, the effective height equation to be eventually derived
improves its formal properties and its predictive power. Perhaps we could
compare the situation with related fields such as e.g.\ the growth of thin
films by physical or chemical vapor deposition (CVD) techniques. In CVD, for
instance (see references in \cite{Cuerno_2007}), the standard continuum
description arises precisely from the coupling between the local (growth)
velocity and the dynamics of the concentration field of the diffusing species
that eventually will stick to the growing aggregate. 

There have been various attempts in the IBS context to combine the
surface dynamics as predicted by a BH-type equation, with the evolution of
relevant surface species, such as adatoms, addimers and surface traps (see
\cite{Erlebacher_1999,Chason_2001} and references therein). However, in this
approach no explicit feedback mechanism is provided from the dynamics of such
species onto the local variation of the surface height so that the dynamics cannot be described by a closed system of equations. On the other extreme, there is a recent proposal in which a full Navier-Stokes (thus, highly coupled) formulation is proposed to describe ripple transverse motion onto a glass surface \cite{Alkemade_2006}. Trying to reach a balance between complexity and completeness in the physical description, one can seek for a formulation that, while simpler than a full hydrodynamic model, still provides an explicit coupling between the surface topography and the evolution of the relevant diffusive fields. For the case of ripple dynamics in the different (macroscopic!) context of aeolian sand dunes \cite{Csahok_2000}, such is the spirit of the so-called ``hydrodynamic'' approach, in which one sets up a system of coupled equations that describe the height of the
eroded substrate profile, $h$, and the thickness of a mobile surface layer, $R$. Although there are relevant
differences between both physical systems ---the size of the structures
is roughly seven orders of magnitude larger in aeolian ripple formation than in IBS systems---, the similarity between the mechanisms of
diffusion and erosion and the analogous form and behavior of the patterns in both systems seem to suggest that
these processes could be modeled by similar formalisms.

This program has been followed in Refs.\cite{Aste_2004,Aste_2005,Castro_2005,Munoz_2005}, where the next  equations are proposed to describe the evolution of the fields $h$ and $R$:
\begin{eqnarray}
&\partial_t R = (1-\phi) \Ge - \Ga - \nabla \cdot \mathbf{J}, \label{ec.R}\\
&\partial_t h = -\Ge+\Ga. \label{ec.h}
\end{eqnarray}
In (\ref{ec.R}) and (\ref{ec.h}), $\Ge(R,h)$ and $\Ga(R,h)$ are, respectively, rates of atom excavation from and
addition to the immobile bulk, $(1-\phi)=\bar{\phi}$ measures the fraction of
eroded atoms that become mobile, and the third term in Eq.\ (\ref{ec.R})
describes motion of mobile material onto the surface. In this way, local redeposition is allowed if $\bar{\phi} \neq 0$ \cite{Kustner_1998}, while the viscous near-surface layer $R$ is provided with a dynamics of its own \cite{Umbach_2001}.

In Refs.\ \cite{Aste_2004,Aste_2005}, a linear dependence of $\Ge$ and $\Ga$
with the local geometry of the surface is considered to study the linear
stability of this system. These studies reveal that depending on the values of
the parameters, the ripple orientation could be aligned in any target direction, one of
the limitations of this model being that the model is not well defined in the absence
of redeposition $\bar{\phi}=0$.

More detailed mechanisms of erosion and addition are explicited in Refs.\ \cite{Castro_2005,Munoz_2005},
  where nonlinear terms for the rate of excavation are introduced.
  These additional mechanisms are seen below to induce richer
pattern dynamics than previous models for the surface height only. Unlike Aste
and Valbusa's and aeolian sand ripples models, in \cite{Munoz_2005}
the momentum in the direction of the projection of the beam directly transmitted
from ions to superficial atoms is neglected. Rather, a diffusive
term for mass transport onto the surface is considered that, in the case of isotropic
amorphous materials, is given by $\mathbf{J}=-D \nabla R$, where $D$ is a
thermally activated constant. Another feature of \cite{Castro_2005,Munoz_2005} is
to consider a non-zero amount of mobile material, $R_{eq}$, even in the absence of
excavation ($\Ge=0$) o redeposition ($\bar{\phi}=0$) which could be thermally induced. This term allows us to write the rate of addition in similar form to the
Gibbs-Thompson expression for surface relaxation via evaporation-condensation, namely,
 \begin{equation}\label{Ga}
\Gamma^{ad}=\gamma_0\left[R-R_{eq}(1-\gamma_{2x}\partial^2_xh-\gamma_{2y}
\partial^2_yh)
\right],
\end{equation}
where $\gamma_0$ is the mean nucleation rate for a flat surface (on the $xy$
plane) and $\gamma_{2x}$, $\gamma_{2y} \geq 0$ describe variations of the
nucleation rate with surface curvatures (and are positive if we assume that
nucleation events are more likely in surface valleys than in protrusions). Note
that, in (\ref{Ga}), the full thickness of the mobile layer is affected by the
shape of the surface.

In \cite{Munoz_2005}, a expression similar to the non-linear generalization (\ref{Eq.complete}) of BH's local erosion velocity is considered for the
rate at which material is sputtered from the bulk. If the beam
direction is in the $xz$ plane, we have,
\begin{eqnarray}\label{Ge}
   \Gamma_{ex} = \alpha_0 \big[ 1 + \alpha_{1x} \partial_x h+ \alpha_{2x} \partial^2_x h
    +\alpha_{2y} \partial^2_y h
    +\alpha_{3x}(\partial_x h)^2+\alpha_{3y} (\partial_y h)^2 \nonumber \\   -(\partial_x h) (\alpha_{4x}
    \partial^2_x h + \alpha_{4y} \partial^2_y h)\big],
\end{eqnarray}
where parameters reflect the dependence of $\Gamma_{ex}$ on the {\em local}
shape of the surface and are function of the angle of incidence, ion and
substrate species, ion flux, energy, and other experimental parameters just as the coefficients of Eq.\ (\ref{kdv_ani}).

\subsubsection{Oblique incidence}

In a weakly nonlinear analysis of this system, the length-scale separation we have already mentioned between the time-scales associated with erosive and diffusive events is seen to play a crucial role. Thus, in particular the diffusive field $R$ relaxes much faster than the target height $h$, making it possible within a multiple-scale formulation to solve perturbatively for the dynamics of the former and derive a closed effective equation for the evolution of the latter which, in the case of oblique incidence \cite{Munoz_2005,Munoz_2007}, reads
\begin{eqnarray}\label{eq.ero}
\partial_t h = \gamma_x \partial_x h + \sum_{i=x,y} \left[
-\nu_i \, \partial^2_i h + \lambda^{(1)}_{i} \, (\partial_i h)^2 + \Omega_ i
\partial_i^2 \partial_x h
+ \xi_i\, (\partial_x h) (\partial^2_i h)\right] \nonumber \\
 +\sum_{i,j=x,y} \left[ -{\mathcal K}_{ij} \partial^2_i \partial^2_j h +
\lambda^{(2)}_{ij}\,
\partial^2_{i} (\partial_j h)^2 \right],\,
\end{eqnarray}
where these new coefficients are related to those of Eqs.\
(\ref{ec.R})-(\ref{Ge}) \cite{Munoz_2005}.

As in the nonlinear continuous theories shown in the previous
section, the reflection symmetry in $\mathbf{\hat x}$ is broken in Eq.\
(\ref{eq.ero}), but it is preserved in the $\mathbf {\hat y}$ axis. Equation (\ref{eq.ero}) generalizes the height equations in \cite{Bradley_1988,Makeev_2002,Park_1999} within BH approach to IBS; the main
difference between (\ref{Eq.complete}) and (\ref {eq.ero}) is that additional terms $ \lambda^{(2)} _ {ij} $ appear in the later. These terms result important to
correctly describe the evolution of the pattern as we will see bellow. To the best of our knowledge,
Eq.\ (\ref{eq.ero}) is new, and indeed has a rich parameter space. Numerical
integration within linear regime retrieves all features of the ripple structure
as predicted by the BH theory, i.e., dependence of the ripple wavelength with
linear terms, and ripple orientation as a function of $\theta$.

%%%%%%%%%%%%%%Figure%%%%%%%%%%%%%%%%%%%%%%%
\begin{figure}[!hmbt]
\begin{center}
\begin{minipage}{0.25\linewidth}
\begin{center}
 \includegraphics[width=\linewidth]{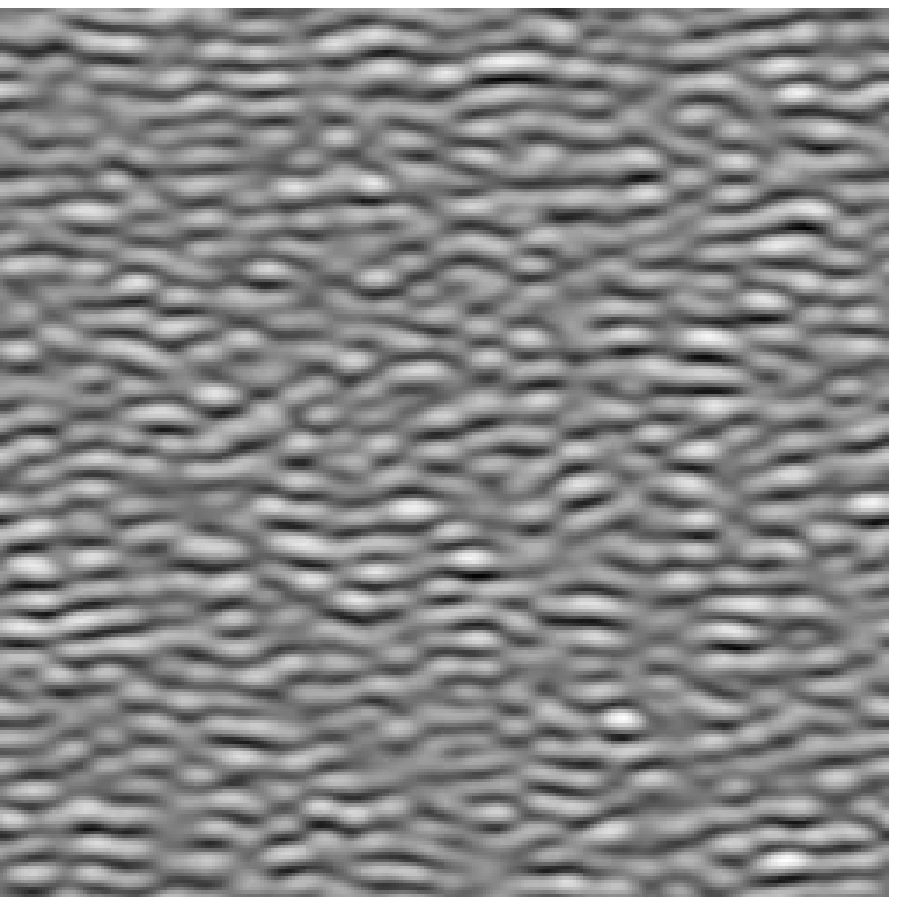}
{\large (a)}
\end{center}
\end{minipage}\hspace*{ 0.03\linewidth}
\begin{minipage}{0.25\linewidth}
 \begin{center}
 \includegraphics[width=\linewidth]{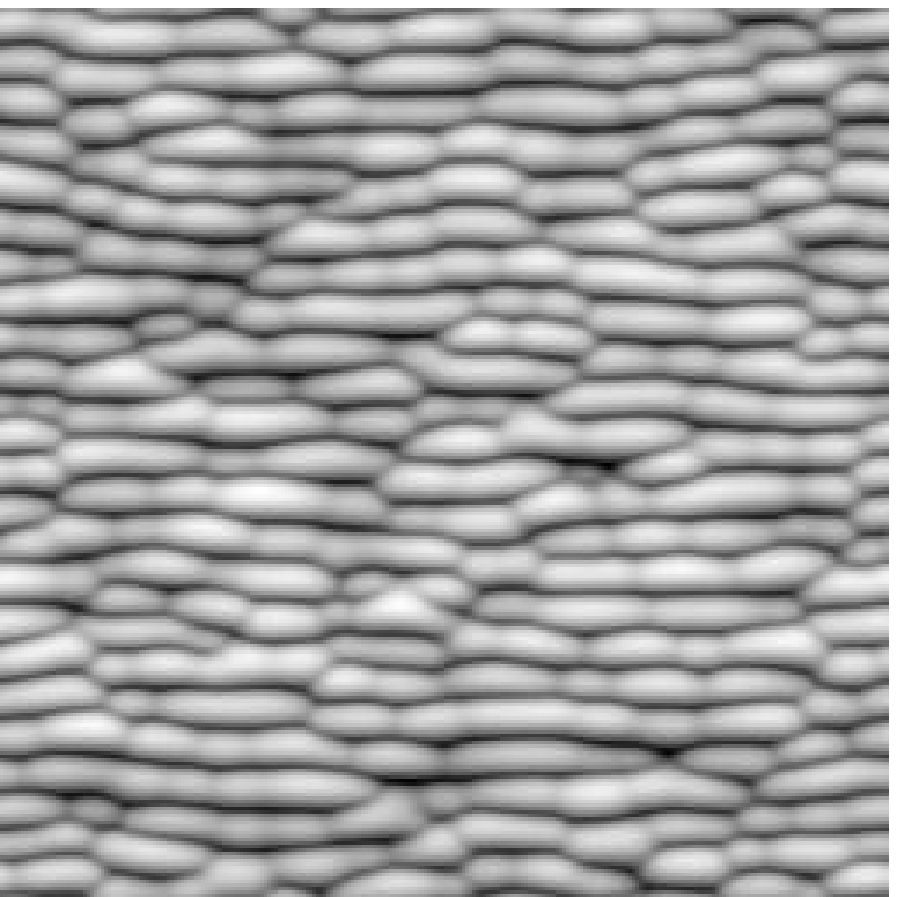}
{\large (b)}
\end{center}
\end{minipage}\hspace*{ 0.03\linewidth}
\begin{minipage}{0.25\linewidth}
\begin{center}
 \includegraphics[width=\linewidth]{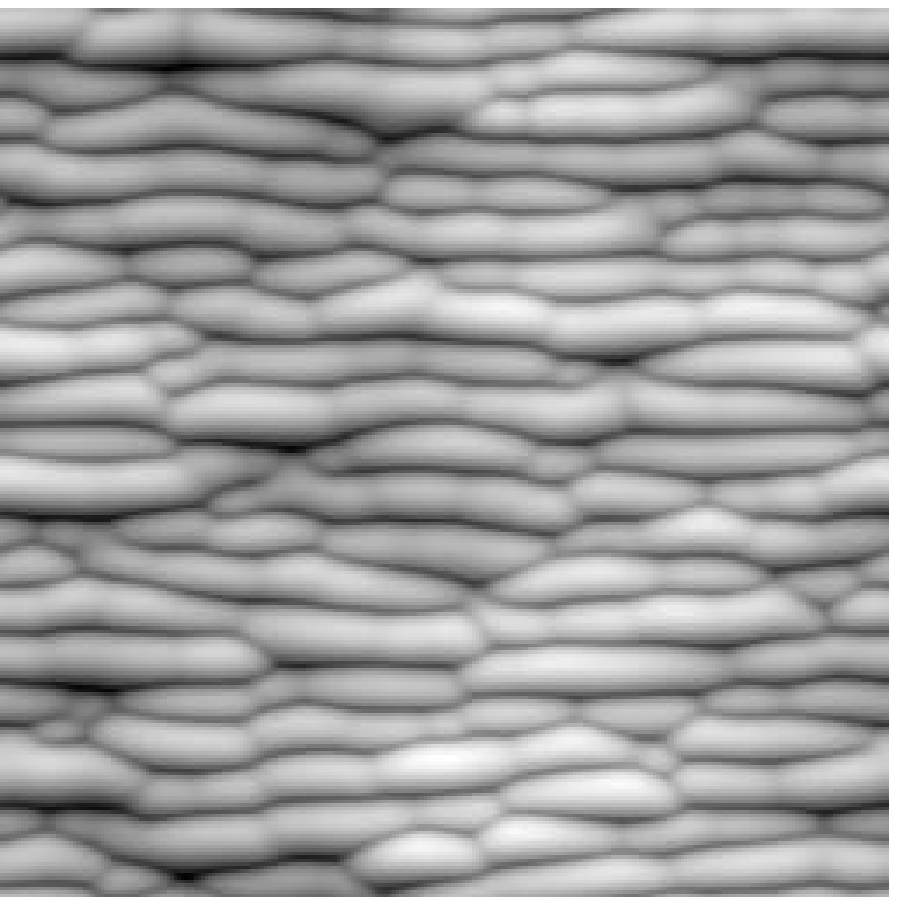}
{\large (c)}
\end{center}
\end{minipage}\vspace{0.03\linewidth}\\
\begin{minipage}{0.27\linewidth}
\begin{center}
 \includegraphics[width=\linewidth]{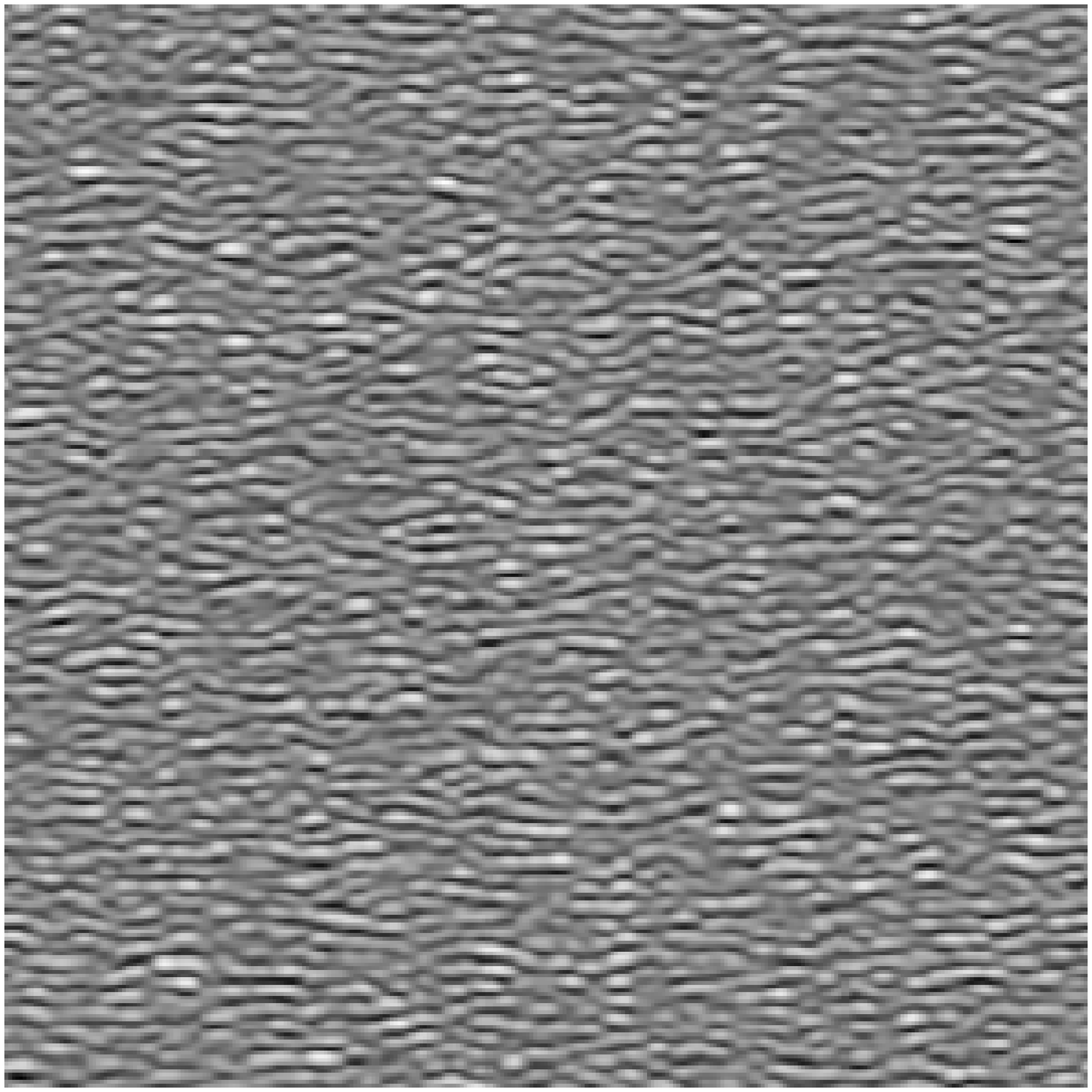}
{\large (d)}
\end{center}
\end{minipage}\hspace*{ 0.01\linewidth}
\begin{minipage}{0.27\linewidth}
 \begin{center}
 \includegraphics[width=\linewidth]{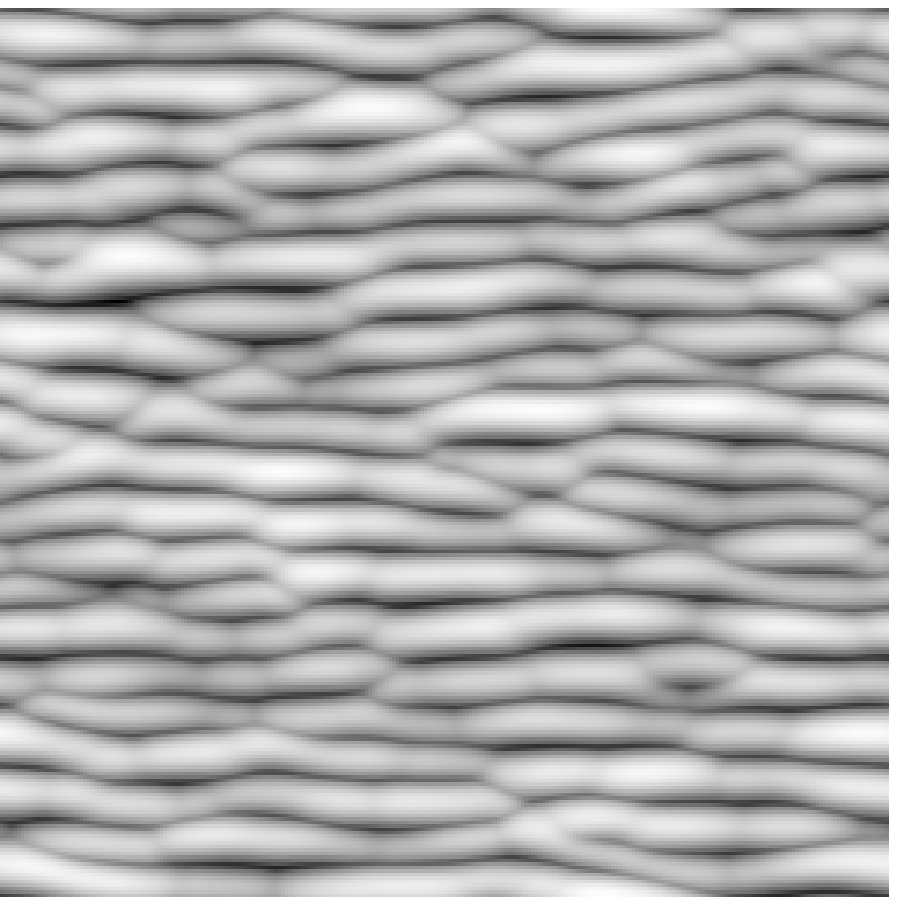}
{\large (e)}
\end{center}
\end{minipage}\hspace*{ 0.01\linewidth}
\begin{minipage}{0.27\linewidth}
\begin{center}
 \includegraphics[width=\linewidth]{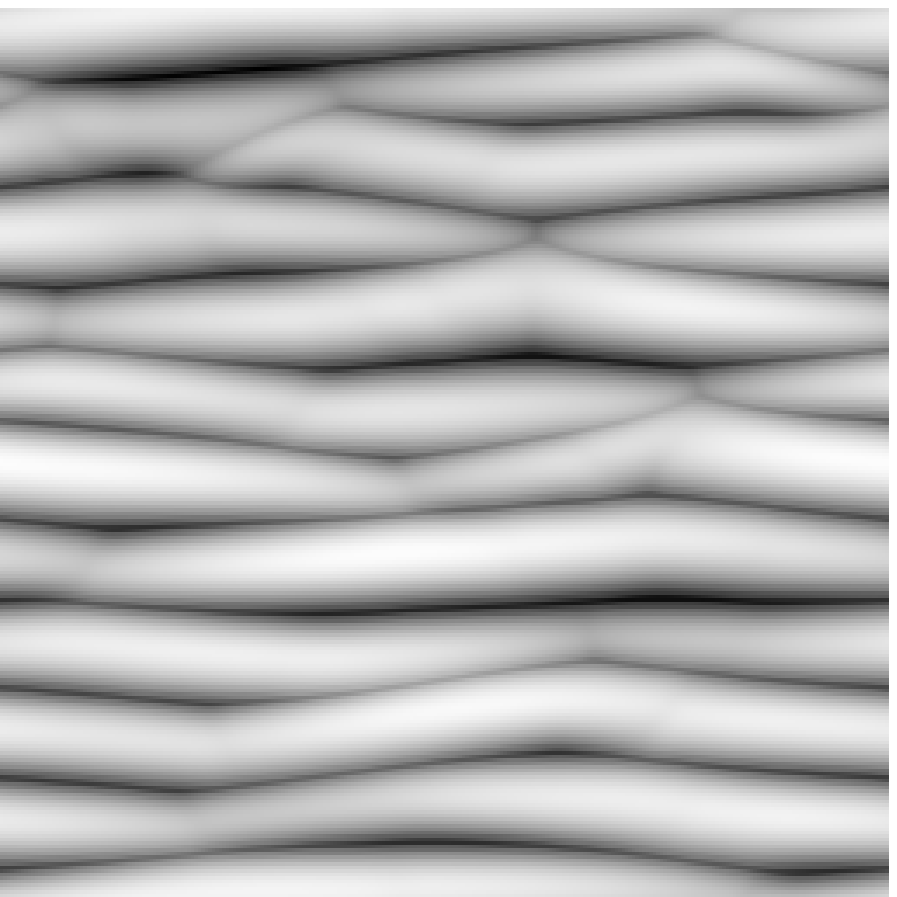}
{\large (f)}
\end{center}
\end{minipage}\vspace{0.03\linewidth}\\
\begin{minipage}{0.27\linewidth}
\begin{center}
 \includegraphics[width=\linewidth]{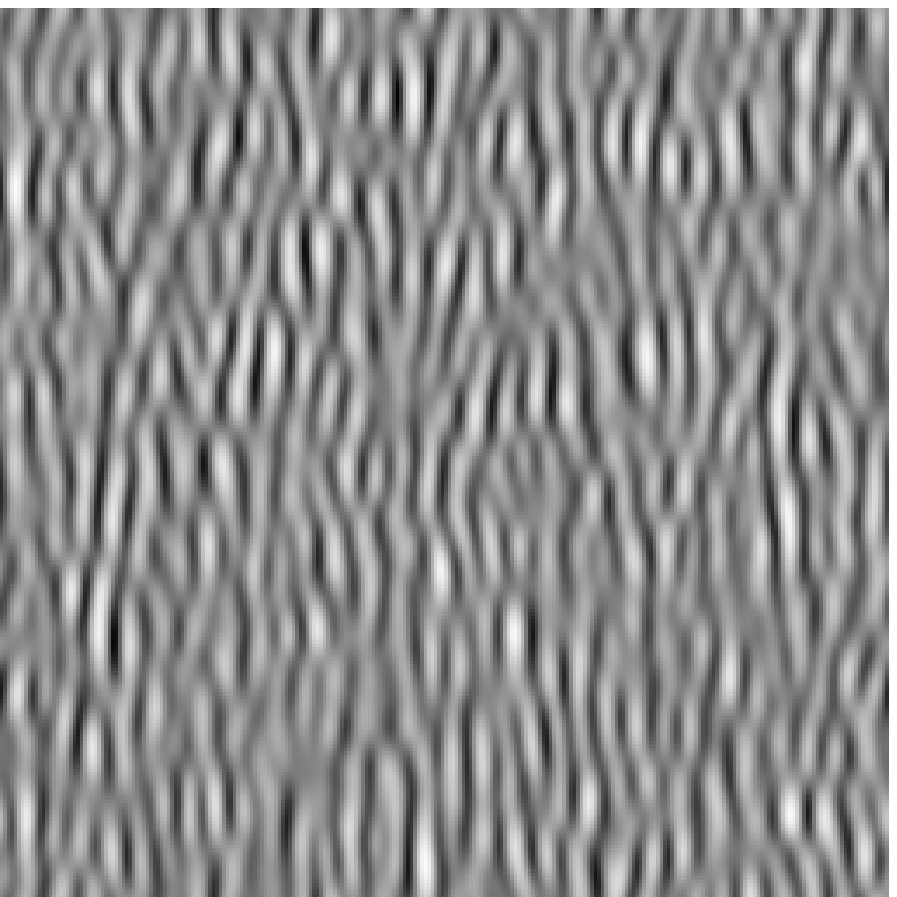}
{\large (g)}
\end{center}
\end{minipage}\hspace*{ 0.01\linewidth}
\begin{minipage}{0.27\linewidth}
 \begin{center}
 \includegraphics[width=\linewidth]{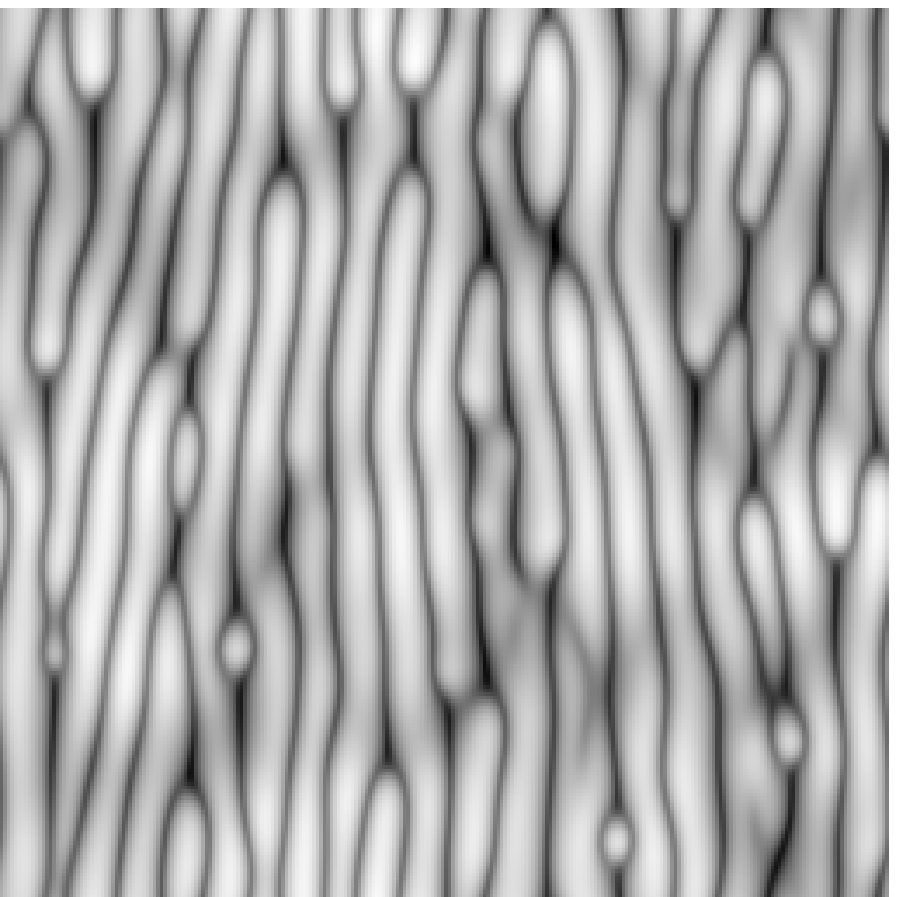}
{\large (h)}
\end{center}
\end{minipage}\hspace*{ 0.01\linewidth}
\begin{minipage}{0.27\linewidth}
\begin{center}
 \includegraphics[width=\linewidth]{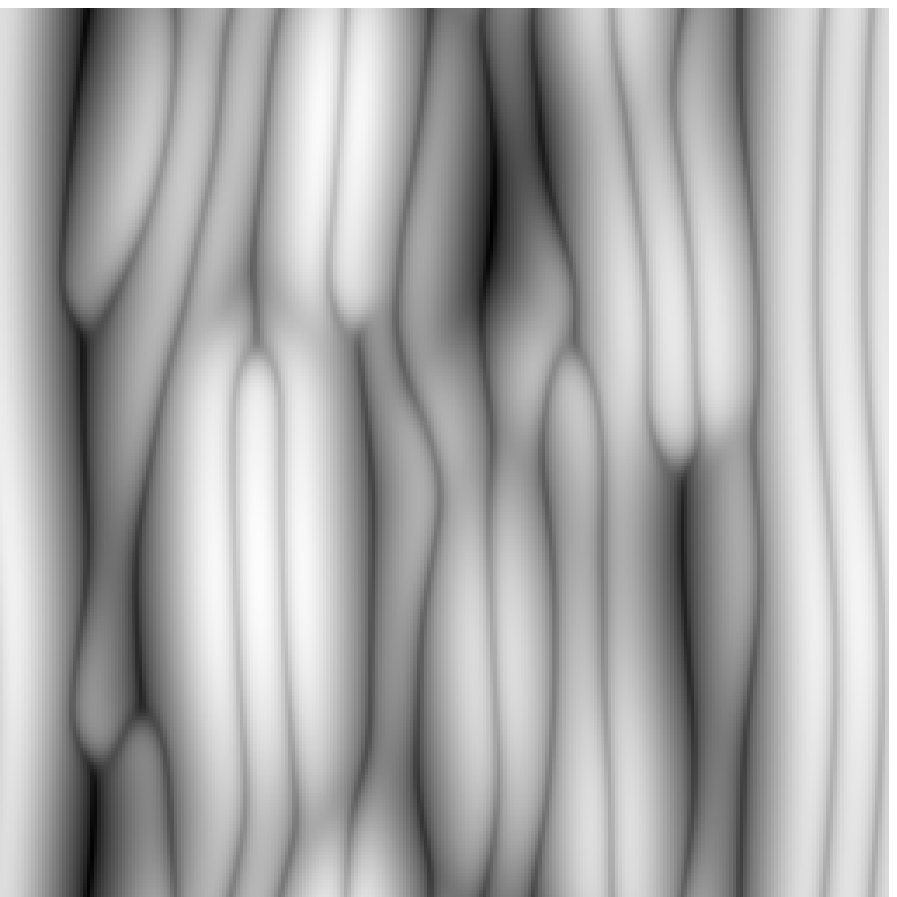}
{\large (i)}
\end{center}
\end{minipage}
\caption{Top views of morphologies described by (\ref{eq.ero}) at $t=10$
(first column), $t=106$ (second column) and $t=953$ (third column) with
$\gamma_x=-0.1$, $\nu_x=1$, $\nu_y=0.1$, $\Omega_x=1$, $\Omega_y=0.5$,
$\xi_i=0.1$, $\lambda_x^{(1)}=1$, $\lambda_y^{(1)}=5$,
$\lambda^{(2)}_{i,j}=-5$, ${\cal K}_{i,j}=1$  (a), (b) and (c). (d), (e) and
(f) with the same parameters as in (a), (b) and (c) except
$\lambda_x^{(1)}=0.1$. (g), (h) and (i) with $\gamma_x=0.1$, $\nu_x=1$,
$\nu_y=-0.95$, $\Omega_i=-0.5$, $\xi_i=0.1$, $\lambda_x^{(1)}=0.1$,
$\lambda_y^{(1)}=1.0$, $\lambda^{(2)}_{i,x}=-0.5$, $\lambda^{(2)}_{i,y}=-5.0$ y
${\cal K}_{i,j}=1$. The $x$ axis is oriented vertically and the $y$ axis
horizontally.}\label{oblicua}
\end{center}
\end{figure}
%%%%%%%%%%%%%%%%%%%%%%%%%%%%%%%%%%%%%%%%

Top views of the morphologies described by Eq.\ (\ref{eq.ero}) at different
times are shown in Fig.\ \ref{oblicua} for different values of the parameters. In all these examples, the ripples are increasing their size in
the course of time while disordering in heights for long distances, whereas the form
of the ripples can vary appreciably depending on parameter values. In some
cases, as in the first and third rows of Fig.\ \ref{oblicua}, the ripples are
disordered longitudinally, whereas in the second row, the ripples are longer
and more straight. In the third row, the ripples are also very disordered in
height and tend to group themselves in domains which contain, approximately,
three ripples each. The ripple coarsening
seen in Fig.\ \ref{oblicua} actually requires the presence of $\lambda^{(2)}_{i,j}$,
whose magnitude and mathematically correct sign are due to describing
redeposition by means of the additional field $R$. When the values of these
coefficients increase relative to $\lambda^{(1)}_i$, coarsening stops later,
and the amplitude and wavelength of the pattern also increase, analogous to the
result for one-dimensional interfaces studied in \cite{Munoz_2006b}. The
coarsening exponent $n$ will take an {\em effective} value that will be larger
the later coarsening stops, and may depend on (simulation) parameter values \cite{Munoz_2005}.

\begin{figure}[!htmb]
\begin{center}
\begin{minipage}{0.5\linewidth}
\begin{center}
 \includegraphics[width=\linewidth]{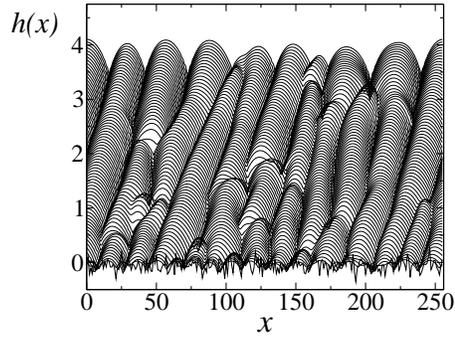}
\end{center}
\end{minipage}\hfill
\begin{minipage}{0.475\linewidth}
\begin{center}
\caption{Cross sections at $y=L/2$ of the two-dimensional surface given by Eq.\ (\ref{eq.ero}) for times between $t=0$ and $t=1500$  separated by a time interval $\bigtriangleup t=20$ with
$\gamma_x=-2$, $\Omega_i=0$, $\nu_x=1$, $\nu_y=0.1$, $\xi_i=3.5$,
$\lambda_x^{(1)}=1$, $\lambda_y^{(1)}=5$, $\lambda^{(2)}_{i,x}=-50$,
$\lambda^{(2)}_{i,y}=-5.0$ and ${\cal K}_{i,j}=1$.} \label{corte}
\end{center}
\end{minipage}
\end{center}
\end{figure}

Asymmetry and transverse in-plane motion of the pattern depends on terms with an odd number of
derivatives in Eq.\ (\ref{eq.ero}). Whereas the linear terms tend to make the
ripples move with a constant velocity, the nonlinear term $\xi_i$ accounts for non-uniform movement of the ripples as observed in some experiments
\cite{Habenicht_2002,Alkemade_2006}. In Fig.\ \ref{corte} we show the
evolution of a transversal section of the surface as described by Eq.\
(\ref{eq.ero}), where the non-uniform movement and the asymmetry of ripples can be appreciated

%%%%%%%%%%%%%%Figure%%%%%%%%%%%%%%%%%%%%%%%
%%%%%%%%%%%%%%%%%%%%%%%%%%%%%%%%%%%%%%%%%%%%%%%%%%%%

\begin{figure}[!htmb]
\begin{center}
\begin{minipage}{0.28\linewidth}
\begin{center}
 \includegraphics[width=\linewidth]{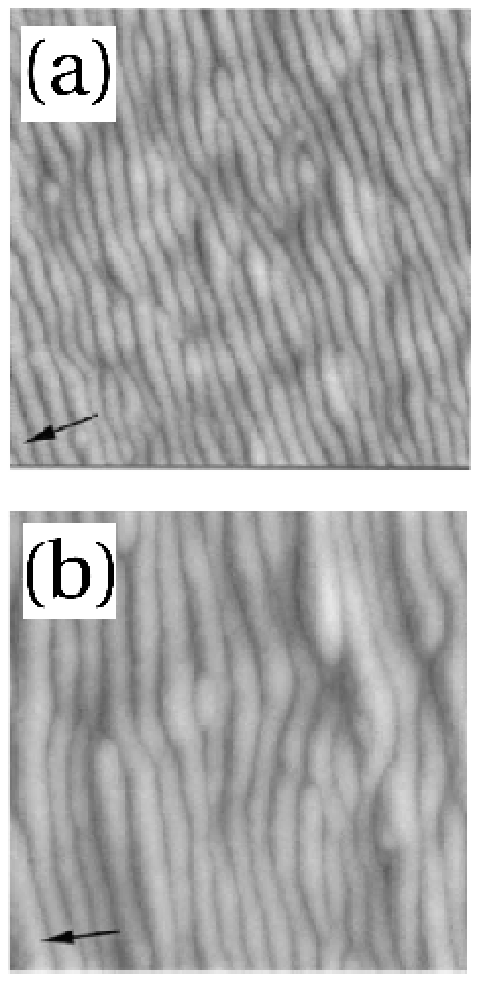}
\end{center}
\end{minipage}\hspace*{0.0175\linewidth}
\begin{minipage}{0.27\linewidth}
\begin{center}
 \includegraphics[width=\linewidth]{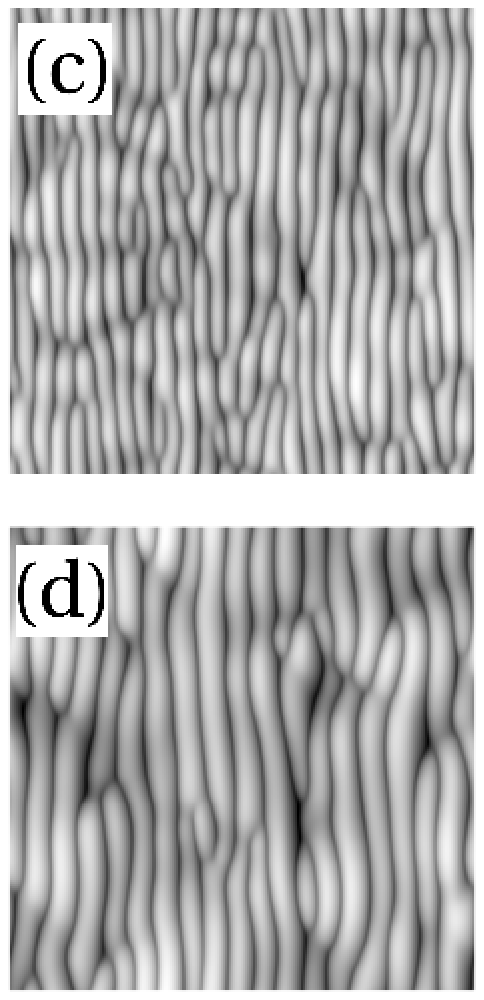}
\end{center}
\end{minipage}
\caption{(a), (b) $1 \times 1$ $\mu$m$^2$ AFM views of fused silica targets irradiated under oblique
incidence after 10 and 60 min. respectively; (c), (d) morphologies obtained by
numerical integration of Eq.\ (\ref{eq.ero}) at times $t_1$ and $6t_1$. Panels (a) and (b) are taken from \cite{Flamm_2001} with permission.} \label{ripples_comp.}
\end{center}
\end{figure}

An example of a comparison of Eq. (\ref{eq.ero}) with experiments is shown in figure \ref{ripples_comp.} where coarsening of the ripples is clearly appreciated. As we see, Eq.\ (\ref{eq.ero}) indeed captures essential properties of the evolution of the experimental topography.
%%%%%%%%%%%%%%Figure%%%%%%%%%%%%%%%%%%%%%%%
%%%%%%%%%%%%%%%%%%%%%%%%%%%%%%%%%%%%%%%%%%%%%%%%%%%%

\subsubsection{Normal incidence}

In the case that the surface is bombarded perpendicularly,  the in-plane asymmetry
introduced by the oblique beam disappears and, for materials that do not
show crystallographically privileged directions, the height equation which describes the
evolution of $h$ reads \cite{Castro_2005,Munoz_2005}
\begin{equation}\label{eq.ero.normal}
  \partial_t h=-\nu \nabla^2 h- {\mathcal K}\nabla^4h+\lambda^{(1)} (\nabla h)^2 +
\lambda^{(2)} \nabla^2 (\nabla h)^2,
\end{equation}
where, again, the coefficients are related to the Sigmund's parameters. A stochastic generalization of this equation has been also proposed in the context of amorphous thin film growth \cite{Raible_2000,Raible_2001}.

An important property of (\ref{eq.ero.normal}) is that, in the absence of redeposited material ($\bar{\phi}=0 $), $\lambda^ {(1)} $ and $ \lambda^ {(2)}$
have the same signs, precisely as occurred in the case of the purely erosive approach
\cite{Kim_2004} discussed in the previous section. This makes Eq.\ (\ref{eq.ero.normal}) nonlinearly unstable \cite{Raible_2001,Castro_2005b}.
Hence, the description with a single height field seems to feature an intrinsic problem
which is solved, if a fraction of redeposited material is large enough, through the
two-field ``hydrodynamic'' approach.

In order to reduce the number of parameters and simplify the analysis of
(\ref{eq.ero.normal}), we rescale $x$ and $y$, $t$, and $h$ by $(K/\nu)^{1/2},
K/\nu^2$ and $\nu/\lambda_1$, respectively, resulting into a single-parameter
equation as done in the one-dimensional counterpart of (\ref{eq.ero.normal})
studied in \cite{Munoz_2006b}. This reads
\begin{equation}\label{eq.ero_red_normal}
\partial_t h=- \nabla^2 h- \nabla^4h+(\nabla h)^2 -r \nabla^2 (\nabla h)^2,
\end{equation}
where $r=-(\nu\lambda^{(2)})/(K \lambda^{(1)})$ is a positive parameter which
allows us to perform a numerical analysis of the complete parameter space of Eq.\
(\ref{eq.ero.normal}) through an appropriate rescaling.

%%%%%%%%%%%%%%Figure%%%%%%%%%%%%%%%%%%%%%%%
\begin{figure}[!htmbp]
\begin{center}
\begin{minipage}{0.4\linewidth}
\begin{center}
 \includegraphics[width=\linewidth]{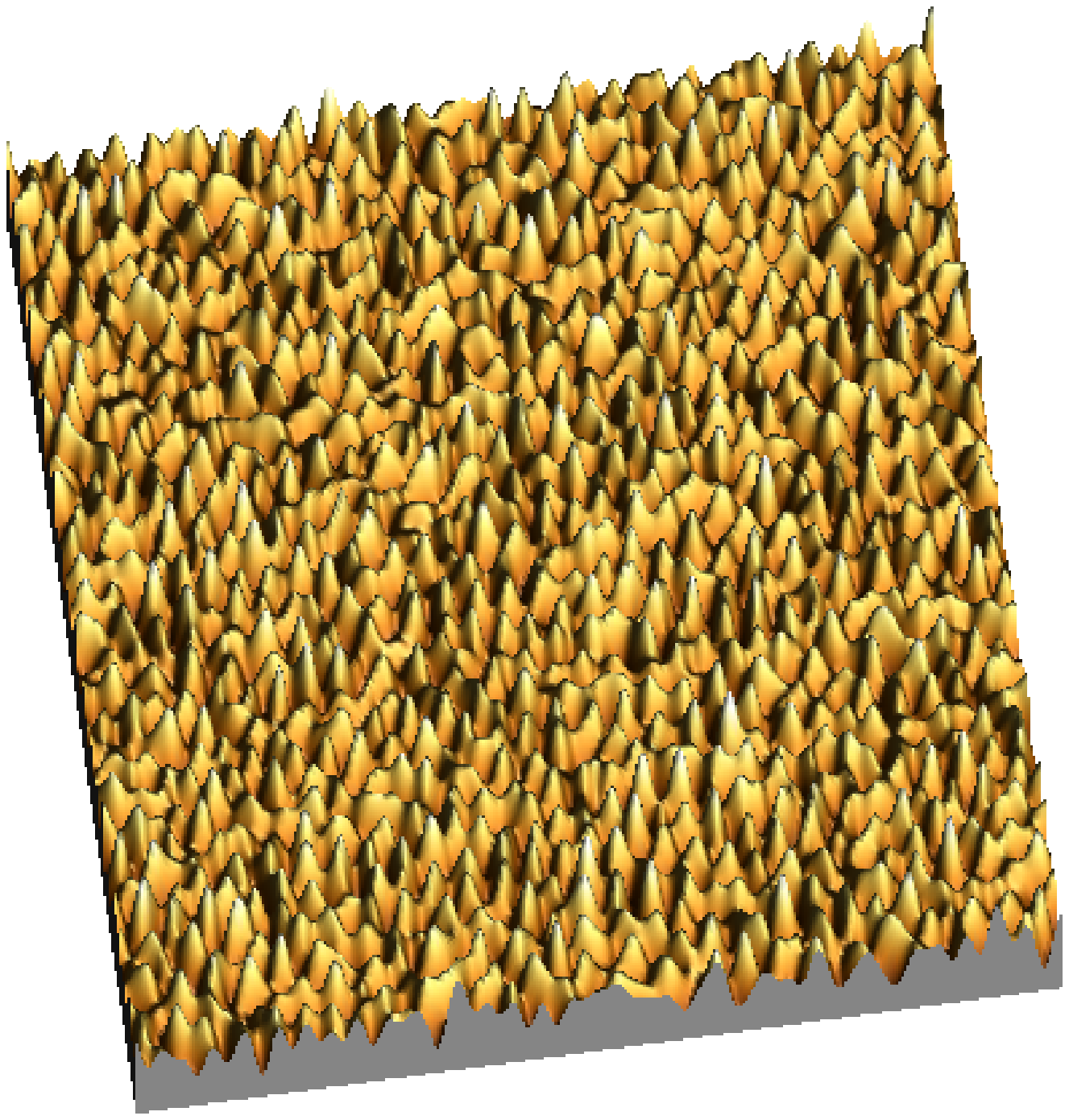}
{\large(a)}
\end{center}
\end{minipage}\hspace{0.075\linewidth}
\begin{minipage}{0.4\linewidth}
\begin{center}
 \includegraphics[width=\linewidth]{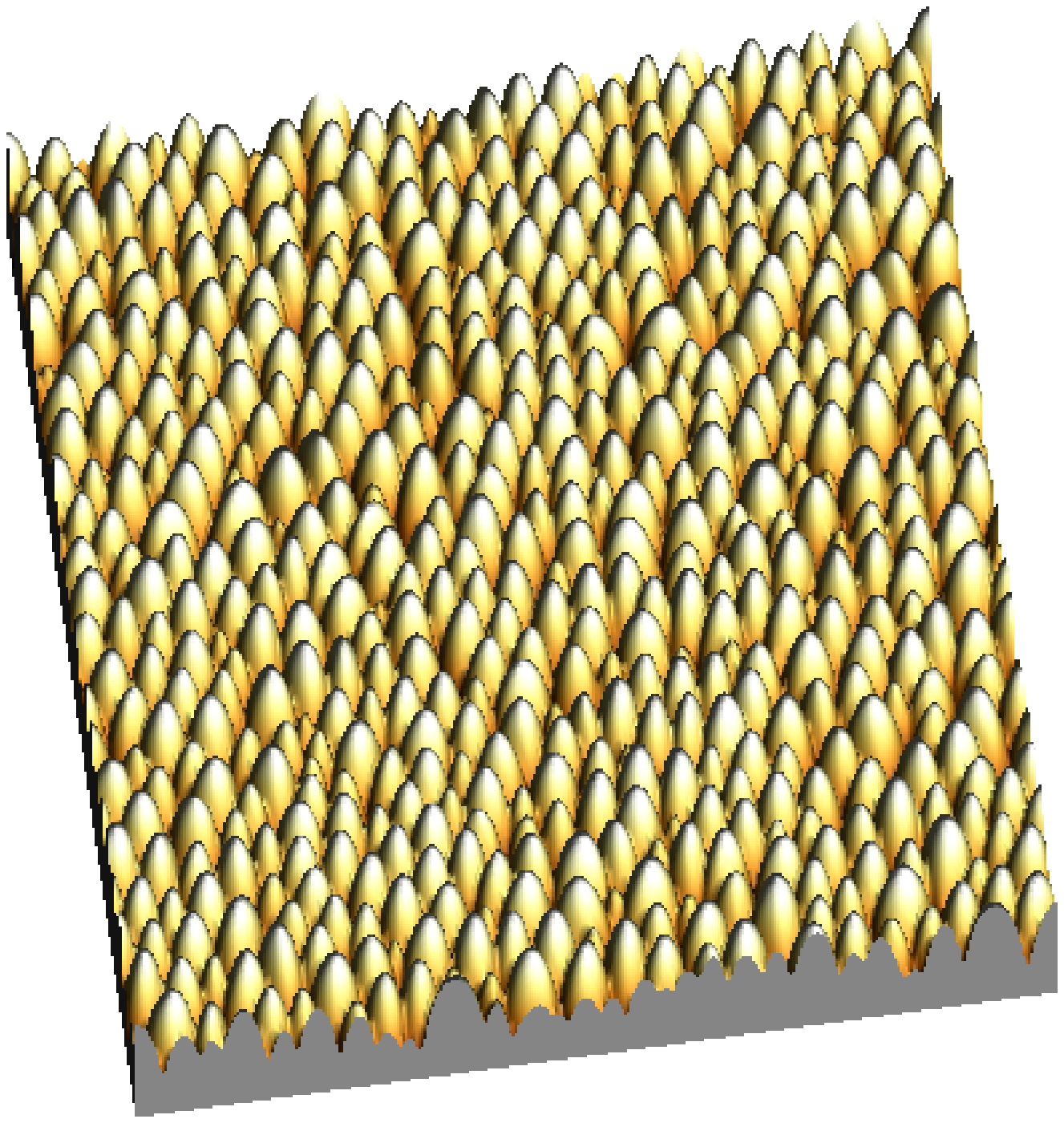}
{\large(b)}
\end{center}
\end{minipage}\\
\begin{minipage}{0.4\linewidth}
\begin{center}
 \includegraphics[width=\linewidth]{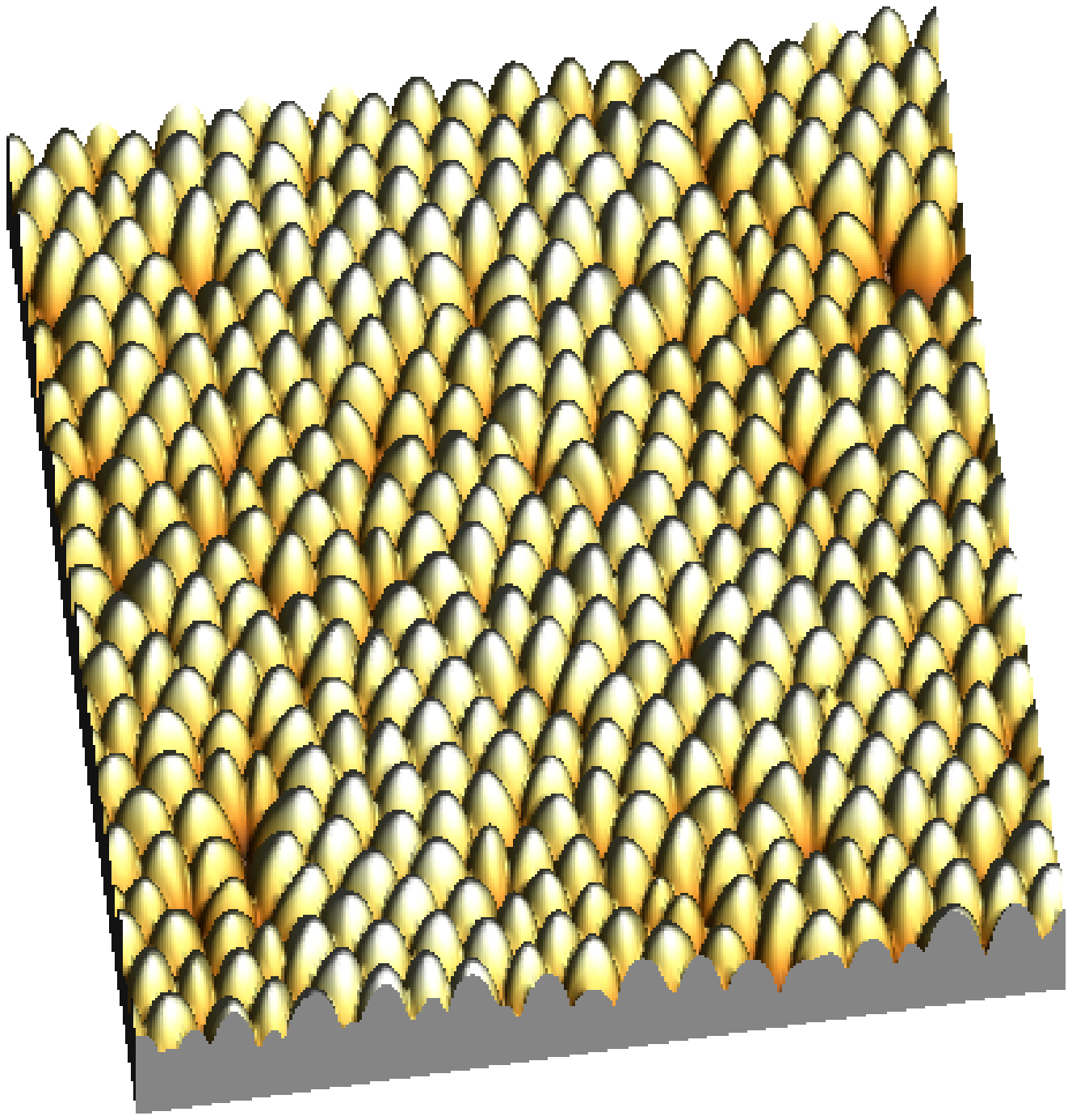}
{\large(c)}
\end{center}
\end{minipage}\hspace{0.075\linewidth}
\begin{minipage}{0.4\linewidth}
 \begin{center}
\caption{ Three-dimensional views of the surface obtained by numerical
integration of Eq.\ (\ref{eq.ero_red_normal}) for $r=5$ at different times: (a)
$t=10$; (b) $t=50$; (c) $t=205$.}\label{normal_r=5}
\end{center}
\end{minipage}
\end{center}
\end{figure}
%%%%%%%%%%%%%%%%%%%%%%%%%%%%%%%%%%%%%%%%

%%%%%%%%%%%%%%Figure%%%%%%%%%%%%%%%%%%%%%%%
\begin{figure}[!htmbp]
\begin{center}
\begin{minipage}{0.25\linewidth}
\begin{center}
 \includegraphics[width=\linewidth]{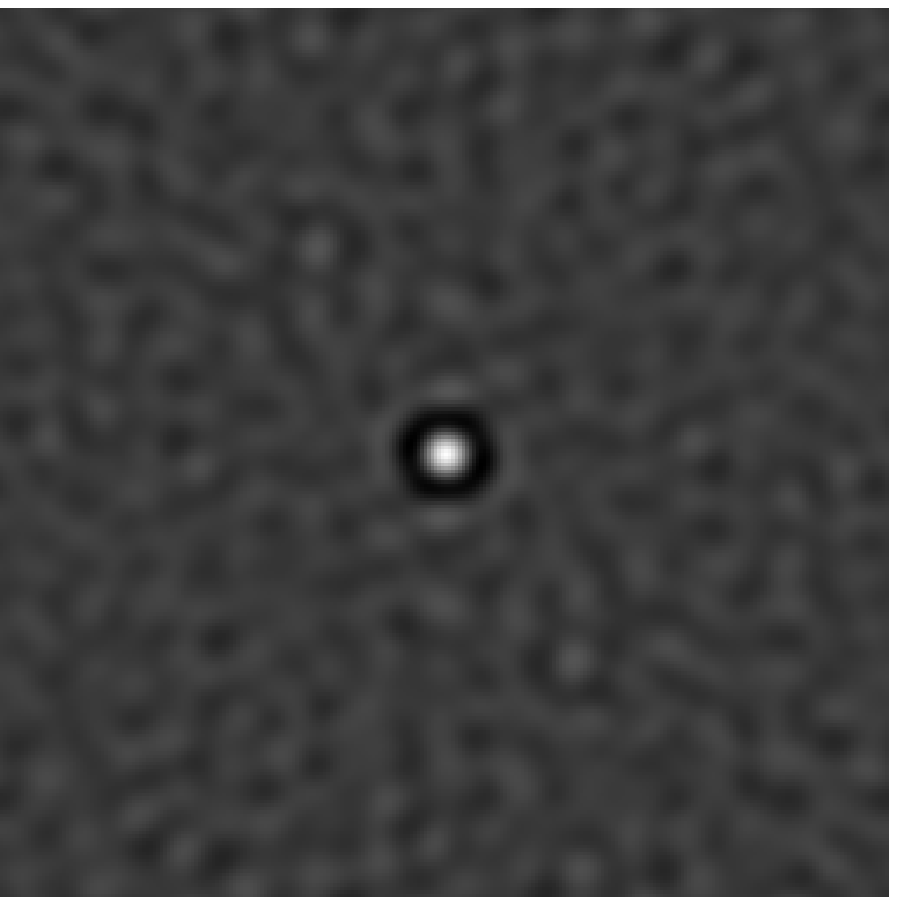}
{\large(a)}
\end{center}
\end{minipage}\hspace*{ 0.03\linewidth}
\begin{minipage}{0.25\linewidth}
\begin{center}
 \includegraphics[width=\linewidth]{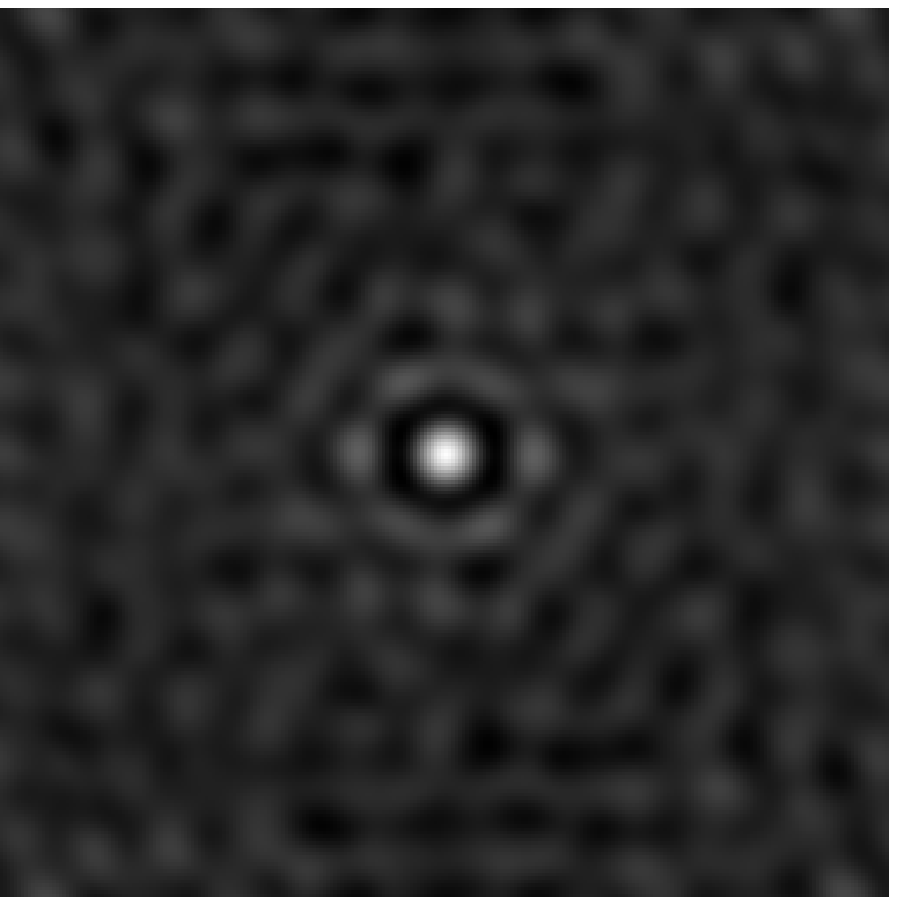}
{\large(b)}
\end{center}
\end{minipage}\hspace*{ 0.03\linewidth}
\begin{minipage}{0.25\linewidth}
\begin{center}
 \includegraphics[width=\linewidth]{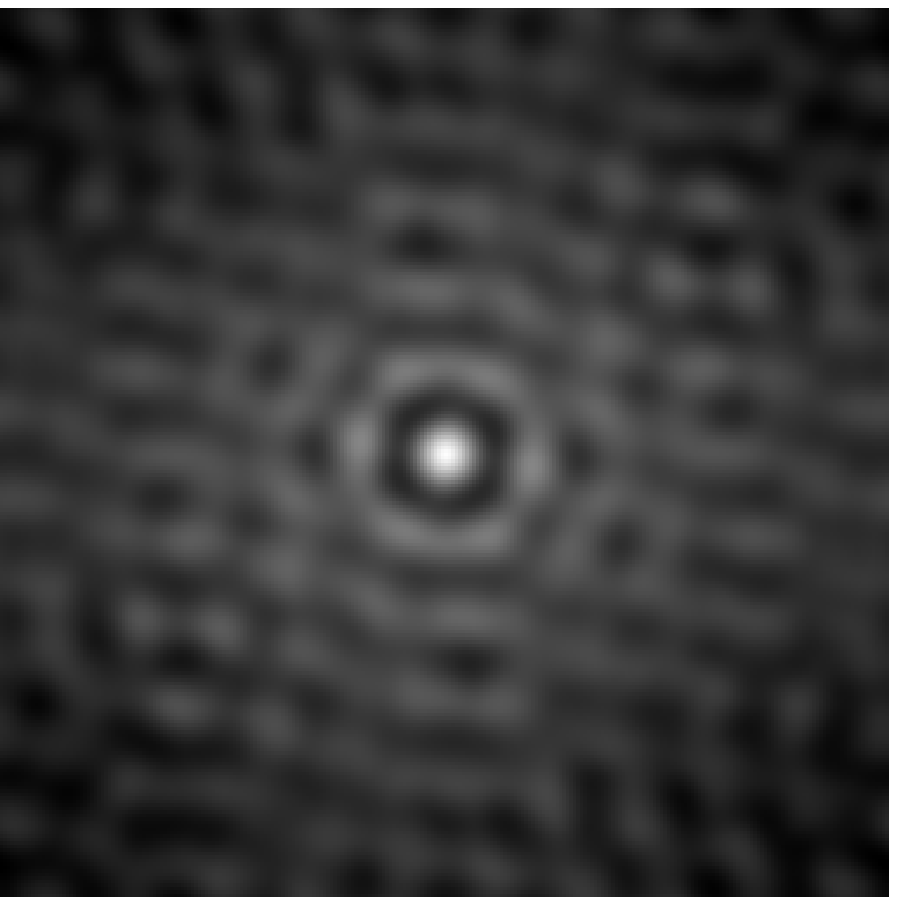}
{\large(c)}
\end{center}
\end{minipage}

\caption{Height autocorrelation corresponding to the panels of Fig.\
\ref{normal_r=5}}\label{autocorrelacion}

\end{center}
\end{figure}
%%%%%%%%%%%%%%%%%%%%%%%%%%%%%%%%%%%%%%%%

\begin{figure}[!hbtmp]
\begin{center}
\begin{minipage}{0.5\linewidth}
\begin{center}
 \includegraphics[width=\linewidth]{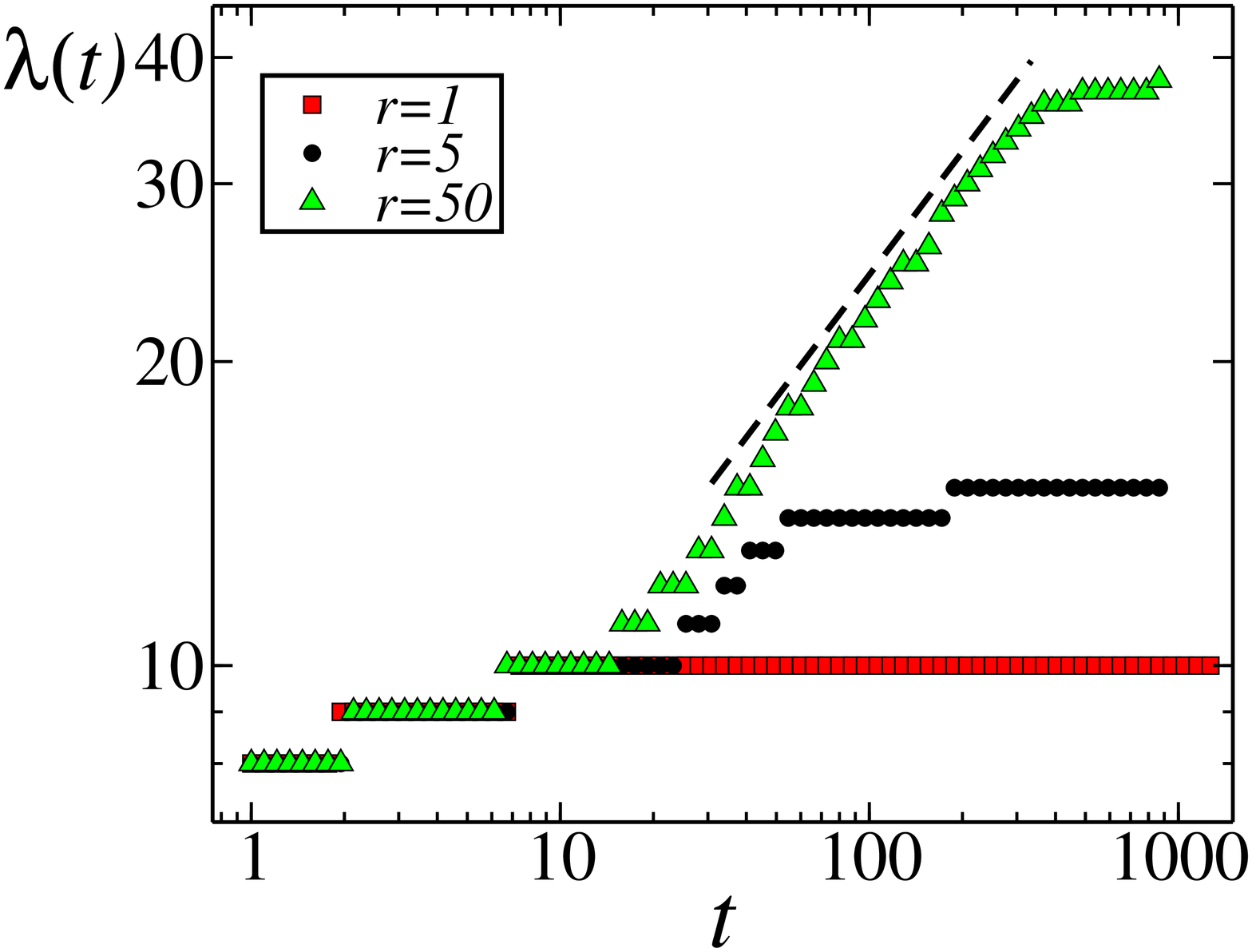}
\end{center}
\end{minipage}\hfill
\begin{minipage}{0.45\linewidth}
 \begin{center}
\caption{Temporal evolution of the pattern wavelength, $\lambda$, given by
Eq.\  (\ref{eq.ero_red_normal}) for different values of $r$. For $r=50$, the dashed line
represents the fit to a power law $\lambda \sim t^{0.40}$.}\label{normal_Lc}
\end{center}
\end{minipage}
\end{center}
\end{figure}

In Figs.\ \ref{normal_r=5} and \ref{autocorrelacion} the evolution of the
height profile and corresponding auto-correlation functions are shown for
$r=5$. Starting from an initial random distribution, a periodic surface
structure with a wavelength of about the maximum of the linear dispersion
relation arises and the amplitude of $h$ increases. Then, a coarsening process occurs and dots grow in width and height, the total number of them
decreasing. The apparent coarsening is quantified in the plot of $\lambda(t)$ shown in
Fig.\ \ref{normal_Lc} where the saturation of ripple wavelength at long times
can be observed. Simultaneously, with dot coarsening, the pattern
increase its in-plane order and eventually leading to an hexagonal dot array as shown in Fig.\ \ref{autocorrelacion}.
%%%%%%%%%%%%%%Figure%%%%%%%%%%%%%%%%%%%%%%%
%%%%%%%%%%%%%%%%%%%%%%%%%%%%%%%%%%%%%%%%

\begin{figure}[!hbtmp]
\begin{center}
\begin{minipage}{0.56\linewidth}
\begin{center}
 \includegraphics[width=\linewidth]{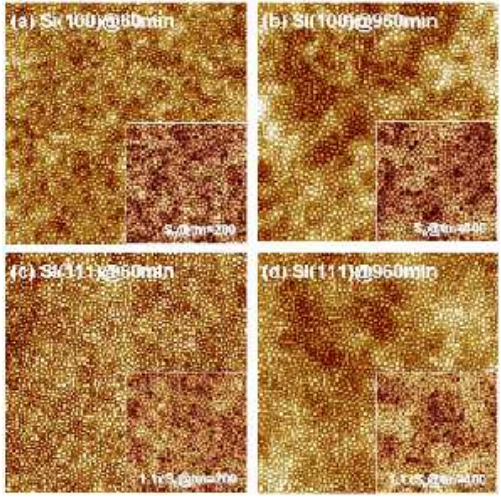}
\end{center}
\end{minipage}\hfill
\begin{minipage}{0.42\linewidth}
 \begin{center}
\caption{AFM $3\times3$ $\mu$m$^2$ images of Si(001) and Si(111) surfaces
sputtered with 1.2 keV Ar$^+$ at normal incidence for different times (see labels). Insets: 2D
views (length scales in arbitrary units) from numerical integration of Eq.\
(\ref{eq.ero.normal}) for different simulation times and surfaces with low
($\nu=0.68$, $\mathcal{K}=1$, $\lambda^{(1)}=0.028$, and $\lambda^{(2)}=0.136$)
and high ($\nu=0.75$, $\mathcal{K}=1$, $\lambda^{(1)}=0.031$, and
$\lambda^{(2)}=0.15$) erosion rates. Reprinted from \cite{Gago_2006} with permission.}\label{gago-2006}
\end{center}
\end{minipage}
\end{center}
\end{figure}

As an example of comparison of this equation with specific experiments, in
this paragraph we show the results recently obtained in \cite{Gago_2006} where an
array of nanodots is obtained over a silicon substrate. In this work, the
time evolution of nanodots is considered on Si(001) and Si(111) targets irradiated at normal incidence with  Ar$^+$ ions of 1.2 keV. In figures
\ref{gago-2006} are shown characteristic AFM images of Si(001) and Si(111) surfaces eroded
at different times. The pattern is similar in both experiments: in both cases
dots of 5-7 nm height and 40-60 nm width are formed and group into short-range hexagonal
order. For times larger than 20 minutes, the surface disorders in heights for larger distances (of the order
of 6 nm in height and about 500 nm of lateral distance). This long-range disorder (kinetic roughening) increases in the
course of the time. In this experiment, the rate of erosion was determined
experimentally in all the processed samples by partially masking them during
the sputtering process and measuring the resulting step edge height with a
profilometer. It was observed that the sputtering rate (SR) was $10\%$ higher
for Si(111) than for Si(001). In order to understand the differences found
experimentally between the two surface orientations, we have integrated Eq.
(\ref{eq.ero.normal}) numerically considering this fact. Thus, as the model
neglects target crystallinity, we have used this relative difference in SR to
simulate the IBS pattern evolution in two model systems. Insets of Figs.\
\ref{gago-2006}(a) and \ref{gago-2006}(b) display the two-dimensional 2D
top-view images from the simulations for surfaces with a given SR at two
different simulation times (time and length units are arbitrary). Similarly,
insets of Figs. \ref{gago-2006}(c)  and \ref{gago-2006}(d) correspond to the
same simulations for a surface with a $10\%$ higher SR, thus representing the
Si(111) orientation. The simulations reproduce in both cases the experimental
coarsening and occurrence of a long-wavelength corrugation. In order to
quantify the pattern features, the PSD functions
corresponding to simulations were systematically analyzed. In Fig.\ \ref{Fig3_6} is shown that the
pattern wavelength saturates earlier for the surface with higher SR, but the
lower SR case attains a larger final dot size. In addition, it is also shown that
the correlation length also saturates earlier for a higher SR
surface. Remarkably, both AFM and simulation results agree in estimating ordered
domains to contain roughly three nanodots. This agreement between simulations
and experimental observations allows us to conclude e.g. that the experimental
differences observed between the pattern evolution on Si(111) and Si(001)
surfaces are due to their different SRs.
%%%%%%%%%%%%%%Figure%%%%%%%%%%%%%%%%%%%%%%%
%%%%%%%%%%%%%%%%%%%%%%%%%%%%%%%%%%%%%%%%

\subsubsection{Rotating substrate}

Following the ideas of Ref.\ \cite{Bradley_1996}, in Ref.\ \cite{Munoz_2005} we obtained the evolution equation for the surface in case the substrate is
rotating simultaneously while bombarding. This reads
\begin{equation}\label{eq.rotante}
 \partial_t h = - \nu_{\rm r} \nabla^2 h -
 {\mathcal K}_{\rm r} \nabla^4 h +\lambda^{(1)}_{\rm r} (\nabla h)^2 +
\lambda^{(2)}_{\rm r} \nabla^2(\nabla h)^2 +\lambda^{(3)}_{\rm r} \, \nabla
\cdot \left[ \left(\nabla^2 h\right) \nabla h \right],
\end{equation}
where a relationship between these parameters and those in Eq.\
(\ref{eq.ero}) exists.

Preliminar simulations of Eq.\ (\ref{eq.rotante}) \cite{Munoz_2007} show that this equation
also presents interrupted coarsening and an ordered array of dots, in which the
effect in the morphology evolution of the new term $\lambda^{(3)}_{\rm r}$  is similar to that of $\lambda^{(2)}_{\rm r}$.

\subsubsection{Comparison between continuum  models}

The continuum description of nanopattern formation by IBS is currently still
open and making progress. As seen in this Section, to date there are several
alternative descriptions available that share some of their predictions, while
differing in several other aspects. It is also true that theoretical work is
not even complete yet ---in terms of analyzing systematically the dependences
of morphological properties with physical parameters such as e.g.\ temperature
or ion energy--- due to two main reasons: (i) the main interfacial equations
are nonlinear, thus not easily amenable to analytical solutions, while some of
the physically interesting features such as the stationary pattern wavelength
depend crucially on non-linear effects; (ii) the parameter spaces of these
models are frequently large, particularly in the oblique-incidence case.
Actually, some of the interface equations that we have been considering seem to
be new in the wider contexts of Non-equilibrium Systems and Non-linear Dynamics
\cite{Munoz_2006b,Castro_2007,Cuerno_2007} even to the extent of providing
examples of thus far unknown behaviors for problems of high current interest,
such as coarsening phenomena \cite{Politi_2006}.

As a partial summary, and for the sake of comparison, we have collected in
table \ref{table_teor} the main available morphological predictions of the
models that we have considered in a larger detail in the text. After comparison
with the analogous experimental tables \ref{table_exp1} and \ref{table_exp2},
the two-field model leading to Eqs.\ (\ref{eq.ero}), (\ref{eq.ero.normal}) and (\ref{eq.rotante}) seems to
date the theoretical description that can account for a larger range of
experimental behaviors from within a single framework. Naturally, more work is
still needed in order to reach a still more complete description of these
physical systems with a larger quantitative predictive power.

\begin{center}
\begin{sidewaystable}\label{table_teor}
\begin{tabular}{|c|c|c|c|c|c|}\hline\hline

& 
$\begin{array}{c}
\textrm{BH} \\ \cite{Bradley_1988,Feix_2005}
 \end{array}$
& 
$\begin{array}{c}
\textrm{Anisotropic KS} \\ \cite{Cuerno_1995,Kim_2005,Makeev_1997,Makeev_2002}
 \end{array}$ 
& 
$\begin{array}{c}
\textrm{Damped KS} \\ 
 \cite{Facsko_2004,Vogel_2005,Vogel_2006} 
\end{array}$ 
& 

$\begin{array}{c}
\textrm{Nonlinear two-field model} \\ 
 \cite{Castro_2005,Munoz_2005}
\end{array}$ 

 \\\hline

$\lambda$ 
&   
$\begin{array}{c}
 \lambda(T)\sim T^{-1/2}\e^{-\Delta T/2K_BT}\\ \\
\lambda(\Phi) \sim \Phi^{-1/2}\\ \\
\lambda(t)\sim \textrm{const.} \\ \\
\lambda(E)\sim E^{-1/2}\\
\end{array}$
& 
$\begin{array}{c}
 \lambda(T)\sim \left\{\begin{array}{l} \textrm{const. (low $T$)}\\
T^{-1/2}\e^{-\Delta T/2K_BT}\\ \textrm{(high $T$)}
\end{array}
\right. \\

\lambda(\Phi)\sim \left\{\begin{array}{l} \textrm{const. (low $T$)}\\
\Phi^{-1/2} \textrm{(high $T$)}
\end{array}
\right. \\

\lambda(t)\sim \textrm{const.}\\

\lambda(E)\sim \left\{\begin{array}{l} E \textrm{ (low $T$)}\\
E^{-1/2} \textrm{(high $T$)}
\end{array}
\right.

\end{array}$

& 
$\lambda(t)\sim \textrm{const.} $
& 

$\begin{array}{c}
 \lambda(T)\sim \left\{\begin{array}{l} \textrm{const. (low $T$)}\\
T^{-1/2}\e^{-\Delta T/2K_BT} \\ \textrm{(high $T$)}
\end{array}
\right. \\

\lambda(\Phi)\sim \left\{\begin{array}{l} \textrm{const. (low $T$)}\\
\Phi^{-1/2} \textrm{(high $T$)}
\end{array}
\right. \\

\lambda(t)\sim \left\{\begin{array}{l} 
\textrm{const. (short }t) \\ 
t^n \textrm{ (intermediate } t)\\
 \textrm{const. (large }t)
\end{array}
\right. \\

\lambda(E)\sim \left\{\begin{array}{l} E \textrm{ (low $T$)}\\
E^{-1/2} \textrm{(high $T$)}
\end{array}
\right.

\end{array}$

\\\hline

$W$
&NM
&
$\begin{array}{c}

 W(t)\sim \left\{\begin{array}{l} 
 \e^{\omega t}\\ \textrm{ (short $t$})\\
 t^\beta \\ \textrm{(intermediate $t$)}\\
\textrm{const.} \\ \textrm{(large $t$)}\\
\end{array}
\right. \\
\end{array}$

&
$\begin{array}{c}

 W(t)\sim \left\{\begin{array}{l} 
 \e^{\omega t}\\ \textrm{ (short $t$})\\
 t^\beta \\ \textrm{(intermediate $t$)}\\
\textrm{const.} \\ \textrm{(large $t$)}\\
\end{array}
\right. \\
\end{array}$
&

$\begin{array}{c}

 W(t)\sim \left\{\begin{array}{l} 
 \e^{\omega t}\\ \textrm{ (short $t$})\\
 t^\beta \\ \textrm{(intermediate $t$)}\\
\textrm{const.} \\ \textrm{(large $t$)}\\
\end{array}
\right. \\
\end{array}$
\\\hline

In-plane order
&Disordered
&Disordered
&Hexagonal
&Hexagonal

\\\hline

$v$
&
Uniform
&
Non-uniform
&Uniform
&Non-uniform

\\\hline

Kinetic roughening
&NM
&Yes
&No
&Yes

\\\hline

\end{tabular}
\caption{Summary of main morphological predictions for some of the continuum models discussed in the text. NM stands for {\em Not measurable} because there is not a
well-defined
saturation value of the roughness. }
\end{sidewaystable}
\end{center}

\section{Applications of IBS patterned surfaces}

As mentioned in the Introduction, surface nanopatterning is the subject of
intense research driven by the current road to miniaturization and the broad
range of eventual applications. Therefore, the formation of self-organized
patterns on surfaces by IBS is relevant from a fundamental point of view but
also from a practical point of view. Whereas the understanding of the
patterning process has been studied deeply from the theoretical and
experimental points of view, the applications of these patters have been
scarcely addressed and a further technological effort is required for their
implementation in real devices. Here, we summarize some of the applications
launched in the literature.

One of the strengths of the IBS nanopatterning process comes from its
universality since, as shown in section 3, it is applicable to metallic,
semiconducting or insulating surfaces. However, in comparison with other
techniques, the major advantages of this technique rely on its simplicity, high
output (fast) and scalable patterned area. Among the pattern characteristics,
we can highlight the tunable wavelength and presence of ordering.

The first applications of IBS patterns were based on the production of
micro-scale ripple patterns. In this case, the usage of these patterns for
optical interference gratings was addressed \cite{Johnson_1979}. Recently,
optical activation of the surface due to the pattern formation has also been
reported \cite{Mussi_2006} which, apart from the surface morphology, may be
also linked directly to the ion induced defects.

Perhaps the most important and direct application of IBS patterning relies with
the method based on the deposition of a given material layer on top of the
target material intended to be patterned. This layer is then ion sputtered,
this process inducing a nanopattern on top of it that, by continuing the
sputtering process, is eventually transferred onto the target substrate. This
strategy was previously used to generate a Ni dot pattern with a typical
wavelength of 250-300 nm and dot height of 13-15 nm \cite{Seoo_2004}. In this
case, the top-layer was a copolymer with self-organized polymer domains. These
domains did show different sputtering rates when ion sputtered. In this case,
these Ni nanopatterns were used to separate DNA molecules.

\begin{figure}
\center
\includegraphics[height=4cm]{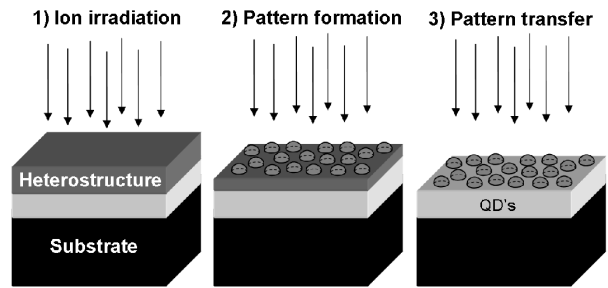}
\caption{Sketch of the fabrication of quantum dot arrays by transferring the
self-organized nanodot pattern induced by IBS to the interface of a
semiconductor heterostructure (adapted from Ref. \cite{Facsko_2000}).}
\label{Fig_5-1} \end{figure}

The interest on IBS patterns was greatly increased by the report of nanodot
pattern formation by Facsko et al. \cite{Facsko_1999} as the distances involved
were clearly in the nanometer range. Therefore, the potential application for
quantum dot (QD) arrays fabrication \cite{Facsko_2000} triggered further
research. This route for QD fabrication is shown schematically in Fig.
\ref{Fig_5-1} and was experimentally proved by the successful photoemission of
confined nanostructures \cite{Facsko_1999}. Despite this result, there are
issues to be improved for the final applicability, such as the crystallinity of
the nanostructures, since ion bombardment induces considerable amorphization on
the surface. This constraint may not play a big role if the nanodot volume is
mostly crystalline. In any case, re-crystallization or damage reduction may be
attained by post-annealing treatments.

\begin{figure}
\center
\includegraphics[height=4cm]{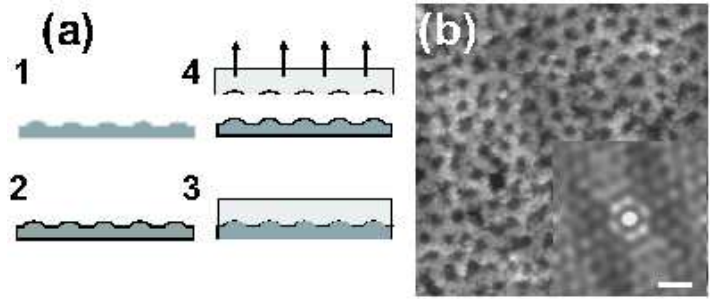}
\caption{(a) Replication method for IBS nanodot pattern transfer consisting in
4 steps: (1) production of the IBS pattern on Si surfaces; (2)
functionalization of the nanostructured Si surface by a OTS monolayer; (3) film
deposition; (4) mechanical film detachment. (b) $750\times 750$ nm$^{2}$ AFM
image of a DLC film surface molded following the method described in (a).
Inset: Two-dimensional auto-correlation corresponding to an area of $545\times
545$ nm$^{2}$. The horizontal bar corresponds to 63 nm.} \label{Fig_5-2}
\end{figure}

The IBS patterns can be also used as molds or masters to transfer or replicate
the pattern to other surfaces with high functionality. This several-step
process may be required when the surface can not be processed directly, such as
the case of polymer materials, or if the desired pattern can not be obtained in
a simple way (for example, when the diffusive regime is operating in metals).
Also, this imprinting path may be interesting for reducing production costs by
faster processing or avoiding the use of vacuum environments. The applicability
of IBS patterned surfaces for molding and replication has been shown by
Azzaroni et al. \cite{Azzaroni_2003} in polymer materials but the method has
been successfully extended to metals \cite{Azzaroni_2004b} and even hard
ceramic materials \cite{Auger_2005}. The replication method is shown
schematically in Fig. \ref{Fig_5-2}a and consists of several sequential steps:
(1) production of the IBS dot pattern on a silicon surface; (2) formation of a
self-assembled monolayer of octadecyltrichlorosilane (OTS) on top of the
nanostructured surface; (3) subsequent growth of the material to be patterned.
This step has to be performed at mild conditions (i.e, low temperatures or
under non-high energetic sputtering conditions) in order to preserve the OTS
layer; (4) thanks to the antisticking properties of the OTS layer the film
deposited on top of it can be easily mechanically detached. Also, due to the
extremely low thickness of the OTS layer the film surface in contact with it
follows the surface morphology of the underlying nanostructured surface. Thus,
the surface morphology of the detached film is the {\itshape negative} of the
IBS nanostructured silicon surface. In Fig. \ref{Fig_5-2}b we show the molded
surface of a diamond-like carbon (DLC) film grown by radio frequency sputtering
following this method. The {\itshape negative} hole structure of the IBS dot
surface is observed while the two-dimensional auto-correlation function (inset)
shows that the short-range hexagonal order is still preserved.

\begin{figure}
\center
\includegraphics[height=3.2cm]{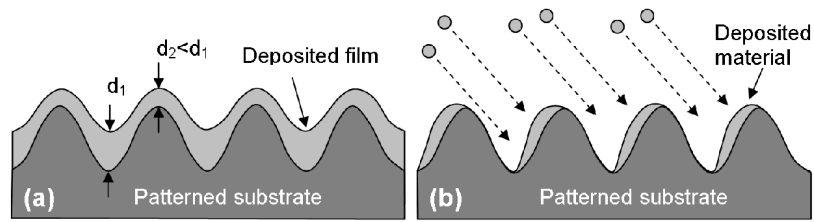}
\caption{Proposals for the formation of magnetic nanostructures using IBS
patterned surfaces as growth templates: (a) non-conformal growth (the different
thickness result in patterned magnetic domains) and (b) shadowed deposition.}
\label{Fig_5-3} \end{figure}

Apart from the direct exploitation of the properties induced in nanopatterned
surfaces, much effort has also been devoted to the use of these surfaces as
{\itshape templates} for further processing. For example, ripple patterns have
been tested as substrates for alignment and manipulation of carbon nanotubes
\cite{Granone_2005}. In this sense, one possible application would be to extend
FIB patterning as a template technique for patterned growth of carbon nanotubes
\cite{Chen_2007} to IBS nanopatterning. Also, the large range of surface
roughness and characteristic wavelengths that are attainable by IBS patterns
open a broad field of bioapplications. For example, surface adhesion of
biological entities (proteins, DNA, $\ldots$) may be controlled for sensing,
diagnostic or biocompatiability enhancement. Moreover, the nanopit patterns
could find applications as supports for catalysis \cite{Schildenberger_2000}.
Another potential application of IBS patterns is for the production of magnetic
nanostructures \cite{Terris_2005}. Moroni and coworkers proved that the
formation of nanoscale ripples by IBS on Co films deeply affect their magnetic
properties \cite{Moroni_2003}. Also, IBS patterns have been used as {\itshape
growth templates} for the production of magnetic nanostructures. For example,
it is well known that the magnetic properties of a superlattice formed by
magnetic materials depend on the final thickness of the stack. Therefore,
irregularities in the thickness due to non conformal growth, i.e. preferential
growth on the cavities (ripples or holes) or on the top of the hills, onto a
patterned surface can be used to define magnetic domains \cite{Chen_2002}.
Another proposal is the use of the surface topography to induce shadowed
deposition under oblique precursors incidence \cite{Teichert_2003}. These
routes are displayed schematically in Fig. \ref{Fig_5-3}.

\begin{figure}
\center
\includegraphics[height=2.8cm]{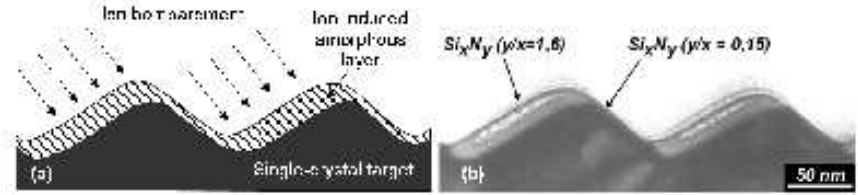}
\caption{(a) Formation of a ripple pattern and inherent modulation in the
amorphized layer as a result of ion bombardment at off-normal incidence. (b)
Cross-sectional TEM image of a ripple pattern produced by N$^{x}$ ion
bombardment of silicon and the formation of different SiN$_{x}$ regions that
are used as implantation barriers. Figure reprinted with permission from
\cite{Smirnov_2003}.} \label{Fig_5-4} \end{figure}

Obviously, the use of patterned surfaces for many applications is not
restricted to IBS patterns. However, the presence of the, mostly undesirable,
amorphous layer is a peculiarity of IBS patterns that can be used for further
applications. For example, the modulation of the amorphous layer in
medium-energy induced ripples was very successfully exploited by Smirnov et al.
\cite{Smirnov_2003} as selective doping implantation barrier. In this case, the
production of ripples on silicon by nitrogen erosion leads to the formation of
amorphous silicon nitride (SiN$_{x}$) regions, as shown in Fig. \ref{Fig_5-4}.
The different nitride layer thicknesses can be used to obtain silicon regions
with different degree of doping through, for instance, reactive ion-etching and
As$^+$ implantation processes.

\section{Open issues}\label{Open_issues}

Despite the large amount of theoretical and experimental work described above,
there are still issues that remain unclear and require further research
efforts. From the experimental point of view, this is especially evident for
the case of nanodot patterns due to their relative novelty. For the case of
nanoripples, in spite of the abundant avaliable data, there are also gaps to
be filled. Perhaps, the least systematically investigated aspects concern
transverse ripple propagation and the relevance of shadowing effects on the
morphological evolution. The former is difficult to address due to its evident
experimental complexity since in-situ, real time monitoring of the ripple
morphology has to be carried out. The second issue, although quite specific, is
interesting because it can provide important data for theory refinement, which
requires exploring systematically the continuum models beyond the customary
small-slope approximation. With regard to nanodot patterns, there is still a
certain lack of systematicity in the studies of the various targets that have
proved amenable to such patterns when subject to IBS. For instance, the role of
preferential sputtering on nanostructuring of hetero-semiconductors has not
been unambiguously assessed yet. In addition, the effect of temperature on the
nanodot morphology or the symmetry of their arrangement is not clear as
different, and in some cases opposite, behaviors have been found. Another issue
that will be interesting to address will be the systematic comparison of the
nanodot pattern properties for the same target material bombarded under normal
incidence and under oblique incidence with simultaneous target rotation. In
this sense, comparison is still incomplete between the IBS patterns produced
with the same equipment on different materials. A related open issue is the
influence of the characteristics of the ion gun optics on the pattern
properties.

Another line of research with important implications in the theoretical
understanding of IBS patterning is that of pattern coarsening. From the data
compiled, it is clear that many systems present coarsening but also there are
others that do not. Thus, an important experimental investigation would be the
study of the physical processes behind pattern coarsening. Once more, the
results of these investigations would contribute to improving the theoretical
models and hence deepen our understanding of IBS patterning formation. In this
connection, an ambitious goal is to clarify the relationship between the more
relevant parameters appearing in the continuum models with the specific
properties of the target materials and ion species. This research would indeed
help to understand quantitatively the different patterns properties found
experimentally, such as for instance the different size and shape of dots found
in Si, GaSb and InP. Moreover, it would be interesting to design specific
experiments in order to verify alternative predictions from the various models.
An example of this can be the experimental verification of recent proposals to
produce novel nanopatterns under specific multi-beam irradiation conditions
\cite{Carter_2004, Carter_2005, Carter_2006}.

Finally, from a more applied point of view, the use of IBS nanostructuring
processes for technological applications remains as an important challenge
since, although several specific applications have been already developed, they
have not yet proved clear advantages with respect to previously existing
alternative routes.

\section*{Acknowledgements}

We are pleased to acknowledge collaborations and exchange with a number of
colleagues, in particular JM Albella, MC Ballesteros, A-L Barab\'asi, M Camero,
T Chini, M Feix, AK Hartmann, R Kree, M Makeev, TH Metzger, O Plantevin, M
Varela and EO Yewande. We would like to thank specially F Alonso for his help
in the ripple experiments on Si at 40 keV shown in Sec.\ \ref{sec:experiments}.

Our work has been partially supported by Spanish grants Nos.\ FIS2006-12253-C06
(-01, -02, -03, -06) from Ministerio de Educaci\'on y Ciencia (MEC),
CCG06-UAM/MAT-0040 from Comunidad Aut\'onoma de Madrid (CAM) and Universidad
Aut\'onoma de Madrid, UC3M-FI-05-007 and CCG06-UC3M/ESP-0668 from CAM and
Universidad Carlos III de Madrid, and S-0505/ESP-0158 from CAM, and by the
Ram\'{o}n y Cajal programme of MEC (RG).

% BibTeX users please use
% \bibliographystyle{}
% \bibliography{}
%
% Non-BibTeX users please follow the syntax
% the syntax of "referenc.tex" for your own citations
%%%%%%%%%%%%%%%%%%%%%%%% referenc.tex %%%%%%%%%%%%%%%%%%%%%%%%%%%%%%
% sample references
% "physics"
%
% Use this file as a template for your own input.
%
%%%%%%%%%%%%%%%%%%%%%%%% Springer-Verlag %%%%%%%%%%%%%%%%%%%%%%%%%%

%
% BibTeX users please use
% \bibliographystyle{}
% \bibliography{}

%%%%%%%%%%%%%%%%%%%%%%%%%%%%%%%%%%%%%%%%%%%%%%%%%%%%%%%%%%%%%%%%%%%%%%  }

%%%%%%%%%%%%%%%%%%%%%%%%%%%%%%%%%%%%%%%%%%%%%%%%%%%%%%%%%%%%%%%%%%%%%%

\printindex
\end{document}